%% file: thesis.tex
\documentclass[
a4paper,
12pt,
bibtotoc, % Include bibliography in contents
liststotoc, % Include lists of figures and tables in contents
idxtotoc, % Include index in contents
onehalfspacing % onehalfspacing or doublespacing
]
{icldt}
\pdfoutput=1
\usepackage[%CVSon, % CVS version in footers
            hyperref, % Use hyperref for links
            ]{ihformat}
\usepackage{bm}            
\def\be{\begin{equation}}
\def\ee{\end{equation}}
\def\beb{\begin{equation*}}
\def\eeb{\end{equation*}}
\def\bea{\begin{eqnarray}}
\def\eea{\end{eqnarray}}
\def\beab{\begin{eqnarray*}}
\def\eeab{\end{eqnarray*}}
\def\nn{\nonumber}
\def \ni {\noindent}

\def\p{\partial}

\def\P{{{\cal{P}}}}
\def\X{{{\cal{X}}}}
\def\X{{{\cal{X}}}}
\def\XB{{{\cal{X}}_{\rm B}}}
\def\S{{\cal{S}}}
\def\T{{\mathfrak{T}}}
\def\H{{\cal H}}
\def\B{{\cal{B}}}
\def\Z{{{\cal{Z}}}}
\def\vp{{{\varphi}}}
\def\dvp{{\delta\varphi}}
\def\w{{\omega}}
\def\cs2{c_{\rm{s}}^2}
\def\dij{\delta_{ij}}

\def \beg {\begin{enumerate}}
\def \en {\end{enumerate}}
\def\fg{{\rm{flat}}}
\def\vpb{\varphi_0}
\def\Pb{P_0}
\def\rhob{\rho_0}
\def\Xb{X_0}
\def\dPn{{\delta P_{\rm{nad}}}}

\def\cph{c_{\rm{ph}}^2}
\def\cs{c_{\rm{s}}^2}
\newcommand\eq[1]{Eq.~(\ref{#1})}
\newcommand\eqs[1]{Eqs.~(\ref{#1})}
\newcommand{\dvph[1]}{\delta\vp_{#1}}

\newcommand{\dU[1]}{{\delta U_{#1}}}

\def\syn{{\rm{syn}}}
\def\com{{\rm{com}}}
\def\fg{{\rm{flat}}}
\def\lg{{\ell}}
\def\udg{{\delta\rho}}
\def\wt{\widetilde}
\def\dT{{\delta{\bf T}_1}}
\def\dTT{{\delta{\bf T}_2}}
\def\drho{{\delta\rho_1}}
\def\drhorho{{\delta\rho_2}}
\def\drhorhorho{{\delta\rho_3}}
\def\dP{{\delta P_1}}
\def\dPP{{\delta P_2}}
\def\dPPP{{\delta P_3}}
\def\ipsi{{\psi_1}}
\def\iipsi{{\psi_2}}
\def\E{{E_1}}
\def\EE{{E_2}}
\def\iphi{{\phi_1}}
\def\iiphi{{\phi_2}}
\def\B{{B_1}}
\def\BB{{B_2}}

\def\dm{{\delta_{\rm c}}}

\def\ep{\epsilon}
\newcommand\gami[1]{{\gamma_{{#1}}^{~i}}}
\newcommand\gamk[1]{{\gamma_{{#1}}^{~k}}}

\begin{document}

% Title page is input not included to remove extra page.
\input{title}

% % % % % % % % % % % % % % % % % % % % % % % % % 
% Abstract
\include{abstract}
% % % % % % % % % % % % % % % % % % % % % % % % %

% % % % % % % % % % % % % % % % % % % % % % % % % 
% Acknowledgements
\include{acknowledgements}

% % % % % % % % % % % % % % % % % % % % % % % % % 

% % % % % % % % % % % % % % % % % % % % % % % % % 
% Table of contents and figures
% % % % % % % % % % % % % % % % % % % % % % % % % 
%\setcounter{tocdepth}{0}
\tableofcontents
\listoffigures
\listoftables
% % % % % % % % % % % % % % % % % % % % % % % % % 

% Set one half spacing now that bulk of the thesis is starting.
\onehalfspacing
%\doublespacing
%\singlespacing
% % % % % % % % % % % % % % % % % % % % % % % % % 

% Include files for each chapter here using 
% 
% Example
\include{intro}

\include{perturbations}

\include{dynamics}
\include{vorticity}

\include{third}

\include{conclusions}

\begin{appendix}
\begin{onehalfspace}
\include{secondij}
\include{thirdeins}
\end{onehalfspace}
\end{appendix}

% % % % % % % % % % % % % % % % % % % % % % % % % 

% If you want to list all the todos that you have put in the document (using
% \addtodo{}) then uncomment the next line:
 %\listoftodos

% % % % % % % % % % % % % % % % % % % % % % % % % % 

% Start single space again for bibliography
\begin{singlespace}
% % % % % % % % % % % % % % % % % % % % % % % % % 

% Bibliography
% Put your bibliography file here
\bibliography{thesis}
% 
% This bibtex style file puts entries in alphabetical order and treats arxiv
% references correctly.
\bibliographystyle{utphys-ih}
% % % % % % % % % % % % % % % % % % % % % % % % % 

\end{singlespace}

% % % % % % % % % % % % % % % % % % % % % % % % % 
\end{document}

%% file: title.tex
% % % % % % % % % % % % % % % % % % % % % % % 
% title.tex - AJC
% % % % % % % % % % % % % % % % % % % % % % % % 
% Redefine CVSRevision for this section
%\renewcommand{\CVSrevision}{\version$Id: title.tex,v 1.1 2009/12/17 17:33:39 ith Exp $}
% 
\college{Queen Mary, }
\department{Mathematical Sciences}
 \supervisor{K.~A.~Malik}
\title{Applications of Cosmological Perturbation Theory}
\author{Adam~J.~Christopherson}
% 
% 
%\hypersetup{pdftitle={Your Title},pdfauthor={Your Name}}
% 

% Change as appropriate
\declaration{%
I hereby certify that this thesis, which is approximately 35,000 words in length,
has been written by me; that it is the record of the work carried out by me at the
Astronomy Unit, Queen Mary, University of London, and that it has not been submitted
in any previous application for a higher degree.
 Parts of this work have been completed
in collaboration with Karim A. Malik and David R. Matravers, and are published in the following
papers:
\begin{itemize}
\item{
{{ A.~J.~Christopherson}} and K.~A.~Malik:
\medskip\\
%\emph{``The non-adiabatic pressure in general scalar field systems,''\\}
 Phys. Lett. B {\bf{675}} (2009) pp. 159-163,
%arXiv:0809.3518 [astro-ph]. 
}
\item{{{A.~J.~Christopherson}}, K.~A.~Malik and D.~R.~Matravers:
\medskip\\
%\emph{``Vorticity generation at second order in cosmological perturbation theory,''\\}
Phys. Rev. D {\bf{79}} 123523 (2009),
%arXiv:0904.0940 [astro-ph.CO]. 
}
\item{{{A.~J.~Christopherson}} and K.~A.~Malik:
\medskip
\\
%\emph{``Practical tools for third order cosmological perturbations,''\\}
JCAP {\bf{11}} (2009) 012,
%arXiv:0909.0942 [astro-ph.CO].
}
\item{{{ A.~J.~Christopherson}}:
\medskip
\\
%\emph{``Gauge conditions in combined dark energy and dark matter systems,''\\}
%JCAP {\bf{11}} (2009) 012,
Phys. Rev. D {\bf{82}} 083515 (2010),
%arXiv:1008.0811 [astro-ph.CO].
}
\item{{{ A.~J.~Christopherson}}, K.~A.~Malik and D.~R.~Matravers:
\medskip\\
%\emph{``Estimating the amount of vorticity generated by cosmological perturbations in the early universe,''\\}
%Phys. Rev. D {\bf{79}} 123523 (2009),
Phys. Rev. D (in press)
}
\item{{{ A.~J.~Christopherson}} and K.~A.~Malik:
\medskip\\
%\emph{``Can Cosmological Perturbations Produce Early Universe Vorticity?''\\}
%Phys. Rev. D {\bf{79}} 123523 (2009),
Invited article, Class. Quantum Grav. {\bf 28} (2011) 114004
%arXiv:1010.4885 [gr-qc].
}
\end{itemize}
I have made a major contribution to all the original research presented in this
thesis.
\vspace{-1cm}
}

\maketitle

%% file: abstract.tex
% % % % % % % % % % % % % % % % % % % % % % % % % 
% abstract.tex - Ian Huston
% $Id: abstract.tex,v 1.2 2009/12/17 17:41:41 ith Exp $
% % % % % % % % % % % % % % % % % % % % % % % % % 
% Redefine CVSRevision for this section
\renewcommand{\CVSrevision}{\version$Id: abstract.tex,v 1.2 2009/12/17 17:41:41 ith Exp $}
\chapter*{Abstract}
\label{ch:abstract}
\addcontentsline{toc}{chapter}{Abstract}
\section*{}
\singlespacing
Cosmological perturbation theory is crucial for our understanding of the universe. The linear theory has been well
understood for some time, however developing and applying the theory beyond linear order is currently at the
forefront of research in theoretical cosmology. 

This thesis studies the applications of perturbation theory to
cosmology and, specifically, to the early universe. Starting with some background material introducing the 
well-tested
 `standard model' of cosmology, we move on to develop the formalism for perturbation theory up to second 
order giving evolution equations for all types of scalar, vector and tensor perturbations, both in gauge dependent and 
gauge invariant form. We then move on to the main result of the thesis, showing that, at second order in perturbation 
theory, vorticity is sourced by a coupling term quadratic in energy density and entropy perturbations. This source 
term implies a qualitative difference to linear order. Thus, while at linear order vorticity decays with the expansion of 
the universe, the same is not true at higher orders. This will have important implications on future measurements of 
the polarisation of the Cosmic Microwave Background, and could give rise to the generation of a primordial seed 
magnetic field. Having derived this qualitative result, we then estimate the scale dependence and magnitude of the 
 vorticity power spectrum, finding, 
for simple power law inputs a small, blue spectrum.

 The final part of this thesis concerns 
higher order perturbation theory, deriving, for the first time, the metric tensor, gauge transformation rules and 
governing equations for fully general third order perturbations. We close with a discussion of natural extensions to 
this work and other possible ideas for off-shooting projects in this continually growing field.

%% file: acknowledgements.tex
% % % % % % % % % % % % % % % % % % % % % % 
% acknowledgements.tex - AJC
% $Id: acknowledgements.tex,v 1.2 2009/12/17 17:41:41 ith Exp $
% % % % % % % % % % % % % % % % % % % % % % 
% Redefine CVSRevision for this section
%\renewcommand{\CVSrevision}{\version$Id: acknowledgements.tex,v 1.2 2009/12/17 17:41:41 ith Exp $}
% 
% 
\chapter*{Acknowledgements}
\label{ch:acknowledgements}
\addcontentsline{toc}{chapter}{Acknowledgements}
I would like to thank Karim Malik for all his help, support and guidance, without
which this truly would not have been possible.
I am very grateful to David Matravers for his valuable input into
an enjoyable collaboration. 
I would like to thank everyone in the cosmology group at Queen Mary, in particular Ian Huston, James Lidsey and Reza Tavakol,
and my fellow students of 301, both past and present. \\

Finally, I would like to thank my parents for their encouragement and Christine
for her constant support, helping me remember that there exists a world outside of science!

\vfill
This work was funded by the Science and Technology Facilities Council (STFC), 
and by the Astronomy Unit and School of Mathematical Sciences at 
Queen Mary, University of London.

\chapter*{}
{\begin{center}
To the loving memory of Ted and Iris.
\end{center}
}

%% file: intro.tex
% % % % % % % % % % % % % % % % % % % % % % % % % % % % 
% intro.tex - AJC
% Introduction
% % % % % % % % % % % % % % % % % % % % % % % % % % % % 
% Redefine CVSRevision for this section. 
% If you don't want to use CVS tags comment out this line
%\renewcommand{\CVSrevision}{\version$Id: chapter.tex,v 1.3 2009/12/17 18:16:48 ith Exp $}

% % % % % % % % % % % % % % % % % % % % % % % % % % % % % % % % 
% =========================================================== %
% % % % % % % % % % % % % % % % % % % % % % % % % % % % % % % % 
\chapter{Introduction}
\label{ch:intro}
% % % % % % % % % % % % % % % % % % % % % % % % % % % % % % % % 
% =========================================================== %
% % % % % % % % % % % % % % % % % % % % % % % % % % % % % % % % 

Over the last few decades cosmology has moved from a mainly theoretical discipline to one 
in which data is of increasing importance. This is promising, since it means that we are no longer
confined to the theorists' playground, but instead have observational data with which to constrain
our models. 

At present the main observable that we have with which to test our theories is the Cosmic Microwave Background (CMB)
radiation. This is radiation that was produced when the universe was around 380,000 years old and had
cooled enough to allow electrons and protons to combine to produce Hydrogen atoms.
 Its first detection in 1964 by Penzias and Wilson was hailed as a firm success of the hot Big Bang
cosmological model, and experiments have been performed in the years after in order to obtain
more details of this radiation. The Cosmic Background Explorer ({\sc COBE}) \cite{COBE}
 and the Wilkinson Microwave Anisotropy
Probe ({\sc WMAP}) \cite{WMAP7} satellites have since probed the anisotropies
of the CMB, finding that it is extremely isotropic (up to one part in 100,000), and the Planck \cite{Planck}
satellite is currently taking data to further increase our wealth of data on the CMB. This observation
of small anisotropies is very much 
in agreement with the theory, which states that quantum perturbations in the field driving inflation
produce small primordial density fluctuations which are then
amplified through gravitational instability to form the structure that exists in the universe today.

In order to study the theoretical framework of the standard cosmological model (see, e.g. Ref.~\cite{Malik:2010zz}, for
 a particularly lucid review),
 one uses cosmological perturbation theory,
which is the main topic of this thesis. The basic idea is quite simple: we model the universe
as a homogeneous `background' which has inhomogeneous perturbations on top. The perturbations
can then be split up order-by-order, with each order being smaller than the one before. 

Early studies of the linear
order theory were mainly done by the following authors. Lifshitz pioneered the early work on perturbations
in Ref.~\cite{Lifshitz:1945du}, which was later extended in Ref.~\cite{Lifshitz:1963ps}, with Bonnor 
considering density perturbations in Ref.~\cite{Bonnor}. This early work by Lifshitz was conducted
in the synchronous gauge, which has since been shown to exhibit gauge artefacts if one is not careful \cite{Press:1980is}. 
Though the authors used intricate 
geometrical arguments in order to remove these unphysical modes, the fact that the calculations cannot
be done easily was far from ideal.
Thus, Hawking \cite{Hawking:1966qi} and Olson \cite{Olson:1976jc} took a different approach to this 
problem and attempted the first fully covariant study of cosmological perturbations, focussing not on
perturbations of the metric tensor, but instead on perturbations of the curvature tensor.
However, the most pioneering work in modern cosmological perturbation theory was completed by Bardeen. In 
Ref.~\cite{Bardeen:1980kt}, Bardeen presented a systematic method for removing the gauge artefacts
by constructing gauge invariant variables. His work focussed on the two metric potentials $\Psi$ and $\Phi$,
which correspond to the two gauge invariant scalar metric perturbations in the longitudinal gauge.
 This work was then followed by the two review
articles by Kodama and Sasaki~\cite{ks} and Mukhanov, Feldman and Brandenberger~\cite{mfb}. These three
articles together arguably form the basis of linear metric cosmological perturbation theory.

An alternative approach to metric perturbation theory is often called the covariant approach. The approach
defines gauge invariant variables using the Stewart-Walker lemma (a perturbation which vanishes in the 
background is gauge invariant \cite{Stewart:1974uz}) and was mostly pioneered by Ellis and collaborators
\cite{Ellis:1989ju, Ellis:1990gi, Ellis:1989jt}. An interesting paper made a first step towards 
relating the covariant approach to
the metric approach which was written by Bruni {\it et~al.}~\cite{Bruni:1992dg}.

Cosmological perturbations can be decomposed into scalar, vector and tensor perturbations, as we will show in
Chapter \ref{ch:perturbations} and, at linear order, the three types of perturbations  decouple from one 
another. The scalar modes are related to density perturbations, vector modes are vortical or 
rotational perturbations, and the tensor modes are  related to gravitational waves. The CMB radiation is polarised, and
the different types of perturbations induce different polarisations. Scalar modes produce only E-mode (or
curl-free) polarisation and tensor modes produce only B-mode (divergence-free) polarisation. Vectors 
produce both, but are usually deemed
negligible, since any produced in the early universe will be inflated away, or will decay
with the expansion of the universe \cite{ks}.

However, once we go beyond linear order, different types of perturbation no longer decouple and so, for example,
vector and tensor perturbations are sourced by scalar modes. This mathematical difference between linear
and higher orders therefore plays an important role in the theory and can result in qualitatively 
different physics beyond linear order which will, in turn, generate different observational signatures. This is,
really, the main reason for extending perturbation theory beyond linear order, which has been studied
by many authors in the last few years 
\cite{Acquaviva:2002ud, Bruni:1996im, Noh:2004bc, Malik2004, Malik:2005cy, Mukhanov1997, 
Nakamura:2003wk, Tomita, Bartolo:2004if, Nakamura:2010yg, Christopherson:2011hn} 
(see Ref.~\cite{MW2008} 
for a recent review and a 
comprehensive list of references on second order cosmological perturbations).
Perturbation theory beyond linear order is the central theme of this thesis.
\\

This thesis is organised as follows. In the remainder of this introductory Chapter we present
the standard model of cosmology in a more detailed sense, briefly introducing inflationary
cosmology, and restating our notation that will be used for the remainder of this thesis
at the end of the Chapter. In Chapter~\ref{ch:perturbations} we introduce the theory
of non-linear cosmological perturbations up to second order. We present the perturbed metric tensor,
and energy momentum tensor for a perfect fluid (i.e. a fluid with no anisotropic stress)
and a scalar field. We consider next the transformation
behaviour of the different perturbations under a gauge transformation, using these to choose gauges
and define gauge invariant variables. Next, we present a discussion
of the thermodynamics of a perfect fluid, discussing the pressure and energy density perturbations 
and the definition of the non-adiabatic pressure perturbation,
which will play a central role in the following work. Finally, to close Chapter \ref{ch:perturbations},
we briefly consider how non-adiabatic pressure perturbations can arise naturally in multiple fluid or 
multi-field inflationary models.

In Chapter~\ref{ch:dynamics} we continue presenting the foundations of cosmological perturbation
theory and present the dynamic and constraint equations up to second order. Starting with the linear order
theory we present the governing equations for scalar, vector and tensor perturbations of a perfect fluid
without fixing a gauge. We then fix a gauge, giving the equations in terms of gauge invariant variables for 
three different gauges: the uniform density, uniform curvature and longitudinal gauges, solving
the equations for the latter two. We then present the Klein-Gordon equation for a scalar field, and highlight the 
important difference between the adiabatic sound speed and the speed with which perturbations travel
for a scalar field system. Finally, we investigate the perturbations of a system containing both dark 
energy and dark matter. Having laid the foundations with the linear theory, we move on to the second
order theory, presenting the governing equations for a perfect fluid coming form energy momentum conservation
and the Einstein equations in gauge dependent form. We then present all equations in the uniform curvature gauge
which we will use in Chapter~\ref{ch:vorticity}, including now only the canonical Klein-Gordon equation. Finally,
in order to connect with the literature, we give the equations for scalars in the Poisson gauge.

Having now developed the tools for second order perturbation theory, in Chapter~\ref{ch:vorticity} we use
the qualitative differences between the linear theory and higher order theory to show that, at second order
in perturbation theory, vorticity is sourced by a coupling between first order energy density and
entropy perturbations. This is analogous to the case of classical fluid mechanics and generalises
Crocco's theorem to an expanding background. To show this, we start by defining the vorticity tensor
in general relativity, and then calculate the vorticity tensor at linear and second order using the fluid
four velocity and the metric tensor defined in Chapter~\ref{ch:perturbations}.  We then compute the evolution
equation for the vorticity, making use of the governing equations in Chapter~\ref{ch:dynamics}. At linear
order vorticity is not sourced, however there exists a non-zero source term at second order when allowing
for fluid with a general equation of state that depends upon both the energy and the entropy. Having derived
this qualitative result, we then give a first quantitative solution, estimating the magnitude and
scale dependence of the induced vorticity using simple input power spectra: the energy density derived
in Chapter~\ref{ch:dynamics}, and using a simple ansatz for the non-adiabatic pressure perturbation. 

In Chapter \ref{ch:third} we extend the formalism from the second order theory to third order, presenting the
gauge transformation rules and constructing gauge invariant variables. Then, considering perfect fluids 
and including all types of perturbation, we present the energy and momentum conservation equations
and give components of the Einstein tensor up to third order. We also give the Klein-Gordon equation for a scalar
field minimally coupled to gravity. 
Finally, to close, we conclude in Chapter \ref{ch:conc} and present possible directions in which
one can extend the work presented in this thesis.

% % % % % % % % % % % % % % % % % % % % % % % % % % % % % % % % 
\section{Standard Cosmology}
\label{sec:standardcosmology}
% % % % % % % % % % % % % % % % % % % % % % % % % % % % % % % % 
% =========================================================== %
% % % % % % % % % % % % % % % % % % % % % % % % % % % % % % % % 

We now introduce some elements of
standard cosmology in a more quantitative sense and, in doing so,
define our notation. The basic
starting point in cosmology is the cosmological principle which states that,
on large enough scales, the universe is  both isotropic and homogeneous. In general 
relativity, geometry is encoded in the metric tensor $g_{\mu\nu}$, or the line element 
$ds^2=g_{\mu\nu}dx^\mu dx^\nu$. The general
line element for an isotropic and homogeneous spacetime, and thus one which
obeys the cosmological principle, takes the form\footnote{Throughout this thesis we use
the so-called `East coast' metric signature $(-+++)$ and the positive $(+++)$ sign convention
in the notation of Misner {\it et al.} \cite{Misner:1974qy}.} \cite{llbook}

\be 
ds^2=-dt^2+a^2(t)\Bigg[\frac{dr^2}{1-Kr^2}+r^2\Big(d\theta^2+\sin^2\theta d\chi^2\Big)\Bigg]\,,
\ee
in spherical coordinates $(t; r,\theta,\chi)$, where $t$ denotes the coordinate time,
 $a(t)$ is a function that depends only on time, and $K$ denotes the global 
curvature of the spatial slices, where $K=1$ denotes a positively curved, or closed, universe,
$K=0$ a flat universe, and $K=-1$ a negatively curved, or open, universe. This is 
the Friedmann-Robertson-Walker (FRW) metric. An alternative
way of representing the FRW metric is 
\be 
ds^2=-dt^2+a^2(t)\gamma_{ij}dx^i dx^j\,,
\ee 
where $\gamma_{ij}$ is the metric tensor on spatial hypersurfaces. 

Since current observations are consistent with a flat, $K=0$ universe, which is also in agreement with inflation, we adopt this
choice henceforth and so write the line element as
\be 
\label{eq:FRWline}
ds^2=-dt^2+a^2(t)\delta_{ij}dx^idx^j\,,
\ee 
where $\delta_{ij}$ denotes the Kronecker delta.

We note here the importance of the function $a(t)$, called the scale factor, in 
an expanding spacetime. We can picture space as a coordinate grid which
expands uniformly with the increase of time. The comoving distance, ${\bm x}$ between
two points, which is just measured by the comoving coordinates, remains 
constant as the universe expands. The physical distance, ${\bm r}$, is proportional to the 
scale factor,
\be 
{\bm r}=a(t){\bm x}\,,
\ee
and so does evolve with time. Thus, an isotropic and homogeneous universe is
characterised not only by its geometry, but also by the evolution of the scale
factor \cite{Dodelson:2003ft}. In order to quantify the expansion rate, we introduce
the Hubble parameter, $H(t)$ defined as
\be 
H(t)=\frac{da/dt}{a}=\frac{\dot{a}}{a}\,,
\ee
where an overdot denotes a derivative with respect to coordinate time, which measures
how rapidly the scale factor changes. 

It is often convenient to use, instead of $t$, the conformal time coordinate $\eta$ defined 
through
\be 
\eta=\int_\infty^t \frac{dt}{a}\,,
\ee
in terms of which the line element (\ref{eq:FRWline}) becomes
\be 
\label{eq:FRWlineeta}
ds^2=a^2(\eta)\Big[-d\eta^2+\delta_{ij}dx^idx^j\Big]\,.
\ee 
In doing this, we have increased the spatial coordinate grid introduced above
to a coordinate grid over the entire spacetime.
We can furthermore define the conformal Hubble parameter
\be 
\H(\eta)=\frac{da/d\eta}{a}=\frac{a'}{a}\,,
\ee
where we have used a prime to denote a derivative with respect to conformal time. Then, the Hubble parameter 
in coordinate and conformal time are related to one another by
\be 
\H = aH\,.
\ee
\\

Having introduced the metric tensor of an FRW universe, we can now go on to discuss
the dynamical equations. In general relativity the curvature of a given spacetime is encoded
in the Riemann curvature tensor, defined as \cite{carroll}
\be 
\label{eq:riemann}
R^\alpha{}_{\mu\beta\nu}
=\Gamma^\alpha_{\mu\nu,\beta}-\Gamma^\alpha_{\mu\beta,\nu}
+\Gamma^\alpha_{\lambda\beta}\Gamma^\lambda_{\mu\nu}
-\Gamma^\alpha_{\lambda\nu}\Gamma^\lambda_{\mu\beta}\,,
\ee
where $\Gamma^\sigma_{\delta\lambda}$ are the Christoffel connection coefficients, 
defined in terms of the  metric tensor and its derivatives as
\be 
\Gamma^\alpha_{\beta\gamma}=\frac{1}{2}g^{\alpha\lambda}(g_{\lambda \beta,\gamma}
+g_{\lambda\gamma,\beta}-g_{\beta\gamma,\lambda})\,,
\ee
where we have introduced the notation $g_{\alpha\beta,\gamma}\equiv\partial_\gamma g_{\alpha\beta}$.
There are two contractions of the Riemann tensor that are particularly useful: the 
Ricci tensor is given by
\be 
R_{\mu\nu}\equiv R^\alpha{}_{\mu\alpha\nu}\,,
\ee
and the Ricci scalar, which is the contraction of the Ricci tensor
\be 
R\equiv g^{\mu\nu}R_{\mu\nu}\,.
\ee
The Riemann tensor obeys the following identity
\be 
\label{eq:bianchi}
\nabla_{[\lambda}R_{\mu\nu]\rho\sigma}=0\,,
\ee
where the square brackets denote anti-symmetrisation over the relevant indices.
Eq.~(\ref{eq:bianchi}) is often called the Bianchi identity. If we introduce the Einstein tensor $G_{\mu\nu}$, 
defined as 
\be 
G_{\mu\nu}=R_{\mu\nu}-\frac{1}{2}g_{\mu\nu}R\,,
\ee
then the  Bianchi identity implies that the divergence of this tensor vanishes identically:
\be 
\label{eq:consG}
\nabla^\mu G_{\mu\nu}=0\,.
\ee
The equation of motion in general relativity is the Einstein equation,
\be 
\label{eq:einstein}
G_{\mu\nu}\equiv R_{\mu\nu}-\frac{1}{2}g_{\mu\nu}R=8\pi G T_{\mu\nu}\,,
\ee
where $G$ is Newton's gravitational constant. $T_{\mu\nu}$ is the energy momentum 
tensor, which describes the energy and momentum of the matter content of the spacetime.
We take the perfect fluid energy momentum tensor which has the following form
\be
\label{eq:emfluid}
{T^\mu}_\nu=(\rhob+\Pb)u_{(0)}^\mu u_{(0)\nu}+\Pb{\delta^\mu}_\nu\,,
\ee
where $\rho$ and $P$ are the energy density and pressure, and $u^\mu$ the fluid four velocity, 
satisfying the constraint $u^\mu u_\mu =-1$. Note that
the subscript `$0$' denotes the 
value of the quantity in the homogeneous and isotropic background: the importance of this notation
will become apparent in Chapter~\ref{ch:perturbations}.
In addition to the Einstein equation, there also exist a set of evolution equations for the matter
variables. These are obtained through the covariant conservation of the energy momentum tensor,\footnote{It should be noted
that this is not the only way one can obtain evolution equations. Instead, varying the action with respect to 
the matter fields will result in the same evolution equations except in the case where the system cannot be described
by an action (e.g., dissipative fluids \cite{Acheson:1990:EFD}).}
\be 
\nabla_\mu T^\mu{}_\nu=0\,.
\ee 
This equation is obtained from the conservation of the Einstein tensor,
Eq.~({\ref{eq:consG}), and by using the Einstein equation, Eq,~(\ref{eq:einstein}).\\

Now, let us consider the FRW spacetime. Using the fluid four velocity
\be 
u_{(0)}^\mu=\frac{1}{a}(1,0)\,, \hskip 1cm u_{(0)\mu}=-a(1,0)\,.
\ee
and the
Christoffel symbols for the (flat) FRW spacetime 
\begin{align}
\Gamma^0_{00}&=\H\,, & \Gamma^0_{ij}&=\H\delta_{ij}\,,&\Gamma^i_{j0}&=\H\delta^i{}_j\,,\\
\Gamma^0_{0i}&=0\,, &\Gamma^i_{00}&=0\,, & \Gamma^i_{jk}&=0\,,
\end{align}
the Einstein equations (\ref{eq:einstein}) are then
\begin{align}
\label{eq:friedmann}
\H^2&=\frac{8\pi G}{3}\rhob a^2\,,\\
\label{eq:raych}
\H'&=-\frac{4\pi G}{3}(\rhob+3\Pb)a^2\,,
\end{align}
where the first equation comes from the 0-0 component and the second from the trace of the spatial Einstein equation.
Eqs.~(\ref{eq:friedmann}) and (\ref{eq:raych}) are called the Friedmann and acceleration equations,
respectively.
The acceleration equation
can be rewritten, by introducing the constant equation of state $\Pb=w\rhob$, as
\be 
\label{eq:acceleration}
\H'=-\frac{4\pi G}{3}(1+3w)\rhob a^2\,,
\ee
where $w$ is the equation of state parameter.
  Energy
conservation gives the continuity equation
\begin{align}
\label{eq:econs}
\rhob'=-3\H(\rhob+\Pb)\,,
\end{align}
which can also be rewritten as
\be 
\rhob'=-3\H(1+w)\rhob\,.
\ee
This can then be integrated to give
\be 
\rhob=\bar{\rhob}\left(\frac{a}{\bar{a}}\right)^{-3(1+w)}\,,
\ee
where an overbar denotes the value of a quantity today. The scale factor and Hubble parameter can then be shown to 
evolve as 
\be 
a=\bar{a}\left(\frac{\eta}{\bar{\eta}}\right)^{2/(1+3w)}\,,
\hspace{1cm}
\H=\frac{2}{1+3w}\eta^{-1}\,.
\ee

We now highlight the evolution of the parameters in the different eras of the universe in the hot Big Bang model. The first is 
the radiation era, where the universe is filled with a fluid of particles moving at (or close to) the speed 
of light, such as photons.\footnote{We should note that, generically, one would expect the initial radiation era to be violently disordered.
However, the existence of the inflationary era guarantees that the radiation era is smooth. We will introduce inflationary cosmology
in the next section.} The equation of state parameter for such a fluid is $w=1/3$, which gives 
\be 
\rhob\propto a^{-4}\,, \hspace{1cm} {\rm and} \hspace{1cm} a\propto\eta\,.
\ee 
The next epoch is the matter domination era, where the universe is filled with collisionless, non-relativistic
particles which better models a universe filled with galaxies.
This matter is called dust, and is well modelled by a pressureless fluid with equation of 
state parameter $w=0$. In this era the energy density and scalar factor evolve, respectively, as
\be 
\rhob\propto a^{-3}\,, \hspace{1cm} {\rm and} \hspace{1cm} a\propto\eta^2\,.
\ee

\section{Inflationary Cosmology}
\label{sec:inflation}

Any summary of modern cosmology would not be complete without a short discussion on 
the inflationary paradigm\footnote{See, e.g., Ref.~\cite{llbook} for a more complete treatment of the classical 
Big Bang problems}, which is
 a crucial part of the standard cosmological model. Historically,
it was introduced in an attempt to solve some outstanding problems in the big bang model. These are:
\begin{itemize}
\item{{\it the Horizon problem}, that CMB radiation coming from areas
of the universe that were never in causal contact are observed
to have the same temperature;}
\item{{\it the Flatness problem}, that in order for the universe
to be so close to flat today, it must have started off very close to flat;}
\item{{\it the Relic problem}, that
no topological relics\footnote{Topological relics, such as magnetic monopoles are generically produced if the symmetry of a 
Grand Unified Theory is restored in the early universe and then broken spontaneously. Such an abundance of relics is higher than observation
allows. See, e.g., Refs.~\cite{Hindmarsh:1994re, vilshell} for more details.}
 are observed, though they are likely produced in the early universe.}
\end{itemize}

In order to solve these problems, inflation was postulated in the early `80s,
simultaneously by Guth \cite{Guth:1980zm}, Starobinsky \cite{Starobinsky:1980te},
Albrecht and Steinhardt \cite{Albrecht:1982wi}
and Linde \cite{Linde:1981mu, Linde:1983gd}. The precise definition of inflation is simple: it is a
period during which the universe undergoes accelerated expansion, i.e.
\be 
\ddot{a}>0\,,
\ee
where we have switched to coordinate time in the section for clarity.
Another definition, equivalent to the first is
\be 
\frac{d}{dt}\left(\frac{1}{aH}\right) < 0 \,.
\ee
This definition is more intuitive, since $1/aH$ is the 
Hubble horizon size, and so inflation is defined as a period in which
the Hubble size is decreasing. This is precisely the condition required to solve the 
flatness problem. Finally, a third equivalent definition of inflation is 
\be 
\rhob+3\Pb<0\,,
\ee
which, since $\rhob$ is always positive from the weak energy condition \cite{hawkingellis},
implies a negative pressure during inflation.

In order to be entirely critical of inflation, we should state that inflation does not completely remove the initial 
flatness problem. To begin an inflationary phase, there must exist a Planck-scale patch of spacetime that
is roughly smooth at an early time. It is hoped that the tuning necessary to achieve this is less than that 
required to suppress the spatial curvature at early times in the absence of an inflationary phase, but
this is not necessarily true; it may be much worse. Similarly, one might argue that inflation does not
entirely solve the relic problem, depending on the model of inflation, since some hybrid models may produce
topological defects as a by-product of their operation. Thus, it is too simplistic to claim that inflation
solves all the problems completely, however any further investigation into this is beyond the remit of this thesis.

 The most popular type of matter which can drive inflation is a scalar field. 
A scalar field has a Lagrangian
density of the form
\be 
{\cal L} = p(X,\vp)\,,
\ee
where $X$ is the relativistic kinetic energy, $X\equiv \frac{1}{2}g^{\mu\nu}\vp_{,\mu}\vp_{,\nu}$ which, for a 
homogeneous field, is then $X=-\frac{1}{2}\dot{\vpb}^2$,
where $\vpb(t)$ is the  scalar field. Inflationary models can be classified according to the form
of the Lagrangian, and whether they contain a single scalar field or multiple fields. The most simple single field 
inflationary model has a Lagrangian
\be 
p(X,\vpb)=X-U(\vpb)\,,
\ee
where $U(\vpb)$ is the potential of the field. This simple model can take very different
forms depending on the choice of this potential function. More exotic models have been
considered more recently which modify the Lagrangian through changing the dependence on $X$
 as well or instead of changing the potential. These models have been of interest since they enable
 us to evade observational bounds placed on the simple, canonical, single field models. 

For the simple, canonical model, then, we can write down a pressure and energy density for the 
scalar field as
\begin{align}
\rhob&=\frac{1}{2}\dot{\vpb}^2+U(\vpb)\,,\\
\Pb &= \frac{1}{2}\dot{\vpb}^2-U(\vpb)\,.
\end{align}
Then, using the energy conservation equation (\ref{eq:econs}) we obtain the Klein-Gordon
equation for the homogeneous field
\be 
\label{eq:kgbkgd}
\ddot{\vpb}+3\H\dot{\vpb}+U_{,\vp}=0\,,
\ee
and the Friedmann equation (\ref{eq:friedmann}) becomes
\be 
\label{eq:friedmannfield}
H^2=\frac{8\pi G}{3}\Big(U(\vpb)+\frac{1}{2}\dot{\vpb}^2\Big)\,.
\ee

A useful approximation exists, which consists of neglecting the first term of 
Eq.~(\ref{eq:kgbkgd}) and the last term of Eq.~(\ref{eq:friedmannfield}), to give
\begin{align}
H^2&\simeq\frac{8\pi G}{3}U\,,\\
3H\dot{\vpb}&\simeq-U_{,\vp}\,,
\end{align}
which, for this approximation to be true, demands two parameters defined as
\begin{align} 
\varepsilon(\vpb) &= 4\pi G \Bigg(\frac{U_{,\vp}}{U}\Bigg)^2\,,\\
\eta(\vpb) &= 8\pi G \,\frac{U_{,\vp\vp}}{U}\,,
\end{align}
to be small (i.e. much less than 1). This approximation is called the {\it slow-roll approximation}, and if it 
holds guarantees that inflation will occur.  In fact, inflation ends when $\varepsilon$
becomes $1$. We do not discuss specific models of
 inflation here, instead pointing the interested reader to one of the 
many textbooks available on the topic \cite{llbook}.

To close this brief section on inflationary cosmology, we consider some observational signatures of 
the early universe. Since this is somewhat beyond the main aim of this thesis, we simply present some results,
and refrain from any derivations. To begin, we use the two point correlator of the linear scalar field perturbation, 
$\langle \dvp_1 \dvp_1\rangle$, defined later on, which one can then relate to the spectrum of the curvature perturbation
that will source the anisotropies in the CMB, e.g. the comoving curvature perturbation 
$\langle {\mathcal R}_1 {\mathcal R}_1 \rangle$, which is defined in Eq.~(\ref{eq:R}). This can then be evolved
forwards using a Boltzmann code \cite{Seljak:1996is, Zaldarriaga:1999ep, Lewis:1999bs},
  to give predictions for the anisotropies in the CMB.

In the notation of the WMAP team \cite{WMAP7}, the primordial spectrum is then taken to be a power
law with amplitude $\Delta_{\mathcal{R}}(k_0)^2$ and spectral index $n_{\rm s}$
\be 
\Delta_{\mathcal{R}}(k)^2=\Delta_{\mathcal{R}}(k_0)^2\Big(\frac{k}{k_0}\Big)^{n_{\rm s}-1}\,,
\ee
where $\Delta_{\mathcal{R}}(k_0)^2=2.38 \times 10^{-9}$ and $n_{\rm s}=0.969$, at the pivot scale
of $k_0=0.002 {\rm Mpc}^{-1}$. The observations are compatible with the mostly adiabatic and Gaussian inflationary 
initial condition, but are incompatible with other a priori equally motivated suggestions, such as density
perturbations induced by topological defects. Constraints on inflationary model building are discussed in 
detail in, e.g., Refs.~\cite{Alabidi:2010sf, Alabidi:2006qa}.
Similarly, we can write the spectrum for tensor perturbations as
\be
\Delta_{{h}}(k)^2=\Delta_{{h}}(k_0)^2\Big(\frac{k}{k_0}\Big)^{n_{\rm T}}\,,
\ee
and we can then define the tensor-scalar ratio, $r$, as \cite{WMAP7}
\be 
r\equiv\frac{\Delta_{{h}}(k_0)^2}{\Delta_{\mathcal{R}}(k_0)^2}\,.
\ee

Inflationary models can then be tested according to their predictions for these observables.
The scalar spectral index is well 
constrained by WMAP observations, and in fact a generic, successful feature of inflation is its ability
to generate a near scale invariant spectrum \cite{llbook}. However, the tensor-scalar ratio is less
constrained, and different inflationary models make varying predictions for $r$
(e.g. Refs.~\cite{Alabidi:2006fu, starobinskii, Alabidi:2010sf, Lyth:1996im} and references therein).
It is hoped that we will be able to narrow the error bars on the tensor-scalar ratio by future 
observations of the polarisation of the CMB from the sky (Refs.~\cite{Planck, CMBPol}) and from the ground
(e.g. Ref.~\cite{Errard:2010bn}) which will in turn
enable us to rule out some models of inflation.

\section{Notation}
To close the introduction, we briefly state some of the notational conventions that will
be used in this thesis. 

\begin{itemize}
\item{We use the mostly positive metric signature, $(-+++)$, and the $(+++)$ convention in the
notation of \cite{Misner:1974qy}.}
\item{Coordinate time is denoted with $t$ and an overdot denotes a derivative with respect to 
coordinate time; the Hubble parameter is $H=\dot{a}/a$.}
\item{Conformal time is denoted by $\eta$ and a prime denotes a derivative
with respect to conformal time; the conformal Hubble parameter is $\H=a'/a$.}
\item{Greek indices $\{\mu,\nu,\ldots\}$ cover the entire spacetime and take the range $\{0\ldots 3\}$.}
\item{Latin indices $\{i,j,\ldots\}$ cover the spatial slice and take the range $\{1\ldots3\}$.}
\item{The index $0$ (as in $u^0$) denotes conformal time, and an index $t$ (as in $u^t$) denotes
coordinate time.}
\item{The order in the perturbative expansion is denoted with a subscript, $\phi_1$, or with 
a subscript enclosed in parentheses if the meaning could be ambiguous, as in $u_{(0)}^\mu$.}
\item{A comma denotes a partial derivative so, e.g.,
\beb
X_{,\mu}\equiv \p_\mu X\equiv\frac{\partial X}{\partial x^\mu}\,, 
\hskip 5mm {\textrm{or, when $X\equiv X(Y)$}},\hskip 5mm
X_{,Y}\equiv \frac{\p X}{\p Y}\,.
\eeb
}
\item{A semicolon denotes a covariant derivative with respect to the full spacetime metric, $g_{\mu\nu}$, i.e.,
\beb 
X_{\nu;\mu}\equiv\nabla_\mu X_\nu\,.
\eeb}
\item{The Lie derivative along a vector field $\xi^\mu$ is denoted $\pounds_\xi$ and takes the following forms for a scalar $\vp$,
a vector, $v_\mu$, and a tensor, $t_{\mu\nu}$:
\bea
\pounds_\xi \vp &=& \xi^\lambda \vp_{,\vp}\,,\\
\pounds_\xi v_\mu &=& v_{\mu,\alpha}\xi^\alpha+v_\alpha\xi^\alpha{}_{,\mu}\,,\\
\pounds_\xi t_{\mu\nu} &=& t_{\mu\nu,\lambda}\xi^\lambda+t_{\mu\lambda}\xi^\lambda{}_{,\nu}+t_{\lambda\nu}\xi^\lambda{}_{,\mu}\,.
\eea
}
\end{itemize}

%% file: perturbations.tex
% % % % % % % % % % % % % % % % % % % % % % % % % % % % 
% perturbations.tex
% cosmological perturbations
% % % % % % % % % % % % % % % % % % % % % % % % % % % % 
% Redefine CVSRevision for this section. 
% If you don't want to use CVS tags comment out this line
%\renewcommand{\CVSrevision}{\version$Id: chapter.tex,v 1.3 2009/12/17 18:16:48 ith Exp $}

% % % % % % % % % % % % % % % % % % % % % % % % % % % % % % % % 
% =========================================================== %
% % % % % % % % % % % % % % % % % % % % % % % % % % % % % % % % 
\chapter{Cosmological Perturbations}
\label{ch:perturbations}
% % % % % % % % % % % % % % % % % % % % % % % % % % % % % % % % 
% =========================================================== %
% % % % % % % % % % % % % % % % % % % % % % % % % % % % % % % % 

As discussed in the introduction, cosmological perturbation theory is an extremely useful, and 
successful, tool to study the universe in which we live. In this Chapter we introduce cosmological
perturbation theory in the metric approach (i.e. where we consider
perturbations to the metric \'a la Bardeen \cite{Bardeen:1980kt})
in a more formal and quantitative manner. We introduce the line element for the 
most general perturbations to FRW, and then define the perturbed energy momentum tensor for
both a perfect fluid and for a scalar field  (with both a canonical and non-canonical action).
We then derive gauge transformations for the different types of perturbation (scalar, vector and tensor),
and use these to define choices of gauge and gauge invariant variables. 
We focus on gauge choice at linear order and give an illustrative example of how to choose a gauge 
at second order.
We close this 
Chapter by considering the thermodynamics of a perfect fluid, defining what we mean
by non-adiabatic, or entropic, perturbations, and briefly discuss how such perturbations can naturally
arise in multi-field or multiple fluid systems.

%%%%%%%%
\section{Metric Tensor}
\label{sec:metric}
%%%%%%%%%

We consider the most general perturbations to the flat
 FRW metric, which gives a line element of the form
\be
ds^2=a^2(\eta)\Big[-(1+2\phi)d\eta^2+2B_idx^id\eta+(\delta_{ij}+2C_{ij})dx^idx^j\Big]\,.
\ee
We choose
a flat background metric because it agrees with observations and  is 
mathematically easier to work with, but should note that all the techniques used in this 
Chapter are valid for a background spacetime with non-zero curvature.

The perturbations of the spatial components of the metric can be further decomposed as
\begin{align}
B_i &= B_{,i}-S_i \,,\\
\label{eq:Cij}
C_{ij} &= -\psi \delta_{ij}+E_{,ij}+F_{(i,j)}+\frac{1}{2}h_{ij}\,.
\end{align}
The perturbations are classified as scalar, vector and tensor perturbations
according to their transformation behaviour on spatial three hypersurfaces \cite{mfb, MW2008}.
 The scalar metric perturbations are $\phi$, the lapse function, $\psi$, the curvature perturbation and $E$ and $B$, 
which make up the scalar shear.
$S_i$ and $F_i$ are divergence free vector 
perturbations,  and $h_{ij}$ is a transverse, traceless tensor perturbation. The perturbations therefore
 obey the following 
relations
\begin{align}
\partial_i S^i&=0\,,\\
\partial_i F^i&=0\,,\\
\partial_i h^{ij}&=0=h^i{}_i\,.
\end{align}

Since the perturbations are inhomogeneous, they depend upon both space and time, e.g.
\be 
\phi\equiv\phi(x^\mu)=\phi(\eta,x^i)\,.
\ee
 The scalar perturbations each contribute one degree of freedom to the perturbed metric tensor, each 
 divergence free vector perturbation has two degrees of freedom, 
 as does the transverse, traceless tensor perturbation, so we see that, in total, there are 10 
 degrees of freedom -- the same as the number of independent components of the perturbed metric tensor. Each 
 perturbation can then be expanded in a series: For example the lapse function is split as
\begin{align}
\phi&=\sum_n\frac{\epsilon^n}{n!}\phi_n\\
&=\epsilon\phi_1+\frac{1}{2}\epsilon^2\phi_2+\frac{1}{3!}\epsilon^3\phi_3+\cdots\,,
\end{align}
where the subscript denotes the order of the perturbation and $\epsilon$ is a fiducial expansion parameter. We will
often omit $\epsilon$ when not required for brevity and write
\be
\phi=\phi_1+\frac{1}{2}\phi_2+\frac{1}{3!}\phi_3+\cdots\,,
\ee
In order to define this expansion uniquely, we choose the first order quantity, $\phi_1$, to have Gaussian statistics.
The series is then
truncated at the required order.
Performing this split then gives us the covariant components of the metric tensor up to second order
\begin{align}
\label{eq:gcov}
g_{00}&=-{a^2}\left(1+2\iphi+\iiphi\right) \,, \\
g_{0i} &= {a^2}\left(2B_{1i}+B_{2i}\right)
 \,, \\
g_{ij} &= {a^2}\left(\delta_{ij}+2C_{1ij}+C_{2ij}\right)\,.
\end{align}
The contravariant components of the metric tensor are  obtained by imposing the constraint 
\be
\label{eq:metricconstraint}
g_{\mu\nu}g^{\nu\lambda}={\delta_\mu}^\lambda\,,
\ee
to the appropriate order. To second order this gives
\begin{align}
\label{eq:gcontra}
g^{00}&=-\frac{1}{a^2}\Big(1-2\iphi-\iiphi+4\iphi^2-B_{1k}B_1^k\Big) \,, \\
g^{0i} &= \frac{1}{a^2}\Big(B_1^i+\frac{1}{2}B_2^i-2\iphi B_1^i-2B_{1k} C_1^{ki}\Big)
 \,, \\
g^{ij} &= \frac{1}{a^2}\Big(\delta^{ij}-2C_1^{ij}-C_2^{ij}+4C_1^{ik}C_{1k}{}^j-B_1^iB_1^j\Big)\,.
\end{align}

%%%%%%%%
\section{Energy Momentum Tensor}
\label{sec:emtensor}
%%%%%%%%

The matter content of the universe is described by the energy momentum tensor.
Since General Relativity links the geometry of spacetime to its matter content, perturbations
in the metric tensor invoke perturbations in the energy momentum tensor.\footnote{There are, in fact, works
where this is not the case and, for example, only the energy momentum tensor is perturbed. However, in order for the
work to be consistent, one should perturb both the geometry and the matter content of the spacetime.} In this section, we 
outline the perturbed energy momentum tensor for a perfect fluid and a scalar field, to second order in
perturbation theory.

\subsection{Perfect Fluid}

The energy momentum tensor for a perfect fluid, i.e. in the absence of anisotropic stress, is as
presented in the previous section,
\be
\label{eq:emdeffluid}
{T^\mu}_\nu=(\rho+P)u^\mu u_\nu+P{\delta^\mu}_\nu\,,
\ee
where $\rho$ is the energy density, $P$ is the pressure and $u^\mu$ is the four velocity of the fluid. 
Note that, for the purposes of this thesis, we define a `perfect fluid' to be a fluid with a 
diagonal energy momentum tensor as in, e.g., Ref.~\cite{Giovannini:2008zzb}.

The energy density and the pressure are expanded up to second order in perturbation theory in the standard way
\begin{align}
\label{eq:rhoexpand}
\rho &= \rhob+\drho+\frac{1}{2}\drhorho \,,\\
P &= \Pb + \dP +\frac{1}{2}\dPP\,.
\end{align}
The fluid four velocity, which is defined as 
\be
\label{eq:fourveldef}
u^\mu = \frac{d x^\mu}{d \tau}\,,
\ee
where $\tau$ is an affine parameter, here the proper time,
and is subject to the constraint
\be
u^\mu u_\mu =-1\,.
\ee
To second order in perturbation theory the fluid four velocity has the contravariant components
\begin{align}
\label{eq:fourvel1}
u^i &= \frac{1}{a}\left({v_1}^i+\frac{1}{2}{v_2}^i\right) \,, \\
\label{eq:fourvel2}
u^0 &= \frac{1}{a}\left(1-\phi_1-\frac{1}{2}\phi_2+\frac{3}{2}\phi_1^2
+\frac{1}{2}v_{1k}{v_1}^k+v_{1k}{B_1}^k \right)\,, 
\end{align}
and the covariant components,
\begin{align}
\label{eq:fourvel3}
u_i &= a\left(v_{1i}+B_{1i}+\frac{1}{2}(v_{2i}+B_{2i})-\phi_1 B_{1i}+2C_{1ik}{v_1}^k
\right) \,, \\
\label{eq:fourvel4}
u_0 &= -a\left(1+\phi_1+\frac{1}{2}\phi_2-\frac{1}{2}\phi_1^2
+\frac{1}{2}{v_1}^kv_{1k}\right)\,,
\end{align}
where $v^i$ is the spatial three velocity of the fluid. Then, by substituting the components of the four velocity along with the expansions of the energy density and pressure into Eq.~(\ref{eq:emfluid}), we obtain the components of the energy momentum tensor, up to second order
\begin{align}
\label{eq:em1}
{T^0}_0 &= -(\rhob+\Pb)({v_1}^k+{B_1}^k)v_{1k}-(\rhob+\drho+\frac{1}{2}\drhorho) \,,\\
{T^0}_i &= (\rhob+\Pb)\left(v_{1i}+B_{1i}+\frac{1}{2}(v_{2i}+B_{2i})-\phi_1(v_{1i}+2B_{1i})+2C_{1ik}{v_1}^k\right)\nonumber\\
\label{eq:em2}
&\qquad +(\drho+\dP)(v_{1i}+B_{1i})\,,\\
\label{eq:em3}
{T^i}_j &= \left(\Pb+\dP+\frac{1}{2}\dPP\right){\delta^i}_j+(\rhob+\Pb){v_1}^i(v_{1j}+B_{1j})\,.
\end{align}

\subsection{Scalar Field}

As shown in Section \ref{sec:inflation}, scalar fields play an important role in modern cosmology through
the theory of inflation.
The energy momentum tensor for a scalar field minimally coupled to gravity is defined through the Lagrangian 
density. For a canonical scalar field, the Lagrangian takes the form
\be
\label{eq:lagcan}
\mathcal{L}=-\frac{1}{2}g^{\mu\nu}\varphi_{,\mu}\varphi_{,\nu}-U(\varphi)\,,
\ee
where $U(\varphi)$ is the potential energy of the scalar field. 
The variational energy momentum tensor (or Hilbert stress-energy tensor) is then defined as \cite{llbook}
\be
\label{eq:emdef}
T_{\mu\nu}\equiv-2\frac{\partial\mathcal{L}}{\partial g^{\mu\nu}}+g_{\mu\nu}\mathcal{L}\,,
\ee
and, for a scalar field $\varphi$, we obtain
\be
{T^{\mu}}_{\nu}=g^{\mu\lambda}\varphi_{,\lambda}\varphi_{,\nu}
-{\delta^\mu}_\nu\left(U(\varphi)+\frac{1}{2}g^{\alpha\beta}\varphi_{,\alpha}\varphi_{,\beta}\right)\,.
\ee
Expanding the scalar field in the usual way, and using the definition of the metric tensor given in Section~\ref{sec:metric},
 gives the components of 
the energy momentum tensor for a scalar field up to second order in perturbation theory (see, e.g., Refs.~\cite{Noh:2004bc, Malik:2005cy, Acquaviva:2002ud})
\begin{align}
{T^0}_0 &= -\frac{\vpb'^2}{2a^2}\Big[
1-2\iphi+2\frac{\dvp_1'}{\vpb'}+4\iphi^2-B_{1k}B_1^k+2\iphi\dvp_1'
-\left(\frac{\dvp_1'}{\vpb'}\right)^2\\
&\qquad\qquad+\frac{\dvp_{1,k}\dvp_{1,}^k}{\vpb'^2}-\iiphi
-\frac{\dvp_2'}{\vpb'}\Big]
-U(\vpb)-U_{,\vp}\dvp_1-U_{,\vp\vp}\dvp_1^2-U_{,\vp}\dvp_2\,,\nn
\\
{T^i}_0 &=\frac{\vpb'^2}{a^2}\Big[B_1^i+\frac{\dvp_{1,}{}^j}{\vpb'}-2\iphi B_1^i-2B_{1k}C_1^{ki}
+2\frac{B_1^i\dvp_1'}{\vpb'}
+\frac{\dvp_1'\dvp_{1,}{}^i}{\vpb'^2}\nn\\
&\qquad\qquad-2\frac{\dvp_{1,j}C_1^{ij}}{\vpb'}+\frac{1}{2}\Big(B_2^i
+\frac{\dvp_{2,}{}^i}{\vpb'}\Big)\Big]\,,
\end{align}
\begin{align}
\label{eq:em0ifield}
{T^0}_i &=-\frac{\vpb'}{a^2}\Big[\dvp_{1,i}+\frac{\dvp_1'\dvp_{1,i}}{\vpb'}-2\iphi\dvp_{1,i}
+\frac{1}{2}\dvp_{2,i}\Big]\,,\\
{T^i}_j &=\frac{\vpb'^2}{a^2}\Big[\frac{B_1^i\dvp_{1,j}}{\vpb'}+\frac{\dvp_{1,}{}^i\dvp_{1,j}}{\vpb'^2}\Big]
-\delta^i{}_j\Big\{U(\vpb)+U_{,\vp}\dvp_1+U_{,\vp\vp}\dvp_1^2+U_{,\vp}\dvp_2 \nn\\
&\qquad
-\frac{\vpb'^2}{2a^2}\Big[1-2\iphi+2\frac{\dvp_1'}{\vpb'}+4\iphi^2-B_{1k}B_1^k+4\frac{\iphi\dvp_1'}{\vpb'}
-2\frac{B_1^k\dvp_{1,k}}{\vpb'}\nn\\
&\qquad\qquad\qquad-\frac{\dvp_{1,}{}^k\dvp_{1,k}}{\vpb'^2}-\left(\frac{\dvp_1'}{\vpb'}\right)^2
-\iiphi-\frac{\dvp_2'}{\vpb'}\Big]\Big\}\,.
\end{align}

We can write down a pressure and energy density for the scalar field by comparing the components of the
scalar field energy momentum tensor to that of a perfect fluid given in Eqs.~(\ref{eq:em1})-(\ref{eq:em3}).
 In the background this gives, as
shown in Section \ref{sec:inflation},
\begin{align}
\label{eq:fieldfluidvariables}
\rhob =\frac{1}{2a^2}\vpb'^2+U(\vpb)\,, \hskip 1cm \Pb =\frac{1}{2a^2}\vpb'^2-U(\vpb)\,,
\end{align}
and, to linear order,
\begin{align}
\drho &= \frac{1}{a^2}\Big(\dvp_1'\vpb'-\iphi\vpb'^2\Big)+U_{,\vp}\dvp_1\,,\\
\dP &= \frac{1}{a^2}\Big(\dvp_1'\vpb'-\iphi\vpb'^2\Big)-U_{,\vp}\dvp_1\,.
\end{align}
We can also express the fluid velocity in terms of the field by comparing Eq.~(\ref{eq:em2}) to Eq.~(\ref{eq:em0ifield})
to give
\be 
V_1=-\frac{\dvp_1}{\vpb'}\,,
\ee
where $V_1\equiv B_1+v_1$.

There has been a lot of recent interest in scalar fields with non-canonical actions: early work in the realm of the early 
universe such as k-inflation
\cite{ArmendarizPicon:1999rj, Garriga:1999vw}
and more current work including the string theory motivated
Dirac--Born--Infeld (DBI) inflation \cite{Silverstein:2003hf, Alishahiha:2004eh, Seery:2005wm, Lidsey:2007gq,
Huston:2008ku} . Scalar fields with non-canonical actions have also recently been considered as dark energy 
candidates
 (see, e.g. Ref. \cite{Unnikrishnan:2008ki}). These models all modify the Lagrangian (\ref{eq:lagcan}) to the more
 general function\footnote{One could, of course, cook up even more general Lagrangians. However, typically if higher-order
 derivatives of the scalar field are present then the model will exhibit Ostrogradski instabilities (see Ref.~\cite{Durrer:2007re}
 for an outline of Ostrogradski's original argument).}
 \be 
 \label{eq:laggen}
 \mathcal{L}=p(X, \vp)\,,
 \ee
 where $X=-\frac{1}{2} g^{\mu\nu}\vp_{,\mu}\vp_{,\nu}$. The Lagrangian for DBI inflation is then
  given by
 \be 
 p(X,\vp)=-T(\vp)\sqrt{1-2T^{-1}(\vp)X}+T(\vp)-V(\vp)\,,
 \ee
 where $T(\vp)$ and $V(\vp)$ are functions that further specify the model.
 
 The energy momentum tensor for the general Lagrangian (\ref{eq:laggen}) is then obtained from Eq.~(\ref{eq:emdef})
 as
 \be
\label{eq:emgen}
{T^{\mu}}_{\nu}
=p_{,X}g^{\mu\lambda}\vp_{,\lambda}\vp_{,\nu}+{\delta^\mu}_\nu p\,.
\ee
This has the components, up to linear order in perturbation theory, and in coordinate time, $t$, 
\begin{align}
{T^t}_t&=
\left(p_0-2p_{,X}\Xb \right) -\left[\left(p_{,X}+2\Xb p_{,XX}\right)
\delta X_1+2\Xb p_{,X\vp}\dvp_1-p_{,\vp}\dvp\right] \,,\\
{T^t}_i&=-p_{,X}\dot{\vpb}\dvp_{1,i}\,,\\
{T^i}_j&=\left(p_0+p_{,X}\delta X_1+p_{,\vp}\dvp_1\right){\delta^i}_j \,,
\end{align}
where we have defined $\delta X_1 =\dot{\vpb}\dot{\dvp_1}-\iphi\dot{\vpb}^2$, and $\Xb=\frac{1}{2}\dot{\vpb}^2$.
 As in the canonical case,
we can define a pressure and energy density for the scalar field. The background quantities are
\be 
\rhob=2p_{,X}X_0-p\,,\hskip 1cm \Pb=p\,,
\ee
and the linear order quantities are
\begin{align}
\drho&=(p_{,X}+2X_0 p_{,XX})\delta X_1+(2X_0p_{,X\vp}-p_{,\vp})\dvp_1\,,\\
\dP&=p_{,X}\delta X_1+p_{,\vp}\dvp_1\,.
\end{align}

%%%%%%%%
\section{Gauge Transformations at First and Second Order}
\label{sec:gaugetrans}
%%%%%%%%%

Gauge transformations play an important role in cosmological perturbation theory. General relativity
is a theory of differential manifolds with no preferred coordinate charts, and is therefore
required to be covariant under coordinate transformations. However, when we come to 
consider perturbations in general relativity we must, for consistency, consider
perturbations of the spacetime itself. That is, in the language of differential geometry,
we consider a one parameter family of  four-manifolds $M_\epsilon$, embedded
in a five-manifold $N$ \cite{stewart1990, MM2008}.
 Each $M_\epsilon$ represents a spacetime, with the base spacetime,
or unperturbed $\epsilon=0$ manifold, $M_0$. The problem comes in perturbation theory when attempting
to compare two objects which `live' in different spaces. In order to deal with this, we introduce a
point identification map $p_\epsilon: M_0\to M_\epsilon$ which relates points in the perturbed
manifold with those in the background. This correspondence introduces a new vector field, $X$
on $N$, and points which lie on the same integral curve $\gamma$ of $X$ are regarded as being
the same physical point.

However, the choice of this point identification map and, therefore, the vector field $X$ is not
unique. The choice of the correspondence between the points on the $M_0$ and those on
$M_\epsilon$ or, equivalently, the choice of the vector field $X$ is called a {\it choice of gauge}, and
$X$ is then the gauge generator. A gauge transformation then tells us how we move from one 
choice of gauge to another.

There are two approaches to gauge transformations. First, consider a point $p$ on
$M_0$. Two generating vectors $X$ and $Y$ then define a correspondence between
this point $p$ and two different points $s$ and $t$ on $M_\epsilon$. Clearly, then,
these choices induce a coordinate change (gauge transformation) on $M_\epsilon$.
This is known as the passive view. Alternatively, consider a point $p$ on $M_\epsilon$.
We then find a point $s$ on $M_0$ which maps to $p$ under the gauge choice $X$
and a point $t$, also on $M_0$ that maps to $p$ under the choice $Y$. In this case,
a gauge transformation is induced on the background manifold, $M_0$. This is
known as the active view. 
We can think of the active approach as the one in which the 
transformation of the perturbed quantities is evaluated at the same coordinate point,
and the passive approach  where the transformation of the perturbed quantities
is taken at the same physical point.\\

In this section we go on to discuss the active and passive approach briefly
then, adopting the active approach, we derive the gauge transformation rules for scalars,
vectors and tensors up to second order in the perturbations.

%%%%%%%%%%%%%%%%%
\subsection{Active Approach}
%%%%%%%%%%%%%%%

In the active approach to gauge transformations, the exponential map is the starting point \cite{MM2008, Mukhanov1997}. Once the 
generating vector of gauge transformation, $\xi^\mu$, has been specified, we can immediately 
write down how a general tensor ${\bf T}$ transforms. The exponential map is 
\be
\label{eq:exmap}
\widetilde{\bf T} = e^{\pounds_{\xi}}{\bf T}\,,
\ee
where $\pounds_{\xi}$ denotes the Lie derivative with respect to the generating vector $\xi^\mu$ which,
up to second order in perturbation theory, is 
\be
\xi^\mu\equiv \epsilon \xi_1^{\mu}
+\frac{1}{2}\epsilon^2\xi_2^{\mu}+\cdots\,.
\ee
The exponential map is then 
\be
\label{eq:expandex}
\exp(\pounds_{\xi})=1+\epsilon\pounds_{\xi_1}
+\frac{1}{2}\epsilon^2\pounds_{\xi_1}^2
+\frac{1}{2}\epsilon^2\pounds_{\xi_2}+\cdots
\ee
up to second order in perturbation theory. Splitting ${\bf T}$  order by order, we find that the 
tensorial quantities transform at zeroth, first and second order, respectively, as 
\cite{Mukhanov1997, Bruni:1996im}
\begin{align}
\widetilde {{\bf T}_0} &= {\bf T}_0 \,,\\
\label{eq:Ttrans1}
\epsilon\widetilde{\delta{\bf T}_1}
&= \epsilon\delta{\bf T}_1 + \epsilon\pounds_{\xi_1} {\bf T}_0  \,,\\
\epsilon^2\widetilde{\delta{\bf T}_2}
\label{eq:Ttrans2}
&=\epsilon^2\Big( \delta{\bf T}_2 +\pounds_{\xi_2} {\bf T}_0 +\pounds^2_{\xi_1}
{\bf T}_0 + 2\pounds_{\xi_1} \delta{\bf T}_1\Big)\,.
\end{align}

By noting that the Lie derivative acting on a scalar is just the directional derivative, 
$\pounds_{\xi}=\xi^\mu(\p / \p x^\mu)$, 
the exponential map can be applied to the coordinates $x^\mu$ to 
obtain the following relationship
between coordinates at two points, $p$ and $q$\footnote{Note that, in some of the literature, 
a different sign is taken in the 
exponent in this equation to obtain correspondence between the active and passive 
approach at linear order. Since we are not solely working at linear order in this work, there
is no advantage gained by making such a choice.}
\be
\label{defcoordtrans}
{x^\mu}( q)
= e^{\xi^\lambda \frac{\p}{\p x^\lambda}\big|_p} \
x^\mu( p)\,.
\ee 
Expanding this to second order gives
\be
\label{eq:coords}
x^\mu(q)=x^\mu(p)+\epsilon\xi_1^\mu(p)
+\frac{1}{2}\epsilon^2\Big(\xi^\mu_{1,\lambda}(p)\xi_{1}^{\lambda}(p)+\xi_2^\mu(p)\Big)\,.
\ee
This coordinate relationship is not required to perform calculations in the 
active approach, but will be useful for the discussion in Section \ref{sec:passive} below.

%%%%%%%%%%%%%
\subsection{Passive Approach}
\label{sec:passive}
%%%%%%%%%%%%%

A natural starting point for discussing the passive approach to gauge transformations is the 
coordinate relationship \eq{eq:coords} since, in the passive approach, one states the relationship
between two coordinate systems and then calculates how variables change when transforming from one 
coordinate system to the other. However, since in the passive approach quantities are evaluated at
the same physical point we need to rewrite \eq{eq:coords} \cite{MM2008}. Choosing $p$ and
$q$ such that $\widetilde{x^\mu}(q)=x^\mu(p)$, \eq{eq:coords} enables us to write
\begin{align}
\widetilde{x^\mu}(q)&=x^\mu(p)\nn\\
\label{eq:passivetrans}
&=x^\mu(q)-\epsilon\xi_1^\mu(p)
-\frac{1}{2}\epsilon^2\Big[\xi_{1,\lambda}^\mu(p)\xi_1^\lambda(p)
+\xi_2^\mu(p)\Big]\,.
\end{align}
Using the first terms of \eq{eq:coords},
\be
x^\mu(q)=x^\mu(p)+\epsilon\xi_1^\mu(p)\,,
\ee
allows us to Taylor expand $\xi_1^\mu$ as
\begin{align}
\xi_1^\mu(p)&=\xi_1^\mu\big(x^\mu(q)-\epsilon\xi_1^\mu(p)\big)\nn\\
\label{eq:taylorxi1}
&=\xi_1^\mu(q)-\epsilon\xi^\mu_{1,\lambda}(q)\xi_1^\lambda(q)\,,
\end{align}
which is valid up to second order. Substituting \eq{eq:taylorxi1} into 
\eq{eq:passivetrans} gives the relationship between the two coordinate systems 
at the same point, $q$
\be 
\widetilde{x^\mu}(q)=x^\mu(q)-\epsilon\xi_1^\mu(q)
-\frac{1}{2}\epsilon^2\Big[\xi_2^\mu(q)-\xi_{1\lambda}^\mu(q)\xi_1^\lambda(q)\Big]\,.
\ee

Having highlighted the two approaches to gauge transformations we now focus on the active 
approach and give some concrete examples.

%%%%%%%%%%%%
\subsection{Four Scalars}
%%%%%%%%%%%%%

We will now look at the transformation of four scalars, choosing the energy density, $\rho$, which can
be expanded as in \eq{eq:rhoexpand}, 
\be 
\rho=\rhob+\drho+\frac{1}{2}\drhorho+\frac{1}{3!}\delta \rho_3+\cdots\,,
\ee
as an example. 

\subsubsection{First Order}

Before studying the transformation behaviour of perturbations at first order, we split the generating vector
$\xi_1^\mu$ into a scalar temporal part $\alpha_1$ and a spatial scalar and divergence free vector part, 
respectively $\beta_1$ and $\gamma_1{}^i$, as
\be 
\xi_1^\mu=(\alpha_1,\beta_{1,}{}^i+\gamma_1{}^i)\,.
\ee
Since the Lie derivative of a scalar $\rho$ with respect to the vector $\xi^\mu$ is simply
\be 
\pounds_\xi \rho=\xi^\mu \rho_{,\mu}\,,
\ee
from \eq{eq:Ttrans1}, we then find that the energy density transforms, at linear order, as
\be
\label{eq:rhotrans1}
\widetilde{\delta\rho_1}=\delta\rho_1+\rhob'\alpha_1\,.
\ee
We see that, at first order, the gauge transformation is completely determined by the time slicing, $\alpha_1$.\footnote{Note
that $\alpha_1$ does not generate a foliation of spacetime by spatial hypersurfaces -- this is inherited from the 
foliation already present in the background spacetime. Instead, $\alpha_1$ labels the time slicing.}

\subsubsection{Second Order}

At second order we split the generating vector $\xi_2^\mu$ in an analogous way to first order as 
\be 
\xi_2^\mu=(\alpha_2,\beta_{2,}{}^i+\gamma_2{}^i)\,.
\ee
Then, using \eq{eq:Ttrans2}, we find that the second order energy density perturbation transforms as
\be
\label{eq:rhotrans2}
\widetilde{\delta\rho_2}=\delta\rho_2
+\rhob'\alpha_2+\alpha_1(\rhob''\alpha_1+\rhob'\alpha_1'+2\delta\rho_1')
+(2\delta\rho_1+\rhob'\alpha_1)_{,k}(\beta_{1,}{}^k+\gamma_{1}{}^k)\,.
\ee
Thus, at second order, the gauge transformation is only fully determined once the time slicing is specified at first and second 
order ($\alpha_1$ and $\alpha_2$) and the spatial threading (or spatial gauge perturbation) is specified at first order ($\beta_1$ and
 $\gamma_1{}^i$) \cite{MW2008}.

%%%%%%%%%%%%
\subsection{The Metric Tensor}
\label{sec:mettrans}
%%%%%%%%%%%%%

We will now focus on the transformation behaviour of the metric tensor. Again, the starting point is the 
Lie derivative, which for a the metric tensor, $g_{\mu\nu}$, is given by
\begin{align}
\label{eq:lietensor} \pounds_{\xi}g_{\mu\nu}
=g_{\mu\nu,\lambda}\xi^\lambda
+g_{\mu\lambda}\xi^\lambda_{~,\nu}+g_{\lambda\nu}\xi^\lambda_{~,\mu}\,.
\end{align}

\subsubsection{First Order}

At first order, the metric tensor transforms, from Eqs.~(\ref{eq:Ttrans1}) and (\ref{eq:lietensor}) as
\be
\label{eq:metrictransfirst}
\wt{\delta g^{(1)}_{\mu\nu}}=\delta g^{(1)}_{\mu\nu}
+g^{(0)}_{\mu\nu,\lambda}\xi^\lambda_1
+g^{(0)}_{\mu\lambda}\xi^\lambda_{1~,\nu}
+g^{(0)}_{\lambda\nu}\xi^\lambda_{1~,\mu}\,.
\ee
We can obtain the transformation behaviour of each particular metric function
by extracting it, in turn, from the above general expression using the method outlined in
Ref.~\cite{MW2008}. From Eq.~(\ref{eq:lietensor}) we obtain the following transformation behaviour
for $C_{1ij}$
\be 
2\widetilde C_{1ij}=2C_{1ij}+2\H\alpha_1 \delta_{ij}
+\xi_{1i,j}+\xi_{1j,i}\,.
\ee
From this we can extract the transformation behaviour of the spatial metric functions.
Here we do not focus on the details, but instead quote
results. We find that the scalar metric perturbations transform as
\begin{align}
\label{eq:transphi1}
\widetilde{\iphi}&=\iphi+\H\alpha_1+\alpha_1' \,,\\
\label{eq:transpsi1}
\widetilde{\ipsi}&=\ipsi-\H\alpha_1\,, \\
\label{eq:transB1}
\widetilde{B_1}&=B_1-\alpha_1+\beta_1'\,,\\
\label{eq:transE1}
\widetilde{E_1}&=E_1+\beta_1\,,
\end{align}
the vectors as
\begin{align}
\label{eq:transS1}
\widetilde{S_1{}^i}&=S_1{}^i-\gamma_1{}^i{}'\,,\\
\label{eq:transF1}
\widetilde{F_1{}^i}&=F_1{}^i+\gamma_1{}^i\,,
\end{align}
and, as is well known, the tensor component, $h_{1ij}$, is gauge invariant.
Finally, the scalar shear, which is defined as 
\be 
\label{eq:defshear}
\sigma_1\equiv E_1'-B_1\,,
\ee
transforms as
\be 
\label{eq:transshear1}
\widetilde{\sigma_1}=\sigma_1+\alpha_1\,,
\ee 
which will be useful later when we come to define gauges and gauge
invariant variables.

%%%%%%%%%%%%%%%
\subsubsection{Second Order}
%%%%%%%%%%%%%%%%%

At second order we obtain the transformation behaviour of the metric tensor from
Eqs.~(\ref{eq:Ttrans2}) and (\ref{eq:lietensor}), noting that 
\be 
\pounds_{\xi_1}^2{\bf T}=\pounds_{\xi_1}(\pounds_{\xi_1}{\bf T})\,.
\ee
 The metric tensor therefore transforms as
\begin{align}
\label{eq:metrictranssecond}
\wt{\delta g^{(2)}_{\mu\nu}}&=\delta g^{(2)}_{\mu\nu}
+g^{(0)}_{\mu\nu,\lambda}\xi^\lambda_2
+g^{(0)}_{\mu\lambda}\xi^\lambda_{2~,\nu}
+g^{(0)}_{\lambda\nu}\xi^\lambda_{2~,\mu}
+2\Big[
\delta g^{(1)}_{\mu\nu,\lambda}\xi^\lambda_1
+\delta g^{(1)}_{\mu\lambda}\xi^\lambda_{1~,\nu}
+\delta g^{(1)}_{\lambda\nu}\xi^\lambda_{1~,\mu}
\Big]\nonumber \\
&\quad+g^{(0)}_{\mu\nu,\lambda\alpha}\xi^\lambda_1\xi^\alpha_1
+g^{(0)}_{\mu\nu,\lambda}\xi^\lambda_{1~,\alpha}\xi^\alpha_1
+2\Big[
g^{(0)}_{\mu\lambda,\alpha} \xi^\alpha_1\xi^\lambda_{1~,\nu}
+g^{(0)}_{\lambda\nu,\alpha} \xi^\alpha_1\xi^\lambda_{1~,\mu}
+g^{(0)}_{\lambda\alpha}  \xi^\lambda_{1~,\mu} \xi^\alpha_{1~,\nu}
\Big]
\nonumber \\
&\quad+g^{(0)}_{\mu\lambda}\left(
\xi^\lambda_{1~,\nu\alpha}\xi^\alpha_1
+\xi^\lambda_{1~,\alpha}\xi^\alpha_{1,~\nu}
\right)
+g^{(0)}_{\lambda\nu}\left(
\xi^\lambda_{1~,\mu\alpha}\xi^\alpha_1
+\xi^\lambda_{1~,\alpha}\xi^\alpha_{1,~\mu}
\right)\,,
\end{align}
from which we can extract, as at first order, the transformation behaviour of individual metric perturbation
functions. It is a little trickier to obtain the transformation behaviour of $\psi_2$ from this expression. However, 
we note that the expression in Eq.~(\ref{eq:metrictranssecond}) gives the transformation of $C_{2ij}$, 
namely,\footnote{In the following and for the rest of this section,
we do not split up the spatial part of the gauge transformation
generating vector into scalar and vector parts for brevity. We denote the spatial part of $\xi^\mu$ by
$\xi^i$.}
\be
\label{eq:C2ijtrans}
2\widetilde C_{2ij}=2C_{2ij}+2\H\alpha_2 \delta_{ij}
+\xi_{2i,j}+\xi_{2j,i}+\X_{ij}\,,
\ee
where $\X_{ij}$ contains terms quadratic in the first order
perturbations and is defined below in Eq.~(\ref{Xijdef}),
and so we need intermediate methods in order to extract the transformation of a particular component. Again, we
we do not go into the unnecessary details here, but instead refer the interested reader to Ref.~\cite{MW2008}. 
After this calculation we obtain
\be
\label{transpsi2}
\wt\psi_2=\psi_2-\H\alpha_2-\frac{1}{4}\X^k_{~k}
+\frac{1}{4}\nabla^{-2} \X^{ij}_{~,ij}\,.
\ee
%

%For ease of presentation we only take  scalar
%perturbations into account and work on large scales, where we can
%neglect gradients. Then, $\X_{2ij}$ takes the simple form
%%
%\be
%\label{eq:X2ijdef}
%\X_{2ij}\equiv
%2\Big[\left(\H^2+\frac{a''}{a}\right)\alpha_1^2
%+\H\alpha_1\alpha_1'\Big] \delta_{ij}
%%
%+4\alpha_1\left(C_{1ij}'+2\H C_{1ij}\right)\,,
%\ee
%%
%and \eq{eq:transpsi2} reduces to
%%
%\be
%\label{eq:transpsi2ls}
%\wt\psi_2=\psi_2-\H\alpha_2-\left(\H^2+\frac{a''}{a}\right)\alpha_1^2
%-\H\alpha_1\alpha_1'+2\alpha_1\left(\psi_1'-\H\psi_1\right)\,.
%\ee

We find that the other second order scalar metric perturbations
transform as  \cite{MW2008}
\begin{align}
\label{transphi2}
\widetilde {\iiphi} &= \iiphi+\H\alpha_2+{\alpha_2}'
+\alpha_1\left[{\alpha_1}''+5\H{\alpha_1}' +\left(\H'+2\H^2
\right)\alpha_1 +4\H\phi_1+2\phi_1'\right] \\
&\quad+2{\alpha_1}'\left({\alpha_1}'+2\phi_1\right)
+\xi_{1k}
\left({\alpha_1}'+\H{\alpha_1}+2\phi_1\right)_{,}^{~k}
+\xi_{1k}'\left[\alpha_{1,}^{~k}-2B_{1k}-{\xi_1^k}'\right]\,, \nn \\
\label{transE2}
\wt E_2&=E_2+\beta_2+\frac{3}{4}\nabla^{-2}\nabla^{-2}\X^{ij}_{~~,ij}
-\frac{1}{4}\nabla^{-2}\X^k_{~k}\,,\\
\label{transB2}
\widetilde B_{2} &= B_{2}-\alpha_2+\beta_2' +\nabla^{-2} \XB{}^k_{~,k}
\,,
\end{align}
where  $\XB_i$ and $\X_{ij}$ are defined as
\begin{align}
\label{defXBi}
\XB_{i}
&\equiv
2\Big[
\left(2\H B_{1i}+B_{1i}'\right)\alpha_1
+B_{1i,k}\xi_1^k-2\phi_1\alpha_{1,i}+B_{1k}\xi_{1,~i}^k
+B_{1i}\alpha_1'+2 C_{1ik}{\xi_{1}^k}'
 \Big]\nonumber\\
&+4\H\alpha_1\left(\xi_{1i}'-\alpha_{1,i}\right)
+\alpha_1'\left(\xi_{1i}'-3\alpha_{1,i}\right)
+\alpha_1\left(\xi_{1i}''-\alpha_{1,i}'\right)\nonumber\\
&+{\xi_{1}^k}'\left(\xi_{1i,k}+2\xi_{1k,i}\right)
+\xi_{1}^k\left(\xi_{1i,k}'-\alpha_{1,ik}\right)
-\alpha_{1,k}\xi_{1,~i}^k\,,
\end{align}
and
\begin{align}
\label{Xijdef}
\X_{ij}&\equiv
2\Big[\left(\H^2+\frac{a''}{a}\right)\alpha_1^2
+\H\left(\alpha_1\alpha_1'+\alpha_{1,k}\xi_{1}^{~k}
\right)\Big] \delta_{ij}+2\left(B_{1i}\alpha_{1,j}+B_{1j}\alpha_{1,i}\right)\nonumber\\
&
+4\Big[\alpha_1\left(C_{1ij}'+2\H C_{1ij}\right)
+C_{1ij,k}\xi_{1}^{~k}+C_{1ik}\xi_{1~~,j}^{~k}
+C_{1kj}\xi_{1~~,i}^{~k}\Big]
\nonumber\\
&
+4\H\alpha_1\left( \xi_{1i,j}+\xi_{1j,i}\right)
-2\alpha_{1,i}\alpha_{1,j}+2\xi_{1k,i}\xi_{1~~,j}^{~k}
+\alpha_1\left( \xi_{1i,j}'+\xi_{1j,i}' \right)
\nonumber\\
&+\xi_{1i,k}\xi_{1~~,j}^{~k}+\xi_{1j,k}\xi_{1~~,i}^{~k}
+\xi_{1i}'\alpha_{1,j}+\xi_{1j}'\alpha_{1,i}
+\left(\xi_{1i,jk}+\xi_{1j,ik}\right)\xi_{1}^{~k}
\,.
\end{align}
Furthermore, the vector perturbations transform
as
\begin{align}
\label{transS2}
\widetilde S_{2i}&=S_{2i}-\gamma_2{}_i{}'-\XB_i+\nabla^{-2}\XB^k_{~,ki}
\,,\\
\label{transFi2}
\wt F_{2i} &= F_{2i}+\gamma_{2i}
+\nabla^{-2}\X_{ik,}^{~~~k}-\nabla^{-2}\nabla^{-2}\X^{kl}_{~~,kli}
\,,
\end{align}
and the tensor perturbation which at second order, unlike at first order, is not gauge invariant, as
\begin{align}
\label{transhij2}
\wt h_{2ij}&= h_{2ij}+\X_{ij}
+\frac{1}{2}\left(\nabla^{-2}\X^{kl}_{~~,kl}-\X^k_{~k}
\right)\delta_{ij}
+\frac{1}{2}\nabla^{-2}\nabla^{-2}\X^{kl}_{~~,klij}\nonumber\\
&+\frac{1}{2}\nabla^{-2}\X^k_{~k,ij}
-\nabla^{-2}\left(\X_{ik,~~~j}^{~~~k}+\X_{jk,~~~i}^{~~~k}
\right)
\,.
\end{align}

%%%%%%%%%%%%%%%
\subsection{Four Vectors}
%%%%%%%%%%%%%%%

Finally, we move on to the transformation behaviour of a four vector. The Lie 
derivative of
a vector, $W_\mu$, is given by
\be 
\label{eq:lievec}
\pounds_\xi W_\mu=W_{\mu,\alpha}\xi^\alpha+W_\alpha\xi^\alpha{}_{,
\mu}\,.
\ee 

\subsubsection{First Order}
 
A four vector transforms at first order, from Eq.~(\ref{eq:lievec}), as
\be 
\widetilde{\delta W}_{1\mu}=\delta W_{1\mu}
+W_{(0)\mu}'\alpha_1+W_{(0)\lambda}\xi^\lambda_{1,\mu}\,,
\ee
which gives, for the specific case of the four-velocity, the transformation rule
\be 
\widetilde{V}_{1i}=V_{1i}-\alpha_{1,i}\,,
\ee 
where the quantity $V_{1i}$ is defined as
\be 
\label{eq:transV}
V_{1i}\equiv v_{1i}+B_{1i}\,.
\ee 
Then, on splitting the velocity perturbation into a vector part and the gradient
of a scalar, as 
\be 
v_{1i}=v_{1i}^{\rm V}+v_{1,i}\,,
\ee
recalling that the metric perturbation $B_{1i}$ can be split up as
\be 
B_{1i}=B_{1,i}-S_{1i}\,,
\ee
and making use of Eqs.~(\ref{eq:transB1}) and (\ref{eq:transS1}), we obtain the 
transformation rules
\begin{align}
\widetilde{v_1}&=v_1-\beta_1'\,, \\
\widetilde{v^{{\rm V} i}_1}&=v^{{\rm V} i}_1-\gamma_1{}^i{}'\,.
\end{align}
 
%%%%%%%%%%%%%%%%%%%%%%%%%
\subsubsection{Second Order}
%%%%%%%%%%%%%%%%%%%%%%%%%

At second order we find that a four vector, $W_{2\mu}$, transforms as
\begin{align}
\label{eq:Wtrans}
\widetilde{\delta W}_{2\mu} &= \delta W_{2\mu}+W_{(0)\mu}'\alpha_2
+W_{(0)0}\alpha_{2,\mu}+W_{(0)\mu}''\alpha_1^2
+W_{(0)\mu}'\alpha_{1,\lambda}\xi_1^\lambda\,,\\
&\qquad
+2W_{(0)0}'\alpha_1\alpha_{1,\mu}
+W_{(0)0}\Big(\xi_1^\lambda\alpha_{1,\mu\lambda}
+\alpha_{1,\lambda}\xi_{1,\mu}^\lambda\Big)
+2\Big(\delta W_{1\mu,\lambda}\xi_1^\lambda
+\delta W_{1\lambda}\xi_{1,\mu}^\lambda\Big)\,.\nn
\end{align}

Focussing again on the fluid four velocity, we obtain from the $i$th component of 
Eq.~(\ref{eq:Wtrans}), the transformation behaviour of $V_{2i}$:
\begin{align}
\widetilde{V_{2i}}
&=V_{2i}-\alpha_{2,i}+4\xi_1^k{}'C_{ik}+2\alpha_1'\Big(\xi_1'+B_{1i}-\frac{3}{2}\alpha_{1,i}\Big)
+\alpha_1(2V_{1i}'-\alpha_{1,i}')\nn\\
&\qquad+2(\xi_{1i,k}+\xi_{1k,i})(\xi_1^k{}'-v_1^k)
+2\H\alpha_1(2B_{1i}-v_{1i}+3\xi_{1i}')\nn\\
&\qquad+\xi_1^k(2V_{1i,k}+\alpha_{1,ik})
+\xi_{1,i}^k(2V_{1k}-\alpha_{1,k})
+2\iphi(\xi_1'-2\alpha_{1,i})\,.
\end{align}
Then, using the transformation behaviour of $B_{2i}$, given by
\be 
\widetilde{B_{2i}}=B_{2i}+\xi_{2i}'-\alpha_{2,i}+\XB_i\,,
\ee
we find that the second order three-velocity, $v_{2i}$, transforms as
\begin{align}
\widetilde{v_{2i}}&=
v_{2i}-\xi_{2i}'+\xi_{1i}'(2\iphi+\alpha_1'+2\H\alpha_1)-\alpha_1\xi_1''-\xi_1^k\xi_{1i,k}'\nn\\
&\qquad +\xi_1^k{}'\xi_{1i,k}+2\alpha_1(v_{1i}'-\H v_{1i})-2v_1^k\xi_{1i,k}+2v_{1i,k}\xi_1^k\,.
\end{align}

%%%%%%%%%%%%%%%%%%%%%%%%%%%%%%%%%%%%%%%
\section{Gauge Choices and Gauge Invariant Variables}
\label{sec:gaugechoice}
%%%%%%%%%%%%%%%%%%%%%%%%%%%%%%%%%%%%%%%

As mentioned previously, a central element to Einstein's theory of general 
relativity is the covariance of the theory under coordinate reparametrisation. 
However, a problem arises when undertaking metric cosmological perturbation
theory since the process of splitting the spacetime into a background and a 
perturbation is not a covariant process (see, e.g., Ref.~\cite{covariance} and Section 2.2 of Ref.~\cite{MM2008})
 Therefore, in doing so, one introduces 
spurious gauge modes so that variables depend upon the coordinate choice. 
Observational quantities should not depend upon the choice of coordinate used,
and therefore this so-called {\it gauge problem} of perturbation theory 
seemingly would introduce confusion and erroneous results. However,
as long as one is careful to remove the gauge modes, this will not pose 
us a problem.

The gauge problem was first `solved' by Bardeen in a consistent way in Ref.~\cite{Bardeen:1980kt}, and
has been studied in much detail, and extended beyond linear order, in the decades since. 
The solution lies with the introduction of gauge invariant variables, that is,
variables which no longer change under a gauge transformation. Bardeen constructed
two such variables for scalar perturbations, which happen to coincide with the 
lapse function and curvature perturbation in the longitudinal gauge 
(see Section.~\ref{sec:long}). However, the systematic approach can be extended to other
gauges: one simply inspects the gauge transformation rules presented in the previous
sections, and chooses coordinates such that, e.g., two of the scalar metric perturbations 
are zero. This enables one to remove the gauge dependencies $\alpha_1$ and $\beta_1$,
 rendering the other scalar perturbations gauge invariant. 
 Similarly, this can be extended beyond scalar perturbations, and 
 one can inspect the transformation rules for a vector perturbation, setting it
 to zero, and thus removing the dependency on $\gamma_1^i$.
 
 Finally, let us emphasise the `gauge issue' by considering degrees of freedom of the metric.
In four dimensions, the metric starts with 16 degrees of freedom: 6 are lost because of
symmetry, 4 more because of coordinate invariance ({\emph{gauge}} choice) of the metric and
4 to do with the Hamiltonian constraints, which arise when writing down the field equations
 -- also called a {\emph{gauge}} choice. This leaves 2 degrees of freedom for the two polarisations of the graviton. 
But this is so far not related to perturbation theory.
By perturbing the metric, one introduces further degrees of freedom which are
not addressed by the above -- the option to change the {\emph{gauge}}, or map.
These are the degrees of freedom addressed by the perturbation {\emph{gauge}}
choice.

This paragraph does well to highlight that `gauge' is a well used term in theoretical physics, and 
is often used to mean different (albeit closely related) things. When discussing cosmological perturbation
theory, and for the rest of this thesis, we reserve the phrase `choice of gauge' to mean a specification of
the mapping between the background and the perturbed spacetimes.\\

In this section we concentrate on the linear theory, present the definitions of 
various gauges commonly used throughout the literature, and  define some gauge 
invariant variables. We then give an example of how the theory works at second order,
by describing the uniform curvature gauge. This section is mainly a review, and more details
can be found in Ref.~\cite{MW2008}.

%%%%%%%%%%%
\subsection{Uniform Curvature Gauge}
\label{sec:flatgauge}
%%%%%%%%%%%

A possible choice of gauge is  one in which the spatial metric is unperturbed. 
At linear order, this amounts to setting $\widetilde{\E}=\widetilde{\ipsi}=0$ and $\widetilde{F_{1i}}=0$.
This specifies the gauge generating vector, $\xi^\mu_1$, using Eqs.~(\ref{eq:transpsi1}) and (\ref{eq:transE1}),
as
\begin{align}
\widetilde{\ipsi}&=\ipsi-\H \alpha_1=0 \\
\label{eq:alpha1flat}
&\Rightarrow \alpha_{1\fg}=\frac{\ipsi}{\H}\,,
\end{align}
\begin{align}
\widetilde{\E}&=\E+\beta_1=0\\
\label{eq:beta1flat}
&\Rightarrow \beta_{1\fg}=-\E\,,
\end{align}
and
\begin{align}
\widetilde{F_1{}^i}&=F_1{}^i+\gamma_1{}^i=0\\
\label{eq:gamma1flat}
&\Rightarrow \gamma_{1\fg}{}^i=-F_1{}^i\,.
\end{align}
The other scalars in this gauge, which are then gauge invariant are,
from Eqs.~(\ref{eq:transphi1}) and (\ref{eq:transB1}),
\begin{align}
\widetilde{\phi_{1\fg}}&=\iphi+\ipsi+\left(\frac{\ipsi}{\H}\right)'\,,\\
\widetilde{B_{1\fg}}&=\B-\frac{\ipsi}{\H}-\E'\,.
\end{align}
The gauge invariant vector metric perturbation is, from Eq.~(\ref{eq:transS1}),
\be 
\widetilde{S_{1\fg}{}^i}=S_1{}^i+F_{1,}{}^i\,.
\ee

Note that the scalar field perturbation on spatially flat hypersurfaces is the gauge invariant
Sasaki-Mukhanov variable \cite{Sasaki:1986hm, Mukhanov:1988jd}, often denoted by $\mathcal{Q}$,
\be 
\mathcal{Q}\equiv\widetilde{\dvp_{1\fg}}=\dvp_1+\vpb'\frac{\ipsi}{\H}\,.
\ee
%

%%%%%%%%%%
\subsection{Longitudinal (Poisson) Gauge}
\label{sec:long}
%%%%%%%%%%

The longitudinal gauge is the gauge in which the shear, $\sigma$, vanishes. It is also known as the conformal Newtonian
or orthogonal zero-shear gauge. Its extension to include vector and tensors is called the Poisson gauge.
This gauge is commonly used in the literature, since the remaining gauge invariant scalars in this gauge are
 the variables introduced by Bardeen \cite{Bardeen:1980kt}.

At linear order the temporal part of the gauge generating vector is specified by the choice $\widetilde{\sigma_1}=0$
as, using Eq.~(\ref{eq:transshear1})
\begin{align}
\widetilde{\sigma_1}&=\sigma_1+\alpha_1=0\\
&\Rightarrow \alpha_{1\lg}=-\sigma_1\,,
\end{align}
where the subscript $\lg$ denotes the value in the longitudinal gauge.
The generating vector is fully specified, for scalar perturbations, by making the gauge choice 
$\widetilde{\E}=0$ (and hence $\widetilde{\B}=0$) as
\begin{align}
\widetilde{\E}&=\E+\beta_1=0\\
&\Rightarrow \beta_{1\lg}=-\E\,.
\end{align}
The other two scalar metric perturbations in this gauge are then 
\begin{align}
\widetilde{\phi_{1\lg}}&=\iphi-\H\sigma_1-\sigma_1'\,,\\
\widetilde{\psi_{1\lg}}&=\ipsi+\H\sigma_1\,,
\end{align}
which, by using the definition of the shear, Eq.~(\ref{eq:defshear}), give
\begin{align}
\widetilde{\phi_{1\lg}}&=\iphi-\H(\E'-\B)-(\E'-\B)'\,,\\
\widetilde{\psi_{1\lg}}&=\ipsi+\H(\E'-\B)\,.
\end{align}
These are then identified with the two Bardeen potentials, $\Phi_1$ and 
$\Psi_1$, respectively (or, in Bardeen's notation, $\Phi_{A}Q^{(0)}$ and 
$-\Phi_{H}Q^{(0)}$).

In the Poisson gauge, the generalisation of the longitudinal gauge beyond scalar 
perturbations, the spatial vector component of the gauge transformation generating vector
by demanding that $\widetilde{S_1{}^i}=0$ gives, using Eq.~(\ref{eq:transS1}),
\begin{align}
\widetilde{S_1{}^i}&=S_1{}^i+\gamma_1{}^i{}'=0\\
&\Rightarrow \gamma_{1\lg}{}^i=\int S_1{}^i d\eta +{\mathcal{F}}_1{}^i{}(x^j)\,,
\end{align}
where ${\mathcal{F}}_1{}^i$ is an arbitrary constant three-vector. Thus, when having fixed the
Poisson gauge, there still exists some residual freedom in this choice of constant vector. 

The remaining gauge invariant vector metric perturbation in the Poisson gauge is then
\be 
\widetilde{F_{1\lg}{}^i}=F_1{}^i+\int S_1{}^i d\eta +{\mathcal{F}}_1{}^i{}(x^j)\,.
\ee

%%%%%%%%%%%%%%%%
\subsection{Uniform Density Gauge}
\label{sec:udg}
%%%%%%%%%%%%%%%%

As an alternative to the gauges above, we can define a gauge with respect to the matter
perturbations. One example is the 
 uniform density gauge which is based upon choosing a spacetime foliation such that the 
density perturbation vanishes.

At first order we can fix $\alpha_1$ by demanding that $\widetilde{\drho}=0$. Using 
Eq.~(\ref{eq:rhotrans1}) we obtain
\begin{align}
\widetilde{\drho}&=\drho+\rhob'\alpha_1=0\\
&\Rightarrow \alpha_{1\udg}=-\frac{\drho}{\rhob'}\,.
\end{align}
We still have the freedom to choose the spatial scalar part of the gauge transformation
generating vector, which can be done unambiguously by choosing, e.g., $\wt{\E}=0$.

An especially interesting variable is the curvature perturbation in
this gauge, $\zeta_1$, which is a gauge invariant variable and defined as
\be 
\label{eq:zeta1}
-\zeta_1\equiv\widetilde{\psi_{1\udg}}=\ipsi+\H\frac{\drho}{\rhob'}\,,
\ee
where the sign is chosen to agree with that in Ref.~\cite{Bardeen:1983qw}\footnote{See Ref.~\cite{Wands:2010af}
for a detailed comparison of different sign conventions and notation used for the curvature perturbations
in different papers.}. 
This quantity will come in useful in later chapters because it is conserved on large scales for an
adiabatic system, as will be shown in Section~\ref{sec:dynamicsudg}.

%%%%%%%%%%%%
\subsection{Synchronous Gauge}
%%%%%%%%%%%%%

The synchronous gauge was popular in early work on cosmological perturbation theory, and 
was introduced by Lifshitz in the groundbreaking work of Ref.~\cite{Lifshitz:1945du}. It is 
characterised by $\wt{\iphi}=0=\wt{B_{1i}}$, so that the $g_{00}$ and $g_{0i}$ components 
of the metric are left 
unperturbed and any perturbation away from FRW is confined to the spatial part of the metric.
It can be thought of physically as the gauge in which
$\eta$ defines proper time for all comoving observers. This gauge is also used in many modern 
Boltzmann codes such as CMBFAST \cite{Seljak:1996is}, and is discussed in detail, and compared
 to the longitudinal gauge in Ref.~\cite{Ma:1995ey}.

The synchronous gauge condition fixes the temporal gauge function through
\begin{align}
\wt{\iphi}&=\iphi+\H\alpha_1+\alpha_1'=0 \\
&\Rightarrow  a\iphi+a'\alpha_{1\syn}+a\alpha_{1\syn}'=a\iphi+(a\alpha_{1\syn})'=0\\
&\Rightarrow \alpha_{1\syn}=-\frac{1}{a}\Big(\int a\iphi d\eta - {\cal{C}}(x^i)\Big)\,,
\end{align}
and the spatial gauge functions as 
\begin{align}
\beta_{1\syn}&=\int (\alpha_{1\syn}-B_1)d\eta + {\cal{D}}(x^i)\,,\\
\gamma_{1\syn}^i&=\int S_1^i d\eta + {\mathcal{E}}^i(x^k)\,.
\end{align}
The function ${\cal{D}}_{,i}(x^k)+{\cal{E}}_i(x^k)$ affects the labelling of the initial spatial
hypersurface. However, the function ${\cal{C}}(x^k)$ affects the scalar perturbations, and so the
synchronous gauge does not determine the time-slicing unambiguously. It is therefore not possible
to define gauge invariant variables from the metric in this gauge, since the remaining scalars (for example the 
curvature perturbation), have spurious gauge dependence:
\be 
\wt{\psi_{1\syn}}=\ipsi+\frac{\H}{a}\Big(\int a\iphi d\eta - {\cal{C}}(x^k)\Big)\,.
\ee
In Lifshitz' original work, the gauge mode was removed using symmetry arguments. 
Nowadays a systematic approach is used to remove this gauge mode, and a further gauge condition
is taken, setting the perturbation in the three velocity of the dark matter fluid to zero. Then,
\begin{align}
&\wt{v_{1 \rm cdm, \, \syn}}=v_1-\beta_{1\syn}'=0\\
&\Rightarrow a(v_1+\B)+\int a\iphi d\eta - {\cal{C}}(x^i)=0\,.
\end{align}
However, we then refer to the field equations, Eq.~(\ref{eq:mmtmlin}) which, in the synchronous gauge guarantees
that for the cold dark matter perturbation (where $\cs=0=\delta P_1$), $V_{1{\rm cdm}}=a(v_1+\B)|_{\rm cdm}$
is a constant (in conformal time). Thus, we obtain
\be 
 {\cal{C}}(x^i)=a(v_1+\B)\Big|_{\rm cdm}\,,
 \ee
 which removes the gauge freedom. Note that we can only choose the dark matter fluid with
 which to define the gauge since it is pressureless. The same does not hold true for a fluid with 
 non-zero pressure. In order to see this we need to use an equation that we will derive in full detail
 later from the momentum conservation, Eq.~(\ref{eq:mmtmconsscal}) which, in coordinate time, is
 \be 
 \dot{V_{1}}-3\H\cs V_{1}+\frac{\cs\drho}{\rhob+\Pb}+\phi_1 =0\,. 
 \ee
In the synchronous gauge, $B_1=0=\iphi$, so this becomes
\be 
\label{eq:synch1}
 \dot{av_1}-3\H\cs a v_1+\frac{\cs\drho}{\rhob+\Pb} =0\,.
 \ee
 If we then set the three-velocity to zero, in order to specify the threading, Eq.~(\ref{eq:synch1}) becomes
 \be 
\frac{\cs\drho}{\rhob+\Pb} =0\,,
 \ee
which states that the fluid is pressureless. Therefore, we can only define the synchronous gauge as
comoving with respect to a pressureless fluid, e.g. cold dark matter, and not with respect to a fluid 
with pressure.

%%%%%%%%%%%%%%
\subsection{Comoving Gauge}
%%%%%%%%%%%%%%

Another example of a gauge defined by the gauge transformation of a matter variable
is the comoving gauge. This is defined by choosing the gauge such that the three-velocity
of the fluid vanishes, $\wt v_{1i}=0$. Furthermore, this choice implies that $\wt V_{1i}=0$. Then,
\begin{align}
&\wt V_1 = V_1-\alpha_1=0\\
\Rightarrow  \,\, &\alpha_{1\com}=V_1=v_1+B_1\,,
\end{align}
and
\begin{align}
&\wt v_1=v_1-\beta_1'=0\\
\Rightarrow \,\, &\beta_{1\com}=\int v_1 d\eta + {\cal{F}}(x^k)\,,
\end{align}
where ${\cal{F}}(x^k)$ denotes a residual gauge freedom. However, note that scalar perturbations are 
all independent of ${\cal{F}}(x^k)$. One of the remaining scalars is the curvature peturbation 
on comoving hypersurfaces, which is quite popular in the literature, and often denoted ${\cal{R}}$:
\be 
\label{eq:R}
{\cal{R}}\equiv\wt {\psi_{1\com}}=\ipsi-\H V_1\,.
\ee

%%%%%%%%
\subsection{Beyond Linear Order}
%%%%%%%%%

At second order the procedure is very much the same as at linear order, and the various gauges are defined in an
analogous way to linear order. Of course, the expressions obtained are much longer, due to the
fact that the second order gauge transformations contain many more terms than those at first order. 
Since we do not intend this work to be a comprehensive summary of gauge choice and 
gauge invariant variables (this topic has already been covered in full, gory detail in Ref.~\cite{MW2008}),
we instead sketch how  the gauge choice and construction of gauge invariant
variables work at second order for one choice of gauge: the uniform curvature gauge.

As at linear order, detailed above in Section \ref{sec:flatgauge}, we 
determine the 
components 
of the gauge transformation generating vector through the conditions $\widetilde{\iipsi}=\widetilde{\EE}=0$
and $\widetilde{F_{2i}}={\bm 0}$. The first gives, from Eq.~(\ref{transpsi2}),
\be 
\label{eq:alpha2flat}
\alpha_{2\fg}=\frac{\iipsi}{\H}
+\frac{1}{4\H}\Big(\nabla^{-2}{\mathcal{X}}_{\fg}^{ij}{}_{,ij}-{\mathcal{X}}_{\fg}^k{}_k\Big)\,,
\ee
where ${\mathcal{X}}_{\fg ij}$ is defined as Eq.~(\ref{Xijdef}) with the first order generators given above
in Eqs.~(\ref{eq:alpha1flat}), (\ref{eq:beta1flat}) and (\ref{eq:gamma1flat}), and the second condition gives
\be 
\beta_{2\fg}=-\EE-\frac{3}{4}\nabla^{-2}\nabla^{-2}\X_{\fg}^{ij}{}_{,ij}
+\frac{1}{4}\nabla^{-2}\X_{\fg}^k{}_{k}\,.
\ee
Finally, imposing the condition $\wt{F_{2i}}={\bm 0}$, gives
\be 
\gamma_{2\fg i}= -F_{2i}
-\nabla^{-2}\X_{\fg ik,}{}^{k}+\nabla^{-2}\nabla^{-2}\X_{\fg}^{kl}{}_{,kli}
\,.
\ee

As an example of a gauge invariant variable at second order, we present the 
lapse function in the uniform curvature gauge, $\phi_{2\fg}$. Using 
Eq.~(\ref{transphi2}) along with Eqs.~(\ref{eq:alpha1flat}) and (\ref{eq:alpha2flat}),
we obtain
\begin{align}
\widetilde{\phi_{2\fg}}&=\iiphi+\frac{{\mathcal A}_2}{\H}
+\frac{1}{\H}\Bigg[\Bigg(\H-\frac{\H'}{\H}\Bigg)+\partial_\eta\Bigg]\Big[\nabla^{-2}\X_{\fg}{}^{kl}_{,kl}
-\X_{\fg}^k{}_{k}\Big]\nn\\
&+\frac{1}{\H^2}(\ipsi''\ipsi+2\ipsi^2)+\Bigg(2-\frac{\H''}{\H^3}\Bigg)\ipsi^2
+\frac{1}{\H}\Bigg(5-6\frac{\H'}{\H^2}\Bigg)\ipsi\ipsi'+\frac{2}{\H}\iphi'\ipsi\nn\\
&+\frac{4}{\H}{\mathcal A}_1\iphi+\frac{1}{\H}\Big[{\mathcal A}_1+2\H\iphi\Big]_{,k}\xi_{1\fg}^k
+\frac{1}{\H}\Big[{\mathcal A}_{1,k}-2\H B_{1ik}\Big]\xi_{1\fg}^k{}'\,,
\end{align}
where we have defined
\be 
{\mathcal A}_{(n)}=\psi_{(n)}'+\Bigg(\H-\frac{\H'}{\H}\Bigg)\psi_{(n)}\,.
\ee

%%%%%%%%%%%%%%%%
\section{Thermodynamics of a Perfect Fluid}
%%%%%%%%%%%%%%%%%

Considerable physical insight can be gained by studying the thermodynamic properties of a system 
\cite{Landau_B80, LopezMonsalvo:2010ut}. 
In this section, we study a single perfect fluid system. Such a system is fully characterised by three 
variables, of which only two are independent. Here we choose the energy density, $\rho$ and the entropy,
$S$, as independent variables, with the pressure being given by the equation of state $P=P(\rho,S)$. The pressure
perturbation can then be expanded, at linear order in perturbation theory, as 
\be 
\label{eq:dPexpand}
\dP=\frac{\p P}{\p S}\Big|_\rho\delta S_1+\frac{\p P}{\p \rho}\Big|_S\drho\,.
\ee
This can be cast in the more familiar form 
\bea
\label{eq:dpsplit}
\dP=\dPn_1+\cs\drho\,,
\eea
by introducing the adiabatic sound speed
\be 
\label{eq:cs2def}
\cs\equiv\left.\frac{\p P}{\p \rho}\right|_{S}\,,
\ee
and by defining the non-adiabatic pressure perturbation, which is proportional to the perturbation in the entropy,
as \cite{Wands2000}
\be
\label{eq:dPndef} 
\dPn_1\equiv\left.\frac{\p P}{\p S}\right|_{\rho}\delta S_1\,.
\ee
Note that $\dPn_1$ is gauge invariant. This can be shown by considering the gauge transformation
for the energy density perturbation, Eq.~(\ref{eq:rhotrans1}), along with the analogous equation for the pressure
perturbation. 
One can extend Eq.~(\ref{eq:dPexpand}) to higher order by simply not truncating
the expansion at linear order, that is
\begin{align}
\label{eq:dP2}
\delta P &= \frac{\p P}{\p S}\delta S 
+ \frac{\p P}{\p \rho}\delta\rho  \nonumber\\
&+\frac{1}{2}\Bigg[
\frac{\p^2 P}{\p S^2}\delta S^2 
+\frac{\p^2 P}{\p \rho\p S}\delta\rho \delta S 
+\frac{\p^2 P}{\p \rho^2}\delta\rho^2\Bigg]+\ldots\,.
\end{align}
The entropy, or non-adiabatic pressure perturbation at second order,
for example, is then found from \eq{eq:dP2}, as \cite{Malik2004}
\be
\delta P_{2\rm{nad}}
=\delta P_2-\cs\delta\rho_2-\frac{\p \cs}{\p\rho}\ \delta\rho_1^2\,.
\ee

%%%%%%%%%%%%
\subsection{Entropy or Non-Adiabatic Perturbations from Inflation}
\label{sec:entropyinfl}
%%%%%%%%%%%

One way in which a non-adiabatic pressure perturbation can be
generated is through the relative entropy perturbation between two or
more fluids or scalar fields. For example, the relative entropy or
isocurvature perturbation, at first order, between two fluids denoted
with subscripts $A$ and $B$ is \cite{Malik:2002jb} (dropping here the subscripts
denoting the order, for brevity)
\be 
{\mathcal S}_{AB}=3\H\Bigg(\frac{\delta\rho_B}{\rho_{0B}'}-\frac{\delta\rho_A}{\rho_{0A}'}\Bigg)\,.
\ee
In a system consisting of multiple fluids, the non-adiabatic pressure perturbation is 
split as \cite{ks,MW2008}
\be 
\delta P_{\rm nad}=\delta P_{\rm intr}+\delta P_{\rm rel}\,,
\ee
where the first term is the contribution from the intrinsic entropy perturbation
of each fluid, and the second term is due to the relative entropy perturbation between each fluid,
${\mathcal S}_{AB}$, and is defined as
\be 
\delta P_{\rm rel}\equiv\frac{1}{6\H \rhob'}
\sum_{A,B}\rho_{0A}'\rho_{0B}'\Big(c_B^2-c_A^2\Big){\mathcal S}_{AB}\,,
\ee
where $c_A^2$ and $c_B^2$ are the adiabatic sound speed of each
fluid. Thus, for a multiple fluid system, even when the intrinsic
entropy perturbation is zero for each fluid, there is a non-vanishing
overall non-adiabatic pressure perturbation. This can be extended to
the case of scalar fields by using standard techniques of treating the
fields as fluids.
Much recent work has been focussed on the discussion of entropy, or isocurvature perturbations 
in multi-field inflationary models. See, e.g. Refs.~\cite{Gordon:2000hv, Polarski:1994rz, Lyth:2001nq, 
Alabidi:2010ba, Byrnes:2010em, Sasaki:2008uc, Langlois:2008vk, Bartolo:2001rt, Linde:1985yf, Linde:1996gt,
Langlois:1999dw, Kofman:1985zx, Gao:2009qy, Lalak:2007vi, Byrnes:2006fr, Langlois:2010dz} 
and references therein.

%% file: dynamics.tex
% % % % % % % % % % % % % % % % % % % % % % % % % % % % 
% dynamics.tex AJC
% dynamics and solutions, background, first order
% % % % % % % % % % % % % % % % % % % % % % % % % % % % 

% % % % % % % % % % % % % % % % % % % % % % % % % % % % % % % % 
% =========================================================== %
% % % % % % % % % % % % % % % % % % % % % % % % % % % % % % % % 
\chapter{Dynamics and Constraints}
\label{ch:dynamics}
% % % % % % % % % % % % % % % % % % % % % % % % % % % % % % % % 
% =========================================================== %
% % % % % % % % % % % % % % % % % % % % % % % % % % % % % % % % 

In this Chapter we give the governing equations for perturbations of a FRW universe. 
The background evolution and constraint equations are presented in Chapter~\ref{ch:intro}, so
here we consider the equations at first and second order in perturbation theory. Starting with
the linear order theory we present the governing equations for scalar, vector and tensor
perturbations for a universe filled with a perfect fluid 
without fixing a gauge. We then make three choices of gauge, the uniform density,
uniform curvature and longitudinal gauges, and solve the evolution equations for scalar
perturbations in the latter two case. Next, we present the evolution equation for a scalar
field -- the Klein-Gordon equation -- for a field with both a canonical and non-canonical Lagrangian,
highlighting the importance of the difference between the adiabatic sound speed and the phase
speed for a scalar field system. We finish our discussion of linear perturbations with an
investigation into the perturbations of a system with both a dark matter and a 
dynamical dark energy component.

Having discussed linear order perturbations, we then move on to the second order theory.
We derive the governing equations for a perfect fluid in a perturbed FRW universe from 
energy momentum conservation and the Einstein equations, without fixing gauge. We go on to 
present the equations for scalar and vector perturbations in the uniform curvature gauge,
which will come in useful in Chapter~\ref{ch:vorticity}, and then for scalar perturbations only,
including now the canonical Klein-Gordon equation at second order. Finally, to connect with
other parts of the literature, we give the equations for scalars in the Poisson gauge.

%%%%%%%%%
\section{First Order}
%%%%%%%%

In this section we give the evolution and constraint equations at first order in 
cosmological perturbation theory for a universe filled with a perfect fluid, in a gauge
dependent form,\footnote{By gauge dependent, we mean that 
the equations are presented in a form where the gauge functions
$\alpha, \beta$ and $\gamma^i$ have not yet been specified}
and for all scalar, vector and tensor perturbations, neglecting anisotropic stress.
We also present the equations in some commonly used gauges, present the 
evolution equation for a scalar field -- the Klein-Gordon equation -- for both a canonical and 
non-canonical field, and
solve some of the evolution equations.

First, in gauge dependent form, energy conservation at linear order gives
\be 
\drho'+3\H(\drho+\dP)=(\rhob+\Pb)(3\ipsi'-\nabla^2\E'-v_{1i,}{}^i)\,,
\ee
where $\nabla^2$ denotes the spatial Laplacian, $\nabla^2\equiv\p_k \p^k$,\footnote{Since we are working with a 
background whose spatial submanifold is Euclidean, the position of the Latin indices does not have a meaning. However,
we preserve the position in order to keep with notational conventions such as the summation convention, and for
ease of future generalisation.}
while momentum conservation gives
\be 
\label{eq:mmtmlin}
V_{1i}'+\H(1-3\cs)V_{1i}+\left[\frac{\dP}{\rhob+\Pb}+\phi_1\right]_{,i} =0\,,
\ee
where we have introduced the covariant velocity perturbation as $V_{1i}=v_{1i}+B_{1i}$.
Note, Eq.~(\ref{eq:mmtmlin}) is the analogue of the Euler equation in an expanding background 
(see Chapter~\ref{ch:vorticity}).

The Einstein equations give the energy constraint equation 
\be 
3\H(\ipsi'+\H\phi_1)-\nabla^2(\ipsi+\H\E')+\H \nabla^2 B_1=-4\pi Ga^2\drho\,,
\ee
and the momentum constraint
\be 
\psi_{1,i}'+\frac{1}{4}S_{1i}+\H\phi_{1,i}
=-4\pi Ga^2(\rhob+\Pb)V_{1i}\,.
\ee
Finally, the $(i,j)$ component gives the equation
\begin{align}
E_{1,}{}^i{}_j''+2\H E_{1,}{}^i{}_j'
+\left(\ipsi-\iphi\right)_,{}^i{}_j-\frac{1}{2}(\p_\eta+2\H)\left(2B_{1,}{}^i{}_j-S_1^i{}_{,j}-S_{1j,}{}^i\right)\nn\\
+(\p_\eta^2+2\H\p_\eta-\nabla^2)\left(F_1^{(i}{}_{,j)}+\frac{1}{2}h_1^i{}_j\right)
+\delta^i{}_j\Bigg\{-2\iphi\Big(\H^2-\frac{2a''}{a}\Big)+2\ipsi''\nn\\
+\nabla^2(\iphi-E_1''-\ipsi+B_1')+2\H\Big(2\ipsi'+\nabla^2(B_1-E_1')+\iphi'\Big)\Bigg\}\nn\\
=8\pi Ga^2(\Pb+\dP)\delta^i{}_j\,.
\end{align}
Simplifying now to scalar perturbations, the energy-momentum conservation equations then become
\begin{align}
\label{eq:energyconsscal}
\drho'+3\H(\drho+\dP)=(\rhob+\Pb)(3\ipsi'-\nabla^2(E_1'-v_1))\,,\\
\label{eq:mmtmconsscal}
V_{1}'+\H(1-3\cs)V_{1}+\frac{\dP}{\rhob+\Pb}+\phi_1 =0\,,
\end{align}
and the Einstein equations are
\begin{align}
&3\H(\ipsi'+\H\phi_1)-\nabla^2(\ipsi+\H\E')+\H \nabla^2 B_1=-4\pi Ga^2\drho\,,\\
&\psi_{1}'+\H\phi_{1}
=-4\pi Ga^2(\rhob+\Pb)V_{1}\,,\\
\label{eq:ij}
&E_{1,}{}^i{}_j''+2\H E_{1,}{}^i{}_j'
+\left(\ipsi-\iphi\right)_,{}^i{}_j-B_{1,}{}^i{}_j'-2\H B_{1,}{}^i{}_j
+\delta^i{}_j\Big\{-2\iphi\Big(\H^2-\frac{2a''}{a}\Big)\nn\\
&+\nabla^2(\iphi-E_1''-\ipsi+B_1')+2\ipsi''+2\H\Big(2\ipsi'+\nabla^2(B_1-E_1')+\iphi'\Big)\Big\}\nn\\
&=8\pi Ga^2(\Pb+\dP)\delta^i{}_j\,.
\end{align}
We can obtain from Eq.~(\ref{eq:ij}) two scalar equations. Firstly, by applying the operator $\p_i\p^j$, using the 
method outlined in Ref.~\cite{Malik:2007nd}, we obtain
\begin{align}
\label{eq:didj}
\ipsi''+2\H\ipsi'+\H\iphi'+\iphi\left(\frac{2a''}{a}-\H^2\right)=4\pi Ga^2\dP\,,
\end{align}
and then, taking the trace of Eq.~(\ref{eq:ij}) and using Eq.~(\ref{eq:didj}), we obtain
\begin{align}
\label{eq:shearev}
(B_1-E_1')'+2\H(B_1-E_1')+\iphi-\ipsi=0\,,
\end{align}
which can be recast in terms of the shear ($\sigma_1 \equiv E_1'-B_1$) as
\be 
\sigma_1'+2\H\sigma_1-\iphi+\ipsi=0\,.
\ee

%%%%%%%%%%%%%
\subsection{Uniform Density Gauge}
\label{sec:dynamicsudg}
%%%%%%%%%%%%

We now work in the uniform density gauge which, as we saw in the previous chapter, is specified by 
$\wt{\drho}=0$, and consider only scalar perturbations.  Eq.~(\ref{eq:energyconsscal}) evaluated in this gauge is
\be 
3\H\wt{\delta P_{1\udg}}=(\rhob+\Pb)\Big[3\wt{\psi_{1\udg}}'-\nabla^2(\wt{E_{1\udg}}'-\wt{v_{1\udg}})\Big]\,.
\ee
Noting that $\wt{\psi_{1\udg}}\equiv-\zeta_1$ is the gauge invariant curvature perturbation
on uniform density hypersurfaces, introduced in Eq.~(\ref{eq:zeta1}), and 
$\wt{\delta P_{1\udg}}=\delta P_{\rm nad 1}$, this then becomes
\be 
\label{eq:evzeta}
3\zeta_1'=-\frac{3\H}{(\rhob+\Pb)}\delta P_{\rm nad 1}-\nabla^2(\wt{E_{1\udg}'}-\wt{v_{1\udg}})\,.
\ee
Recall, as mentioned in section \ref{sec:udg}, that the uniform density gauge condition does 
not specify the spatial gauge function, $\beta_1$. The transformations of $\E$ and 
$v_1$ do not depend upon the temporal gauge function, so $\wt{E_{1\udg}}=\wt{E}$, and similarly
for $v_1$. However, the combination $\wt{\E}'-\wt{v_1}$ is a gauge invariant variable,
the three-velocity in the longitudinal gauge, and so Eq.~(\ref{eq:evzeta}) can be written solely in terms of 
gauge invariant variables as 
\be 
3\zeta_1'=-\frac{3\H}{(\rhob+\Pb)}\delta P_{\rm nad 1}-\nabla^2 v_{1\lg}\,.
\ee

We then recover the known result that, on large scales, the evolution of the curvature
perturbation is proportional to the non-adiabatic pressure perturbation:
\be 
\zeta_1'=-\frac{\H}{(\rhob+\Pb)}\delta P_{\rm nad 1}\,.
\ee
Thus, the curvature perturbation on uniform density hypersurfaces is conserved on large
scales for adiabatic perturbations such as those from a single fluid or a single scalar
field \cite{Wands2000}.

Let us now consider whether  a general scalar field can support non-adiabatic perturbations.
We can calculate the non-adiabatic pressure perturbations for a general scalar field with 
Lagrangian ${\mathcal{L}}\equiv{\mathcal{L}}(X,\vp)$, Eq.~(\ref{eq:laggen}). Noting that
the pressure and energy density  are both functions of $X$ and $\vp$,
and so their perturbations can be expanded in a series as
\be 
\label{eq:dpscal}
\dP=\frac{\p P}{\p \vp}\dvp_1+\frac{\p P}{\p X}\delta X_1\,,
\ee
and
\be 
\label{eq:drhoscal}
\drho=\frac{\p \rho}{\p \vp}\dvp_1+\frac{\p \rho}{\p X}\delta X_1\,,
\ee
the non-adiabatic pressure perturbation can then be obtained by substituting Eqs. (\ref{eq:dpscal})
and (\ref{eq:drhoscal}) into Eq.~(\ref{eq:dPexpand}), giving
\be 
\label{eq:dpnadgen}
\delta P_{\rm nad 1}=\rho_{,\vp}\Bigg(\frac{P_{\, \vp}}{\rho_{,\vp}}-\cs\Bigg)\dvp_1
+\rho_{,X}\Bigg(\frac{P_{\, X}}{\rho_{,X}}-\cs\Bigg)\delta X_1\,.
 \ee
In such a system and on large scales, the relationship between
$\dvp_1$ and $\delta X_1$ is \cite{nonad}
\be 
\delta X_1=\ddot{\vpb}\dvp_1\,,
\ee
which, when substituted into Eq.~(\ref{eq:dpnadgen}), along with the definition
for the adiabatic sound speed, $\cs$, yields 
\be 
\delta P_{\rm nad 1}=0\,.
\ee
Therefore, the curvature perturbation on uniform density hypersurfaces is conserved on 
large scales in an expanding universe not only for a canonical scalar field, but for any scalar field with Lagrangian
(\ref{eq:laggen}).\footnote{Note that, in a contracting universe this is not true (see, e.g., Ref.~\cite{Khoury:2008wj}).}
 This result is in agreement with Ref.~\cite{Langlois:2008mn}.

%%%%%%%%%%%%
\subsection{Uniform Curvature Gauge}
\label{sec:dynamicsflat}
%%%%%%%%%%%%%

Now, returning to the full equations, we neglect tensor perturbations 
and work in uniform curvature gauge, where $\wt\E=\wt\ipsi=\wt F_1^i=0$. The 
energy conservation
equation then becomes\footnote{We omit the tildes and gauge subscripts for brevity.}
\be 
\drho'+3\H(\drho+\dP)+(\rhob+\Pb)v_{1i,}{}^i=0\,,
\ee
while momentum conservation gives
\be 
\label{eq:mmtmconflat}
V_{1i}'+\H(1-3\cs)V_{1i}+\left[\frac{\dP}{\rhob+\Pb}+\phi_1\right]_{,i} =0\,.
\ee

Let us now consider the dynamics of scalar, linear perturbations in the uniform curvature gauge. 
The energy-momentum conservation equations then reduce to
\begin{align}
\label{eq:enconsflat}
\drho'+3\H(\drho+\dP)=(\rhob+\Pb)\nabla^2 v_1\,,\\
\label{eq:mmtmconsflat}
V_{1}'+\H(1-3\cs)V_{1}+\frac{\dP}{\rhob+\Pb}+\phi_1 =0\,,
\end{align}
and the Einstein equations give
\begin{align}
\label{eq:scal00}
&3\H^2\phi_1+\H \nabla^2 B_1=-4\pi Ga^2\drho\,,\\
\label{eq:scal0i}
&\H\phi_{1}
=-4\pi Ga^2(\rhob+\Pb)V_{1}\,,\\
&\H\iphi'+\iphi\left(\frac{2a''}{a}-\H^2\right)=4\pi Ga^2\dP\,,\\
&B_1'+2\H B_1+\iphi=0\,.
\end{align}
We will now solve this set of equations for the energy density perturbation of a perfect fluid.
 Let us first rewrite the equations in terms
of the `new' velocity perturbation
\be
{\mathcal{V}_1}\equiv\left(\rho_{0}+P_{0}\right)\left(v_1+B_1\right)\,.
\ee
This enables us to write Eq.~(\ref{eq:scal0i}), using Eq.~(\ref{eq:friedmann}), the background Friedmann
equation, as
\be 
\iphi=-\frac{3}{2}\frac{\H}{\rhob}{\mathcal{V}_1}\,,
\ee 
and using this, Eq.~(\ref{eq:scal00}) becomes
\be 
\nabla^2 B_1=\frac{9}{2}\frac{\H^2}{\rhob}{\mathcal{V}_1}-\frac{3}{2}\frac{\H}{\rhob}\drho\,.
\ee 
Then the evolution equations, Eqs.~(\ref{eq:enconsflat}) and (\ref{eq:mmtmconsflat}) are
\begin{align}
\label{eq:drho1}
&\delta\rho_1'+\frac{3}{2}\H\left(3+w\right)\delta\rho_1+3\H\dP
- k^2{\mathcal{V}_1}-\frac{9}{2}\H^2\left(1+w\right){\mathcal{V}_1}=0\,,\\
\label{eq:dv1}
&{\mathcal{V}_1}'+\frac{\H}{2}\left(5-3w\right){\mathcal{V}_1}+\dP=0\,,
\end{align}
where $w\equiv\Pb/\rhob$ defines the background equation of state,\footnote{We do not demand that $w$
be constant here.} and we are working in
Fourier space, $k$ being the comoving wavenumber.
Equations (\ref{eq:drho1}) and (\ref{eq:dv1}) make up a system of coupled,
linear, ordinary differential equations. Given an equation of state
and initial conditions, this system can be solved immediately (for a
given $k$), numerically.  One can also obtain a qualitative solution
by considering a system of two equations such as this one.  However,
if one wants to solve the system quantitatively, and analytically, it
is easier to rewrite the system as a single second order differential
equation, which we do in the following. We solve \eq{eq:drho1} for ${\mathcal{V}_1}$
and get
\be
\label{eq:defV1}
{\mathcal{V}_1}=\frac{1}{\T}\left(
\delta\rho_1'+\frac{3}{2}\H\left(3+w\right)\delta\rho_1+3\H\delta P_1
\right)\,,
\ee
where
$
\T\equiv k^2+\frac{9}{2}\H^2\left(1+w\right)\,.
$

After some further algebraic manipulations of \eq{eq:defV1}, and using
\eq{eq:dv1}, we arrive at the desired evolution equation
\begin{align}
\label{eq:master}
\delta\rho_1''+\left(7\H-\frac{\T'}{\T}\right)\delta\rho_1'
&+\frac{3}{2}\H\left(3+w\right)\left[
\frac{\H'}{\H}+\frac{w'}{(3+w)}-\frac{\T'}{\T}+\frac{1}{2}\H\left(5-3w\right)
\right]\delta\rho_1\nonumber \\
&+3\H\delta P_1'
+\left[
3\H\left(\frac{\H'}{\H}-\frac{\T'}{\T}\right)
+\frac{3}{2}\H^2\left(5-3w\right)+\T
\right]\delta P_1=0\,.
\end{align}
Equation (\ref{eq:master}) is a linear differential equation, of second
order in (conformal) time. It is valid on all scales, for a single
fluid with any (time dependent) equation of state. Furthermore, it
assumes nothing more than a perfect fluid and hence allows for non-zero
non-adiabatic pressure perturbations.

Having derived a general governing equation (\ref{eq:master}) valid in
any epoch, we now restrict our analysis to radiation domination, where
the background equation of state parameter is $w=\frac{1}{3}$ and the
adiabatic sound speed is $\cs=\frac{1}{3}$.  We
recall from Eq.~(\ref{eq:dpsplit}) that the first order pressure perturbation can be expanded as
\be
\label{dP1split}
\delta P_1\equiv \cs \delta\rho_1 +\dPn_1\,,\nn
\ee
where $\dPn_1$ is the non-adiabatic pressure perturbation, which is proportional to the 
perturbation in the entropy, and is defined in Eq.~(\ref{eq:dPndef}).
Then, the general governing equation, \eq{eq:master}, becomes, during radiation 
domination, where $\H\propto 1/ \eta$
%we then get
%
\begin{align}
&\delta\rho_1''+4\H\left(2+\frac{3\H^2}{k^2+6\H^2}\right)\delta\rho_1'
+\left(6\H^2+\frac{1}{3}(k^2+6\H^2)+\frac{72\H^4}{k^2+6\H^2}\right)
\delta\rho_1\nonumber\\
&
\label{eq:masterrad}
\qquad\qquad\qquad\qquad\qquad\qquad+3\H\dPn_1'
+\left[9\H^2+36\frac{\H^4}{k^2+6\H^2}+k^2
\right]\dPn_1
=0\,.
\end{align}
For the case of zero non-adiabatic pressure perturbations the second
line in \eq{eq:masterrad} vanishes, and the resulting equation can be
solved directly using the Frobenius method, to give \cite{vortest}
\begin{align}
\delta\rho_1({\bm k}, \eta)&=C_1({\bm k})\eta^{-5}\Big[\cos\left(\frac{k\eta}{\sqrt{3}}\right)k\eta-2\sin\left(\frac{k\eta}{\sqrt{3}}\right)\sqrt{3}\Big]\nn\\
&
+C_2({\bm k})\eta^{-5}\Big[2\cos\left(\frac{k\eta}{\sqrt{3}}\right)\sqrt{3}+\sin\left(\frac{k\eta}{\sqrt{3}}\right)k\eta\Big]\,,
\end{align}
where $C_1$ and $C_2$ are functions of the wave-vector, ${\bm k}$, and $k$ is the wavenumber, $k\equiv|{\bm k}|$. For
small $k{\eta}$, the trigonometric functions can be expanded in power
series giving, to leading order, the approximation
\be
\label{eq:drho1highest}
\drho({\bm k},{\eta})\simeq A({\bm k})k{\eta}^{-4}+B({\bm k}){\eta}^{-5}\,,
\ee
for some $A({\bm k})$ and $B({\bm k})$, determined by the initial
conditions.

In order to solve for a non vanishing non-adiabatic pressure, we make
the ansatz that the non-adiabatic pressure grows as the decaying
branch of the density perturbation in \eq{eq:drho1}, i.e.,
\be
\label{eq:ansatz}
\dPn_1({\bm k}, \eta)= D({\bm k}) k^\lambda{\eta}^{-5}\,.
\ee
This assumption is well motivated, since we would expect the
non-adiabatic pressure to decay faster than the energy density, and in fact observations are consistent with
a very small entropy perturbation today. Furthermore, if one is considering a relative 
entropy perturbation between more than one fluid or field, as time increases the system will equilibriate and one
species will dominate.\footnote{Of course, given a specific model of the early universe this relative entropy 
perturbation can be calculated.} This
gives the solution
\begin{align}
\label{eq:drho1nonad}
\drho({\bm k}, \eta)  &={{C_1({\bm k})\,{\eta}^{-5} 
\left[ \cos \left( \frac{k\eta}{\sqrt {3}}
 \right) k\eta-2\,\sin \left( \frac{k\eta}{\sqrt {3}} \right) \sqrt {3} 
\right]
}{}}\nn\\
&\qquad+C_2({\bm k})\,\eta^{-5}{{ \left[ 2\,\cos \left(\frac{k\eta}{\sqrt {3}}
 \right) \sqrt {3}+\sin \left(\frac{k\eta}{\sqrt {3}} \right) k\eta \right] {
}}{}}-3D({\bm k})k^\lambda\eta^{-5}\,.
\end{align}

%
%Anticipating results from the following section, we put the vector dependence
%of the $\drho_1$ and $\dP_{\rm{nad}1}$ into Gaussian random variables $\hat{E}({\bm k})$, allowing
%us to write, for example, $\drho_1({\bm k}, \eta)=\drho_1({ k}, \eta)\hat{E}({\bm k}),$
%where $k=\left|{\bm k}\right|$. Then as a further approximation for the scalar $k$-dependent
%source terms, we can expand
%%
%%
%%note that the
%%solutions for $\drho_1$ and $\dP_{\rm{nad}1}$ depending on ${\bm k}$
%%will in the end only depend on $k=\left|{\bm k}\right|$. 
%%As a further approximation we can therefore expand
%Eq.~(\ref{eq:drho1nonad}) to lowest order in $k\eta$, and use the
%aforementioned ansatz for the non-adiabatic pressure perturbation,
%Eq.~(\ref{eq:ansatz}). Taking the functions $A({k})$ and $D({k})$ to
%be power laws in $k$ this gives
%%
%\be
%\label{input_power}
%\drho_1({k}, \eta)%=\bar{A}(k) k \eta ^{-4}
%=A k^{\beta}\eta^{-4}\,, \hspace{1cm}
%\dPn_1({k}, \eta)= D {k}^\alpha{\eta}^{-5}\,,
%\ee
%%
%where ${A}$ and $D$ are yet unspecified amplitudes and $\beta$ and
%$\alpha$ undetermined powers.
%%For the density perturbation, we can set these by relating to
%%$\zeta$, say, later.

%%%%%%%%%%%%
\subsection{Longitudinal Gauge}
%%%%%%%%%%%%

In this section we consider scalar, linear perturbations in the longitudinal gauge in order
to make a connection with the literature. In the longitudinal gauge, $E_1=B_1=0$, and 
$\ipsi=\Psi_1$, $\iphi=\Phi_1$, and so Eq.~(\ref{eq:shearev}) tells us that, in the absence of anisotropic stress,
 $\Phi_1=\Psi_1$.
Taking this into account, the energy-momentum conservation equations become
\begin{align}
\drho'+3\H(\drho+\dP)=(\rhob+\Pb)(3\Phi_1+\nabla^2 v_1)\,,\\
v_{1}'+\H(1-3\cs)v_{1}+\frac{\dP}{\rhob+\Pb}+\Phi_1 =0\,,
\end{align}
and the Einstein equations are
\begin{align}
\label{eq:long1}
&3\H(\Phi_1'+\H\Phi_1)-\nabla^2\Phi_1=-4\pi Ga^2\drho\,,\\
\label{eq:long2}
&\Phi_{1}'+\H\Phi_{1}
=-4\pi Ga^2(\rhob+\Pb)v_{1}\,,\\
\label{eq:long3}
&\Phi_1''+3\H\Phi_1'+\Phi_1\left(\frac{2a''}{a}-\H^2\right)=4\pi Ga^2\dP\,,
\end{align}

Assuming adiabatic perturbations, in which case $\dP=\cs\drho$ and $w=\cs$, we can combine
Eqs.~(\ref{eq:long1}) and (\ref{eq:long3}), using Eqs.~(\ref{eq:friedmann}) and (\ref{eq:acceleration}) to
give
\be 
\Phi_1''+3\H(1+\cs)\Phi_1'+\cs k^2\Phi_1=0\,,
\ee 
which, in radiation domination (where we recall from section \ref{sec:standardcosmology}
 that $\cs=1/3$ and $\H\propto\eta^{-1}$), then becomes
\be 
\eta\Phi_1''+4\Phi_1'+\frac{1}{3}\eta k^2 \Phi_1 =0\,.
\ee
This equation can then be solved (see, e.g., Ref.~\cite{kishorethesis} for the general method) 
to give 
\begin{align}
\Phi_1({\bm k},\eta)&=\frac{\tilde{C_1} ({\bm k})} {(k\eta)^3}
\Bigg[\cos\left(\frac{k\eta}{\sqrt{3}}\right)k\eta
-\sin\left(\frac{k\eta}{\sqrt{3}}\right)\sqrt{3}\Bigg]\nn\\
&\qquad+\frac{\tilde{C_2} ({\bm k})} {(k\eta)^3}
\Bigg[\cos\left(\frac{k\eta} {\sqrt{3}}\right)\sqrt{3}
+\sin\left(\frac{k\eta}{\sqrt{3}}\right)k\eta\Bigg]\,.
\end{align}

% % % % % % % % % % % % % 
\subsection{Scalar Field Evolution and Sound Speeds}
% % % % % % % % % % % % % %

In this section we consider the dynamics of a scalar field.
Many oscillating systems can be described by a wave equation, that is
a second order evolution equation of the form
\be
\label{eq:wave_equ}
\frac{1}{\cph}\ddot{\mathfrak{X}}-\nabla^2{\mathfrak{X}}+F({\mathfrak{X}},\dot{\mathfrak{X}})=0\,,
\ee
where ${\mathfrak{X}}$ is, for example, the velocity potential, 
$F({\mathfrak{X}},\dot{\mathfrak{X}})$ is the damping term, and $\cph$ is the
phase speed, i.e.~the speed with which perturbations travel through
the system \cite{Sommerfeld, landau}. The situation for a scalar field is not 
dissimilar, and the evolution equation in this case is the Klein-Gordon
equation, where ${\mathfrak{X}}=\vp$ is the scalar field. The Klein-Gordon
equation is obtained by substituting the expressions for the energy density and the pressure
into the energy conservation equation. The background equation is given in Section \ref{sec:inflation}
and here we focus on the perturbations.

There is some confusion in the literature on the meaning of adiabatic
sound speed and phase speed. Although this seems not to affect the
results of previous works, and the adiabatic sound speed is often
simply used as a convenient shorthand for $\dot{\Pb}/\dot{\rhob}$, as
defined in Eq.~(\ref{eq:cs2def}), it is often confusingly used to denote the
phase speed. The adiabatic sound speed defined above describes the response
of the pressure to a change in density at constant entropy,
and is the speed with which pressure perturbations travels through a
classical fluid. A more intuitive introduction of the adiabatic sound
speed is using the compressibility \cite{landau},
\be 
\beta\equiv\frac{1}{\rho}\frac{\p\rho}{\p P}\Big|_{S}
=\frac{1}{\rho\cs}\,,
\ee
 which describes the change in density due to a
change in pressure while keeping the entropy constant.

A collection of scalar fields can also be described
as a fluid, but the analogy is not exact. 
Whereas in the fluid case the phase speed $\cph$ and the adiabatic sound speed
$\cs$ are equal, this is not true in the scalar field case and the
speed with which perturbations travel is given \emph{only} by $\cph$.
The phase speed, $\cph$ 
can be read off from the perturbed Klein-Gordon equation governing the
evolution of the scalar field. This is just a damped wave equation,
like \eq{eq:wave_equ}, and so by comparing the coefficients of the
second order temporal, and second order spatial derivatives, we obtain
the phase speed.\\

We first focus on the canonical scalar field with Lagrangian (\ref{eq:lagcan});
the background equation is given in section \ref{sec:inflation}. The equation for linear perturbations
is, in the uniform curvature gauge,
\be 
\label{eq:KGcan}
\dvp_1''+2\H\dvp_1'+2a^2U_{,\vp}\iphi-\vpb'\nabla^2\B
-\nabla^2\dvp_1-\vpb'\iphi'+a^2U_{,\vp\vp}\dvp_1=0\,.
\ee
We can express this in closed form, that is containing only matter perturbations, by replacing the
metric perturbations using the appropriate field equations. Doing so gives (e.g. Ref.~\cite{Malik:2006ir})
\begin{align}
\dvp_1''&+2\H\dvp_1'-\nabla^2\dvp_1\nn\\
&\qquad
+a^2\Bigg[U_{,\vp\vp}+\frac{8\pi G}{\H}\Bigg(2\vpb'U_{,\vp}+\vpb'^2\frac{8\pi G}{\H}U(\vpb)\Bigg)
\Bigg]
\dvp_1=0\,.
\end{align}

Similarly, we can calculate the Klein-Gordon equation for a scalar field with the non-canonical Lagrangian,
(\ref{eq:laggen}).  We obtain
\begin{align}
\label{eq:pertevolution}
\frac{1}{\cph}\ddot{\dvp}+\Bigg[\frac{3H}{\cph}+C_1\Bigg]\dot{\dvp}
+\Bigg[\frac{k^2}{a^2}+C_2\Bigg]\dvp=0\,,
\end{align}
where the coefficients $C_1$ and $C_2$, both functions of $\vp$ and
$X$, are
\begin{align}
\label{eq:pertevolution2}
C_1&=\frac{\cph}{p_{,X}^2}\left[p_{,\vp}-3Hp_{,X}\dot{\vpb}-
p_{,X\vp}\dot{\vpb}^2\right]\left[3p_{,XX}\dot{\vpb}+\dot{\vpb}^3p_{,XXX}\right] \nn\\
&+\frac{1}{p_{,X}}\left[\dot{\vpb}p_{,X\vp}+\dot{\vpb}^3p_{,XX\vp}\right]\,,
\end{align}

\begin{align}
C_2&=\frac{\cph}{p_{,X}^2}\left[p_{,\vp}-3Hp_{,X}\dot{\vpb}
-p_{,X\vp}\dot{\vpb}^2\right]\nn\\
&\times \left[p_{,X\vp}+\dot{\vpb}^2p_{,XX\vp}
-\frac{4\pi G\dot{\vpb}p_{,X}}{H}\left(5\dot{\vpb}^2p_{,XX}+\dot{\vpb}^4p_{,XXX}
+2p_{,X}\right)   \right] \nonumber\\
&-\frac{4\pi G\dot{\vpb}}{H\cph}\left[3H\dot{\vpb}p_{,X}+p_{,\vp}
-\frac{p_{,X}\dot{\vpb}}{H}(3H^2+2\dot{H})+\cph(\dot{\vpb}^4p_{,XX\vp}+p_{,\vp})\right]\nonumber\\
&+\frac{1}{p_{,X}}\left[\dot{\vpb}^2p_{,X\vp\vp}-p_{,\vp\vp}+3H\dot{\vpb}p_{,X\vp}\right]\,.
\end{align}
The constants in this equation of motion can be interpreted physically: $C_1$ is an additional damping term, and $C_2$ affects 
the frequency of the oscillations.
The phase speed is then read off as
\be
\label{eq:cphgeneral}
\cph=\frac{p_{,X}}{p_{,X}+2\Xb p_{,XX}}\,,
\ee 
which, using Eq.~(\ref{eq:fieldfluidvariables}) reduces to \cite{Garriga:1999vw}
\be
\label{def_cph_final}
\cph=\frac{P_{0,X}}{\rho_{0,X}}\,.
\ee
The adiabatic sound speed for the Lagrangian (\ref{eq:laggen}) is
given by
\be
\label{eq:csgeneral}
\cs=\frac{p_{,X}\ddot{\vpb}+p_{,\vp}}
{p_{,X}\ddot{\vpb}-p_{,\vp}+p_{,XX}\dot{\vpb}^2\ddot{\vpb}+
p_{,X\vp}\dot{\vpb}^2} \,.
\ee
Finally, the adiabatic sound speed for the canonical Lagrangian is
\be
\label{eq:csstandard}
\cs=1+\frac{2U_{,\vp}}{3H{\vpb}'} \,,
\ee
and becomes $\cs=-1$ in slow-roll. The phase speed for a canonical
scalar field is $\cph=1$, as can be read off from Eq.~(\ref{eq:KGcan}).\\

Thus we note that, although  in classical fluid
systems the adiabatic sound speed and phase speed are the same,
 they are different in the scalar field models studied here,
with only the phase speed describing the speed with which
perturbations travel.
We emphasise that only the definition for the adiabatic sound speed,
\eq{eq:cs2def}, enters the definition of the pressure
perturbation. Similarly,
in the adiabatic case when $\delta P_{\rm nad 1}=0$, \eq{eq:dPndef} reduces to
$\dP=\cs\delta\rho_1$. Again, this convenient relation between the
pressure and the energy density perturbation is only correct when
using the adiabatic sound speed, as defined in \eq{eq:cs2def}.

%%%%%%%%%%%%%%%%
\subsection{Combined Dark Energy and Dark Matter System}
%%%%%%%%%%%%%%

In this section, as an application of gauge choice and the linear equations, we consider 
perturbations of a combined dark matter and dark energy system.
The importance of considering dark energy perturbations in such a system is still under discussion.
Recently, in Ref.~\cite{Hwang:2009zj}, it was shown that ignoring the dark energy perturbations
results in gauge dependent predictions. The authors showed this by first taking the governing 
equations for the dark matter/dark energy system calculating, using a numerical simulation,
the matter power spectrum in three gauges: the comoving gauge, the uniform curvature gauge
and the uniform expansion gauge. As expected, the same result was obtained for each gauge. They 
then ignored  perturbations to the dark energy component, and repeated the analysis, this time 
obtaining different results for each of the three gauges. They then concluded that it was crucial
to include the dark energy perturbations when analysing such a system.
Thus one cannot use the familiar evolution
equation for the dark matter density contrast, $\dm\equiv \delta\rho_{\rm c} / \rho_{\rm c}$,
\be 
\label{eq:dmstandard}
\ddot{\dm}+2H\dot{\dm}-4\pi G \rho_{\rm c}\dm
=0\,,
\ee
 which is obtained by ignoring dark energy perturbations once the equations have been
put into a gauge comoving with the dark matter.

 In this section we consider a similar question,
and ask whether we can come to the same result without using numerical simulations,
but instead by using the formalism of cosmological perturbation theory. The section contains 
work published in Ref.~\cite{Christopherson2010}.

We consider here the linear governing
equations for scalars only in coordinate time and without fixing a gauge (dropping the 
subscript `1' in this section for ease of presentation). Though some of the equations
from previous sections are replicated here, since we are using a different time coordinate
 in previous sections, we give all equations used for completeness. 
We assume that the dark matter and dark energy are non-interacting, and 
so energy momentum conservation for each fluid is given by
\be 
\label{eq:emcons}
\nabla_\mu T_{(\alpha)}^\mu{}_\nu=0\,.
\ee
Note that this assumption is for brevity, and to allow us to deal with more
manageable equations. The results highlighted in this work will still
hold in the case of interacting fluids for which
the overall energy-momentum tensor is covariantly conserved, but the components
obey
\be 
\nabla_\mu T_{(\alpha)}^\mu{}_\nu=Q_{(\alpha)\nu}\,,
\ee
where $Q_{(\alpha)\nu}$ is the energy-momentum transfer to the $\alpha$th fluid \cite{Hwang:2001fb, 
MW2008}.

Then, we obtain an evolution equation for each fluid from the energy (temporal) component of 
Eq.~(\ref{eq:emcons}),
\begin{align}
\label{eq:evgen}
\dot{\delta\rho}_\alpha+3H(\delta\rho_\alpha+\delta P_\alpha)
+(\rhob+\Pb)\frac{\nabla^2}{a^2}(V_\alpha+\sigma)
%\qquad\qquad\qquad\qquad\qquad\qquad
=3(\rho_\alpha+P_\alpha)\dot{\psi}\,,
\end{align}
where the covariant velocity potential of each fluid is defined as
$
V_\alpha=a(v_\alpha+B)
$.

Considering the dark matter fluid and the dark energy scalar field, respectively, 
Eq.~(\ref{eq:evgen}) then gives
\bea
\label{eq:fluidev}
&&\dot{\dm}=3\dot{\psi}
-\frac{\nabla^2}{a^2}(\sigma+V_{\rm c})\,,\\
\label{eq:fieldev}
&&\ddot{\dvp}+3H\dot{\dvp}+\Big(U_{,\vp\vp}-\frac{\nabla^2}{a^2}\Big)\dvp
=\dot{\vp}\Big(3\dot{\psi}-\frac{\nabla^2}{a^2}\sigma+\dot{\phi}\Big)-2U_{,\vp}\phi\,,
\eea
where we have used the expressions for the energy density perturbation and pressure perturbation of a scalar field given in
 Section \ref{sec:emtensor}, and note that the dark matter is pressureless.
We have also
used the background evolution equation for the dark energy scalar field
\be 
\ddot{\vpb}+3H\dot{\vpb}+U_{,\vp}=0\,,
\ee
the background conservation equation for the dark matter,
\be 
\dot{\rho_{\rm c}}+3\H\rho_{\rm c}=0\,,
\ee
and the expression
\be 
V_\vp=-\frac{\dvp}{\dot{\vpb}}\,.
\ee 
There is also a momentum conservation equation, coming from the 
spatial component of Eq.~(\ref{eq:emcons}), corresponding to each fluid
\be 
\label{eq:mmtmcons}
\dot{V_\alpha}-3H c_\alpha^2 V_\alpha+\phi+\frac{\delta P_\alpha}{\rho_\alpha +P_\alpha}=0\,,
\ee
where $c_\alpha^2=\dot{P_\alpha}/\dot{\rho_\alpha}$.
\\
The Einstein field equations give (from the previous section or, e.g., Ref.~\cite{seminal})
\begin{align}
\label{eq:00}
&3H(\dot{\psi}+H\phi)-\frac{\nabla^2}{a^2}(\psi+ H \sigma)+4\pi G \delta\rho=0\,,\\
\label{eq:0i}
&\dot{\psi}+H\phi+4\pi G(\rhob+\Pb)V=0\,,\\
\label{eq:ijtracefree}
&\dot{\sigma}+H\sigma-\phi+\psi=0\,,\\
\label{eq:ijtrace}
&\ddot{\psi}+3H\dot{\psi}+H\dot{\phi}+(3H^2+2\dot{H})\phi
-4\pi G\delta P=0\,,
\end{align}
where the total matter quantities are defined as the sum of the quantity for each fluid or field, i.e.
\begin{align}
\delta\rho&=\delta\rho_\vp+\delta\rho_{\rm c}\,,\\
\delta P&=\delta P_{\vp}\,,\\
(\rho+P)V&=(\rho_\vp+P_\vp)V_\vp+\rho_{\rm c}V_{\rm c}\,,
\end{align}
and we have used the fact that the dark matter is pressureless, i.e.
$P_{\rm c}=0=\delta P_{\rm c}$.
%Note that, by treating the scalar field as a fluid, we can write 
%the energy density and pressure perturbation for the dark energy, respectively,
%as
%%
%\begin{align}
%\delta\rho_\vp&=\dot{\vp}\dot{\dvp}-\phi\dot{\vp}^2+U_{,\vp}\dvp\,,\\
%\delta P_{\vp}&=\dot{\vp}\dot{\dvp}-\phi\dot{\vp}^2-U_{,\vp}\dvp\,.
%\end{align}
%
%
Introducing a new variable $\Z$, both for notational convenience and
to assist with the following calculations, defined as
\be 
\Z\equiv3(\dot{\psi}+H\phi)-\frac{\nabla^2}{a^2}\sigma\,,
\ee
we can rewrite Eq.~(\ref{eq:ijtrace}), using Eqs.~(\ref{eq:00})
and (\ref{eq:ijtracefree}), as
\be 
\label{eq:zev}
\dot{\Z}+2H\Z+\Big(3\dot{H}+\frac{\nabla^2}{a^2}\Big)\phi
=4\pi G(\delta\rho+3\delta P)
\,.
\ee
Then, from Eq.~(\ref{eq:fluidev}),
\be 
\label{eq:zrel}
\Z=\dot{\dm}+3H\phi+\frac{\nabla^2}{a^2}V_{\rm c}\,,
\ee 
and from Eq.~(\ref{eq:mmtmcons}) for the dark matter fluid (for which $\cs=0$),
\be 
\label{eq:phirel}
\phi=-\dot{V_{\rm c}}\,.
\ee 
Differentiating Eq.~(\ref{eq:zrel}) gives
\be 
\dot{\Z}=\ddot{\dm}+3(H\phi)\dot{}+\frac{\nabla^2}{a^2}
\Big( \dot{V_{\rm c}}-2H V_{\rm c}\Big)\,.
\ee
Substituting this into Eq.~(\ref{eq:zev}) gives 
\begin{align}
\label{eq:fluidmaster}
\ddot{\dm}+2H\dot{\dm}-4\pi G \rho_{\rm c}\dm
&=
8\pi G(2\dot{\vpb}\dot{\dvp}-U_{,\vp}\dvp)\nn\\
\qquad\qquad\qquad
&+\dot{V_{\rm c}}(6(\dot{H}+H^2)+16\pi G \dot{\vpb}^2)+3H\ddot{V_{\rm c}}\,.
\end{align}
The evolution equation for the field is then obtained solely in terms of
matter perturbations from Eq.~(\ref{eq:fieldev})
by using Eqs.~(\ref{eq:phirel}) and (\ref{eq:zrel}):
\begin{align}
\label{eq:fieldmaster}
\ddot{\dvp}+3H\dot{\dvp}+\Big(U_{,\vp\vp}-\frac{\nabla^2}{a^2}\Big)\dvp
=\dot{\vp}\Big(\dot{\dm}+\frac{\nabla^2}{a^2}V_{\rm c}+3H\dot{V_{\rm c}}
+3H\dot{V_{\rm c}}-\ddot{V_{\rm c}}\Big)
+2U_{,\vp}\dot{V_{\rm c}}\,.
\end{align}

It is worth restating that we have derived these equations in a general form without fixing a gauge.
If we set the dark matter velocity to zero, i.e. $V_{\rm c}=0$, then they reduce to those
presented in, for example, 
Refs.~\cite{Hwang:2001fb, Chongchitnan:2008ry} (for the case of a zero energy-momentum transfer).\\

We now want to consider fixing the gauge freedoms in this set of governing equations. In order to
do so, we need to consider the gauge transformations of the variables under the transformations
 \begin{align}
 \widetilde{t}&=t-\alpha\,,\\
 \widetilde{x}^i&=x^i-\beta_,{}^i\,,
 \end{align}
 where the generating vector is then defined as 
 \be 
 \xi^\mu=(\alpha, \beta_{,}{}^i)
 \,.
 \ee
 Note that, since the choice of time coordinate is different to that used in the previous chapter,
 the following gauge transformations will differ from those presented in, e.g., 
Section \ref{sec:mettrans} due to the different
definition of $\xi^\mu$.
 Then,  scalar quantities such as the field perturbation transform as
 \be 
 \label{eq:fieldtrans}
 \widetilde{\dvp}=\dvp+\dot{\vp}\alpha\,,
 \ee
 the density contrast transforms as
 \be 
 \widetilde{\dm}=\dm+\frac{\dot{\rho_{\rm c}}}{\rho_{\rm c}}\alpha\,,
 \ee 
and the components of the velocity potential as
\be 
\label{eq:veltrans}
\widetilde{v}_\alpha=v_\alpha-a\dot{\beta}\,.
\ee
Furthermore, by considering the transformation behaviour of the metric
tensor we obtain the following transformation rules for the scalar metric
perturbations
\begin{align}
\widetilde{\phi}&=\phi+\dot{\alpha} \,,\\
\widetilde{\psi}&=\psi-H\alpha\,, \\
\widetilde{B}&=B-\frac{1}{a}\alpha+a\dot{\beta}\,,\\
\label{eq:transE}
\widetilde{E}&=E+\beta\,,
\end{align}
and so the scalar shear transforms as
\be 
\widetilde{\sigma}=\sigma+\alpha\,.
\ee
Finally Eq.~(\ref{eq:veltrans})  gives the transformation behaviour of
the components of the covariant velocity potential
\be 
\widetilde{V}_\alpha=V_\alpha-\alpha\,.
\ee

Turning now to the case at hand, choosing the gauge in which the perturbation in the 
dark energy field is zero, $\widetilde{\dvp}=0$,
fixes $\alpha$ as 
\be 
\alpha=-\frac{\dvp}{\dot{\vp}}\,.
\ee
Since none of the gauge transformations of the quantities involved in the governing
equations depend upon the threading $\beta$, we do not need to consider fixing it explicitly
here. (Of course, one can rigorously fix $\beta$ by choosing a suitable gauge condition, 
as outlined in Section \ref{sec:gaugechoice}.)
The governing equations in this gauge are then
\begin{align}
\label{eq:un1}
&\ddot{\widehat\dm}+2H\dot{\widehat\dm}-4\pi G \rho_{\rm c}\widehat{\dm}
=
\dot{\widehat{V_{\rm c}}}(6(\dot{H}+H^2)+16\pi G \dot{\vp}^2)+3H\ddot{\widehat{V_{\rm c}}}\,,\\
\label{eq:un2}
&\ddot{\widehat{V_{\rm c}}}-\frac{2U_{,\vp}}{\dot{\vpb}}\dot{\widehat{V_{\rm c}}}
-\frac{\nabla^2}{a^2}\widehat{V_{\rm c}}=-\dot{\widehat\dm}\,,
\end{align}
where the hat denotes that the variables are evaluated in the uniform field fluctuation gauge. 
That is, in this gauge, $\widehat{\dm}$
and $\widehat{V_{\rm c}}$ are gauge invariant variables  defined as
\begin{align}
\widehat{\dm}=\dm-\frac{\dot{\rho_{\rm c}}}{\rho_{\rm c}\dot{\vpb}}\dvp\,,\hspace{1cm}
\widehat{V_{\rm c}}=V_{\rm c}+\frac{\dvp}{\dot{\vpb}}\,.
\end{align}

Alternatively, choosing a gauge comoving with the dark matter,
in which $\widetilde{V_{\rm c}}=0$ fixes the generating vector as
\be 
\alpha=V_{\rm c}\,,
\ee
and reduces the governing equations to
\begin{align}
\label{eq:com1}
&\ddot{\bar{\dm}}+2H\dot{\bar{\dm}}-4\pi G \rho_{\rm c}\bar{\dm}
=
8\pi G(2\dot{\vpb}\dot{\bar{\dvp}}-U_{,\vp}\bar{\dvp})
\,,\\
\label{eq:com2}
&\ddot{\bar{\dvp}}+3H\dot{\bar{\dvp}}+\Big(U_{,\vp\vp}-\frac{\nabla^2}{a^2}\Big)\bar{\dvp}
=\dot{\vpb}\dot{\bar{\dm}}\,,
\end{align}
where the bar denotes variables in the comoving gauge and we have
\begin{align}
\bar{\dm}=\dm+\frac{\rho_{\rm c}}{\dot{\rho_{\rm c}}}V_{\rm c}\,,\hspace{1cm}
\bar{\dvp}=\dvp+\dot{\vpb}V_{\rm c}\,.
\end{align}

By studying the above systems of equations, it is evident that choosing 
the dark energy field perturbation to be zero is a well defined choice of gauge,
reducing the governing equations to Eqs.~(\ref{eq:un1}) and (\ref{eq:un2}). 
Then, having done so, we are no longer allowed the freedom to make another
choice of gauge. Alternatively, choosing a gauge comoving with the dark
matter uses up the gauge freedom, and so we are not permitted to neglect
the perturbation in the dark energy field. In fact, doing so will result in
erroneous gauge dependent results. It is clearest to see why this is the case
by considering the set of governing equations. By making our choice of gauge we
are left with a set of equations which is gauge invariant: that is, performing a gauge 
transformation will leave the set of equations unchanged. However, by neglecting the
perturbation in the dark energy field after having chosen the gauge comoving with the
dark matter amounts to setting the right hand side of Eq.~(\ref{eq:com1}) to zero. This 
resulting equation will, in general, then no longer be gauge invariant.

Thus, we conclude that
 the dark energy perturbation must be considered in a system
containing a mixture of dark matter and dark energy. Our result is 
consistent with that of Ref.~\cite{Hwang:2009zj}, though we have shown this 
by simply using the formalism of cosmological perturbation theory 
instead of relying on more involved 
numerical calculations.

%Thus, we conclude with the same result as that of Ref.~\cite{Hwang:2009zj}
%namely that, if choosing to work in a gauge comoving with the dark matter,
%one must include perturbations of the dark energy scalar field 
%in order for the system of equations to make sense. Otherwise, if one chooses
%to set the perturbation to the dark energy field to zero then one no longer has
%the freedom to choose another gauge condition, and doing so will make any
%results calculated gauge dependent.

%%%%%%%%%%%%%%%%%
\section{Second Order}
\label{sec:eqnsecond}
%%%%%%%%%%%%%%%%%

We can extend the governing equations presented in the previous section to beyond linear order by simply not 
truncating the expansion of each variable after the first term. Doing so, we obtain equations
with similar structure to those at linear order, however with new couplings between different type of perturbation.
In fact, these couplings will turn out to be the reason for the qualitative difference between the linear
and higher order theories. 
In this section, we will present the full second order equations for scalar, vector and tensor perturbations
in a gauge dependent format. These equations have been derived previously, for example in Ref.~\cite{Noh:2004bc}, though 
are derived in a slightly different way, using the ADM split, and so are written in terms of the extrinsic curvature. The
energy conservation equations have also been derived in Ref.~\cite{MW2008}. 

Conservation of energy momentum again gives us two equations: an energy conservation equation
\begin{align}
&\drhorho' +3\H(\drhorho+\dPP)+(\rhob+\Pb)(C_2^i{}_{i}'+v_{2i,}{}^i)
+2(\drho+\dP)_{,i}v_1^i \nn \\
&+2(\drho+\dP)(C_1^i{}_{i}'+v_{1i,}{}^i)+2(\rhob+\Pb)\Big[(V_1^i+v_1^i)V_{1i}'+v_1^i{}_{,i}\iphi\nn\\
&-2C_{1ij}'C_1^{ij}
+v_1^i(C_1^j{}_{j,i}+2\iphi_{,i})+4\H v_1^i(V_{1i}+v_{1i})\Big] =0\,,
\end{align}
and a momentum conservation equation
\begin{align}
&\Big[(\rhob+\Pb)V_{2i}\Big]'+(\rhob+\Pb)(\phi_{2,i}+4\H V_{2i})
+\dPP_{,i}+2\Big[(\drho+\dP)V_{1i}\Big]'\nn \\
&+2(\drho+\dP)(\phi_{1,i}+4\H V_{1i}) -2(\rhob+\Pb)'\Big[(B_{1i}+V_{1i})\iphi -2C_{1ij}v_1^j\Big]\nn\\
&
+2(\rhob+\Pb)\Big[V_{1i}(C_1^j{}_{j}'+v_1^j{}_{,j})-B_{1i}(\iphi'+8\H\iphi)
+(2C_{1ij}v_1^j)'
\nn \\
&+v_1^j(V_{1i,j}-B_{1j,i}+8\H C_{1ij})  -\iphi(V_{1i}'+B_{1i}'+2\iphi_{,i}+4\H v_{1i})\Big]=0\,,
\end{align}
where we have not expanded the terms in $C_{ij}$ in order to keep the equations compact, but recall 
from Eq.~(\ref{eq:Cij}), that
\be 
C_{ij} = -\psi \delta_{ij}+E_{,ij}+F_{(i,j)}+\frac{1}{2}h_{ij}\,.
\ee

The Einstein equations give the $(0-0)$ component
\begin{align}
&\nabla^2C_2^{j}{}_{j}-C_{2ij,}{}^{ij}
+2\H( -C_2^{i'}{}_{i}+B_2^{i}{}_{,i}+3\H\phi_2)
+2C_1^{j}{}_{j,i}(\frac{1}{2}C_1^{k}{}_{k,}{}^{i}-2C_1^{ik}{}_{,k}) \nn\\
& +2B_1^{i}\Big[ C_1^{j'}{}_{j,i}-C_{1ij,}^{'}{}^{j}
+\frac{1}{2}\left(\nabla^2B_{1i}-B_{1j,i}{}^{j}\right)+2\H\left(C^{1j}{}_{j,i}-2C_{1ij,}{}^j-\phi_{1,i}\right)\Big]\nn\\
& +4C_1^{ij}\Big[2C_{1jk,i}{}^{k}-C_1^{k}{}_{k,ij}-\nabla^2C_{1ij} 
 +2\H(C_{1ij}^{'}-B_{1i,j})\Big]
+2C_{1jk,i}(C_1^{ik}{}_{,}{}^{j}-\frac{3}{2}C_1^{jk}{}_{,}{}^{i}) \nn\\
&+2C_1^{i'}{}_{i}(B_{1j,}{}^{j}-\frac{1}{2}C_1^{j'}{}_{j}+4\H\phi)
+4C_1^{ij}{}_{,i}C_{1jk,}{}^{k}+2C_{1ij}^{'}(\frac{1}{2}C_1^{ij'}-B_1^{j}{}_{,}{}^{i})\nn \\
&
+\frac{1}{2}B_{1j,i}(B_1^{i}{}_{,}{}^{j}+B_1^{j}{}_{,}{}^{i})  -6\H^2(4\phi_1^2-B_{1i}B_1^{i})-B_1^{i}{}_{,i}B_{1j,}{}^{j}
-8\H B_1^{i}{}_{,i}\iphi\nn\\
& =-8\pi Ga^2\Big[2(\rhob+\Pb)V_1^kv_{1k}+\drhorho\Big] \,,
\end{align}
and the $(0-i)$ component
\begin{align}
&C_2^{k}{}_{k,i}^{'}
-C_{2ik,}^{'}{}^{k} -\frac{1}{2}\left(B_{2k,i}{}^k-\nabla^2 B_{2i}\right)-2\H \phi_{2,i}
+16\H\phi_{1,i}\iphi 
-2C_1^{j'}{}_{j}\phi_{1,i} \nn \\
&+2C_{1ij}^{'}\left(2C_1^{kj}{}_{,k}-C_1^{k}{}_{k,}{}^{j}+\phi_{1,}{}^{j}\right)
+4C_1^{kj}\Big[C_{1ik,j}^{'}-C_{1jk,i}^{'}+\frac{1}{2}\left(B_{1k,ij}-B_{1i,kj}\right)\Big]\nn\\
& 
+2B_1^{j}\left(C_{1kj,i}{}^{k}-C_1^{k}{}_{k,ij}+C_{1ik,k}{}^{j}-\nabla^2C_{1ij}-2\H B_{1j,i}\right)
-\Big(B_{1i,j}+B_{1j,i}\Big)\phi_{1,}{}^{j} \nn \\
& +2\left(B_{1i,j}-B_{1j,i}\right)\left(\frac{1}{2}C_1^{k}{}_{k,}{}^{j}-C_1^{jk}{}_{,k}\right)
-2C_{1ik,j}\left(B_1^{k}{}_{,}{}^{j}-B_1^{j}{}_{,}{}^{k}\right)
+2B_1^{j}{}_{,j}\phi_{1,i} \nn\\
& +2\iphi\left[B_{1j,i}{}^{j}-\nabla^2B_{1i}+2\left(C_{1ij,}^{'}{}^{j}-C^{1j'}{}_{j,i}\right)\right] -2C_1^{kj'}C_{kj,i}\nn\\
&=16\pi G\Big[\frac{1}{2}V_{2i}-\iphi(V_{1i}+B_{1i})+2C_{1ik}v_1^k
+(\drho+\dP)V_{1i}\Big] \,.
\end{align}

We present the full $(i-j)$ component in Appendix \ref{ch:AppB}.\\

The equations for scalar perturbations only in a gauge dependent form are then obtained
by substituting $C_{ij}=-\psi \delta_{ij}+E_{,ij}$ and $B_i=B_{,i}$, at 
both first and second order, into the above. The energy conservation equation then becomes
\begin{align}
&\drhorho'+3\H(\drhorho+\dPP)+(\rhob+\Pb)\Big(\nabla^2(\EE'+v_2)-3\iipsi'\Big)
+2(\drho+\dP)_{,i}v_{1}{}^i\nn\\
&+2(\drho+\dP)\Big(\nabla^2(\E'+v_1)-3\ipsi'\Big)
+2(\rhob+\Pb)\Big[(V_{1,i}'+4\H v_{1,i})(V_{1,i}+v_{1,i})\nn\\
&+3\ipsi\ipsi'+\nabla^2 v_1\iphi-(\ipsi\nabla^2E)'
+E_{1,ij}'E_{1,}{}^{ij}+v_{1,}{}^i(2\phi_{1,i}-3\psi_{1,i}+\nabla^2E_{1,i})\Big]=0\,,
\end{align}
while the momentum conservation equation is 
\begin{align}
&\Big[(\rhob+\Pb)V_{1,i}\Big]'+(\rhob+\Pb)\Big(\iiphi+4\H V_2\Big)_{,i}+\delta P_{2,i}
+2\Big[V_{1,i}(\drho+\dP)\Big]'\nn\\
&+2(\drho+\dP)\Big(\iphi+4\H V_1\Big)_{,i}
-2(\rhob+\Pb)'\Big[(V_1+\B)_{,i}\iphi-2(E_{1,ij}v_{1,}{}^j-\ipsi v_{1,i})\Big]\nn\\
&+2(\rhob+\Pb)\Big[V_{1,i}\Big(\nabla^2(\E'+v_1)-3\ipsi'\Big)-B_{1,i}(\iphi'+8\H\iphi)
+v_{1,}{}^j(v_{1,ij}+8\H E_{1,ij})\nn\\
&+2\Big(v_{1,}{}^j E_{1,ij}-\ipsi v_{1,i}\Big)'
-\iphi\Big((V_1+B_1)'+2\iphi+4\H v_1\Big)_{,1}-8\H\ipsi v_{1,i}\Big]=0\,.
\end{align}
Turning now to the Einstein equations, the energy constraint is 
\begin{align}
&3\H(\iipsi'+\H\iiphi)+\nabla^2\Big(\H(\BB-\EE')-\iipsi\Big)
+\nabla^2\B\Big(\nabla^2(\E'-\frac{1}{2}\B)-2\ipsi'\Big)\nn\\
&+\B_{,i}\Big(\H(3\H\B_,{}^i-2\nabla^2\E_,{}^i-2(\ipsi+\iphi)_,{}^i)
-2\ipsi_,{}^i{} '\Big)+2\E_,{}^{ij}(\ipsi-2\H\B)_{,ij}\nn\\
&+4\H(\ipsi-\iphi)\Big(3\ipsi'-\nabla^2(\E'-\B)\Big)
+\E_,{}^{ij}{}'\Big(4\H\E+\frac{1}{2}\E'-\B\Big)_{,ij}\nn\\
&+\ipsi'\Big(2\nabla^2(\E'-2\H\E)-3\ipsi')\Big)
+\ipsi_,{}^i(2\nabla^2\E-3\ipsi)_{,i}
+2\nabla^2\ipsi(\nabla^2\E-4\ipsi)\nn\\
&-12\H^2\iphi^2
+\frac{1}{2}\Big(\B_{,ij}\B_,{}^{ij}+\nabla^2\E_{,j}\nabla^2\E_,{}^j
-\E_{,ijk}\E_,{}^{ijk}-\nabla^2\E'\nabla^2\E'\Big)
\nn\\
&=-4\pi Ga^2\Big(2(\rhob+\Pb)V_{1,}{}^kv_{1,k}+\drhorho\Big)\,,
\end{align}
and the momentum constraint
\begin{align}
&\iipsi_{,i}'+\H\iiphi_{,i}-\E_{,ij}'(\ipsi+\iphi+\nabla^2\E)_,{}^j
+\B_{,ij}(2\H\B+\iphi)_,{}^j\nn\\
&-\Big[\ipsi_{,i}(\nabla^2\E-4\ipsi)\Big]'
-\iphi_{,i}\Big(8\H\iphi+2\ipsi'+\nabla^2(\E'-\B)\Big)\nn\\
&-\B_{,j}\ipsi_{,i}{}^j+2\ipsi_,{}^j{}'\E_{,ij}+\E_{,jk}{} '\E_{,i}{}^{jk}
-\ipsi_{,i}'(\nabla^2\E+4\iphi)-\nabla^2\ipsi\B_{,i}\nn\\
&=-4\pi Ga^2\Big[(\rhob+\Pb)\Big(V_{2,i}-2\iphi(V_1+\B)_{,i}
-4(\ipsi v_{,i}-\E_{,ik}v_{1,}{}^k)\Big)\nn\\
&\qquad\qquad\qquad+2(\drho+\dP)V_{1,i}\Big]\,,
\end{align}
while, from the trace of the $i-j$ component, we obtain
\begin{align}
&3\H(2\iipsi+\iiphi)'+\nabla^2(\EE''+2\EE'+2\iipsi-\BB'-\iiphi+2\H\BB)
-3\iiphi\Big(\H^2-2\frac{a''}{a}\Big)+3\iipsi''\nn\\
&+(\ipsi-\iphi)\Big(12(\ipsi''+2\H\ipsi')+4\nabla^2(\iphi+(\B-\E)')+8\H\nabla^2(\B-\E')\Big)\nn\\
&+\E_,{}^{ij}\Big(8\H(\E'-\B)_{,ij}+2\ipsi_{,ij}-4\iphi_{,ij}-4\B_{,ij}'\Big)
+\E_,{}^{ij}\Big(\frac{5}{2}\E_{,ij}'-\B_{,ij}\Big)\nn\\
&+2\nabla^2\E'\Big(4\iphi'-\nabla^2(\E'-2\B)\Big)
+\nabla^2\E_,{}^i\Big(\nabla^2\E_{,i}+2\iphi_{,i}-4\H\B_{,i}-2\B_{,i}'\Big)\nn\\
&+\ipsi_{,}{}^i\Big(2\nabla^2\E_{,i}-4\H\B_{,i}
-2(\ipsi+\iphi)_{,i}-2\B_{,i}'\Big)
-2\iphi'(\nabla^2\B+12\H\iphi)\nn\\
&-2\iphi_{,i}\iphi_,{}^i+2\nabla^2\ipsi(\nabla^2\E-4\ipsi)
+\ipsi'\Big(3\ipsi'-6\iphi'-8\H\nabla^2\E-2\nabla^2(\E'+\B)\Big)\nn\\
&+2\B_,{}^i\Big(\H(3\B'-2\iphi)-3\ipsi'\Big)_{,i}
+\frac{1}{2}\Big(\B_{,ij}\B_,{}^{ij}-\E_{,ijk}\E_,{}^{ijk}-\nabla^2\B\nabla^2\B\Big)\nn\\
&+4(\E_,{}^{ij}\E_{,ij}''-\ipsi ''\nabla^2\E)
+3\Big(\H^2-2\frac{a''}{a}\Big)
 \Big( 4\iphi^2-\B_{,i}\B_,{}^i \Big)\nn\\
 &=4\pi Ga^2\Big(3\dPP+2(\rhob+\Pb)v_{1,}{}^iV_{1,i}\Big)\,.
\end{align}

We can obtain, from the $i-j$ component a fourth Einstein equation, as at linear order,
by applying the operator $\p_i\p^j$. However, since we do not require this equation for the
work in this thesis, as one can always use the energy-momentum conservation equations
in place of the $i-j$ equations, we omit that equation here.

%%%%%%%%%%%%%%%%%%%%%%%%%%%
\subsection{Uniform Curvature Gauge}
%%%%%%%%%%%%%%%%%%%%%%%%%%%

We now present the second order equations in the uniform curvature gauge, where $\wt\psi=\wt E=\wt F_i=0$, at
both first and second orders. We also consider only scalar and vector perturbations in this section,
i.e. we choose to neglect tensor perturbations, and so $C_{ij}=0$, at both first and second order.
We obtain the energy conservation equation
\begin{align}
\label{eq:flatenergy2}
&\drhorho' +3\H(\drhorho+\dPP)+(\rhob+\Pb)v_{2i,}{}^i
+2(\drho+\dP)_{,i}v_1^i +2(\drho+\dP)v_{1i,}{}^i\nn \\
&+2(\rhob+\Pb)\Big[(V_1^i+v_1^i)V_{1i}'+v_1^i{}_{,i}\iphi
+2v_1^i \iphi_{,i}+4\H v_1^i(V_{1i}+v_{1i})\Big] =0\,,
\end{align}
and momentum conservation equation
\begin{align}
\label{eq:flatmomentum2}
&\Big[(\rhob+\Pb)V_{2i}\Big]'+(\rhob+\Pb)(\phi_{2,i}+4\H V_{2i})
+\dPP_{,i}+2\Big[(\drho+\dP)V_{1i}\Big]'\nn \\
&+2(\drho+\dP)(\phi_{1,i}+4\H V_{1i}) -2(\rhob+\Pb)'(B_{1i}+V_{1i})\iphi \nn\\
&
+2(\rhob+\Pb)\Big[V_{1i} v_1^j{}_{,j}-B_{1i}(\iphi'+8\H\iphi)+v_1^j(V_{1i,j}-B_{1j,i}) 
\nn \\
& -\iphi(V_{1i}'+B_{1i}'+2\iphi_{,i}+4\H v_{1i})\Big]=0\,.
\end{align}
The Einstein equations give us an energy constraint
\begin{align}
&
2\H(B_2^{i}{}_{,i}+3\H\phi_2)
 +2B_1^{i}\Big[ \frac{1}{2}\left(\nabla^2B_{1i}-B_{1j,i}{}^{j}\right)-2\H\phi_{1,i}\Big]\nn\\
&
+\frac{1}{2}B_{1j,i}(B_1^{i}{}_{,}{}^{j}+B_1^{j}{}_{,}{}^{i})  -6\H^2(4\phi_1^2-B_{1i}B_1^{i})-B_1^{i}{}_{,i}B_{1j,}{}^{j}
-8\H B_1^{i}{}_{,i}\iphi\nn\\
& =-8\pi Ga^2\Big[2(\rhob+\Pb)V_1^kv_{1k}+\drhorho\Big] \,,
\end{align}
a momentum constraint
\begin{align}
&\frac{1}{2}\left(\nabla^2 B_{2i}-B_{2k,i}{}^k\right)-2\H \phi_{2,i}
+16\H\phi_{1,i}\iphi -4\H B_1^{j}B_{1j,i}
 \nn \\
& 
-\Big(B_{1i,j}+B_{1j,i}\Big)\phi_{1,}{}^{j}
+2B_1^{j}{}_{,j}\phi_{1,i}  +2\iphi(B_{1j,i}{}^{j}-\nabla^2B_{1i}) \nn\\
&=16\pi G\Big[\frac{1}{2}V_{2i}-\iphi(V_{1i}+B_{1i})
+(\drho+\dP)V_{1i}\Big] \,,
\end{align}
and a third equation from the $(i-j)$ component
\begin{align}
&-\frac{1}{2}(B_2^{i'}{}_{,j}+B_{2j,}{}^{i'})-\phi_{2,}{}^i{}_j-\H ( B_2^i{}_{,j}+B_{2j,}{}^i )\nn\\
&+\delta^i{}_j\Big\{2\phi_2\Big(\frac{2a''}{a}-\H^2\Big)+2\H(B_2^k{}_{,k}+\phi_2')
+B_2^{k'}{}_k+\nabla^2\phi_2\Big\}+2\phi_{1,}{}^i\phi_{1,j}\nn\\
&+B_1^k(B_{1j,}{}^i{}_k+B_1^i{}_{,jk}-2B_{1k,}{}^i{}_j)+(B_1^i{}_{,j}+B_{1j,}{}^i)(\iphi'+B_1^k{}_{,k})
-B_{1k,}{}^iB_1^k{}_{,j}-B_{1j,}{}^kB_1^i{}_{,k}\nn\\
&+B_1^i(B_{1k,j}{}^k-\nabla^2 B_{1j}+4\H\phi_{1,j})
+2\iphi[B_{1j,}{}^{i'}+B_1{}^{i'}{}_{,j}+2\phi_{1,}{}^i{}_j+2\H(B_{1j,}{}^i+B_1{}^i{}_{,j})]\nn\\
&+2\delta^i{}_j\Big\{
\Big(\H^2-\frac{2a''}{a}\Big)(4\phi_1^2-B_{1k}B_1{}^k)-2\phi_1[B_1^{k'}{}_{,k}-\nabla^2\phi_1
+2\H(2\phi_1'+B_1^k{}_{,k})]\nn\\
&
B_1^k[\nabla^2B_{1k}-B_{1l,k}{}^l+2\H(B_{1k}'-\phi_{1,k})]
-\frac{1}{4}(2B_1^k{}_{,k}B_{1l,}{}^l-B_{1l,k}B_1^k{}_,{}^l-3B_1^l{}_{,k}B_{1l,}{}^k)\nn\\
&
-\phi_1'B_1^k{}_{,k}-\phi_{1,k}\phi_{1,}{}^k
\Big\}
=8\pi G\Big[2(\rhob+\Pb)v_1^iV_{1j}+\delta^i{}_j\dPP\Big]\,.
\end{align}

Considering scalar perturbations only, energy conservation gives
\begin{align}
&\drhorho'+3\H(\drhorho+\dPP)+(\rhob+\Pb)\nabla^2v_2
+2(\drho+\dP)_{,i}v_{1}{}^i\nn\\
&2(\drho+\dP)\nabla^2v+2(\rhob+\Pb)\Big[(V_{1,i}'+4\H v_{1,i})(V_{1,i}+v_{1,i})\nn\\
&+\nabla^2 v_1\iphi+2v_{1,}{}^i\phi_{1,i}\Big]=0\,,
\end{align}
and momentum conservation gives
\begin{align}
&\Big[(\rhob+\Pb)V_{1,i}\Big]'+(\rhob+\Pb)\Big(\iiphi+4\H V_2\Big)_{,i}+\delta P_{2,i}
+2\Big[V_{1,i}(\drho+\dP)\Big]'\nn\\
&+2(\drho+\dP)\Big(\iphi+4\H V_1\Big)_{,i}
-2(\rhob+\Pb)'(V_1+\B)_{,i}\iphi\nn\\
&+2(\rhob+\Pb)\Big[V_{1,i}\nabla^2v_1-B_{1,i}(\iphi'+8\H\iphi)
+v_{1,}{}^jv_{1,ij}\nn\\
&-\iphi\Big((V_1+B_1)'+2\iphi+4\H v_1\Big)_{,1}\Big]=0\,.
\end{align}

Finally, the Einstein equations for scalars only in the uniform curvature gauge are
\begin{align}
&3\H^2\iiphi+\nabla^2\H\BB-\iipsi
-\frac{1}{2}(\nabla^2\B)^2
+\H\B_{,i}(3\H\B_,{}^i-2\iphi_,{}^i)
+4\H\iphi\B\nn\\
&-12\H^2\iphi^2
+\frac{1}{2}\B_{,ij}\B_,{}^{ij}
=-4\pi Ga^2\Big(2(\rhob+\Pb)V_{1,}{}^kv_{1,k}+\drhorho\Big)\,,
\end{align}
\begin{align}
&\H\iiphi_{,i}
+\B_{,ij}(2\H\B+\iphi)_,{}^j
-\iphi_{,i}\Big(8\H\iphi-\nabla^2\B\Big)\nn\\
&=-4\pi Ga^2\Big[(\rhob+\Pb)\Big(V_{2,i}-2\iphi(V_1+\B)_{,i}
\Big)+2(\drho+\dP)V_{1,i}\Big]\,,
\end{align}
and
\begin{align}
&3\H\iiphi'+\nabla^2(2\H\BB-\BB'-\iiphi)
%-3\iiphi\Big(\H^2-2\frac{a''}{a}\Big)
%
-\iphi\Big(4\nabla^2(\iphi+\B')+8\H\nabla^2\B\Big)\nn\\
&-2\iphi'(\nabla^2\B+12\H\iphi)-2\iphi_{,i}\iphi_,{}^i
+2\H\B_,{}^i\Big(3\B'-2\iphi\Big)_{,i}\nn\\
&+\frac{1}{2}\Big(\B_{,ij}\B_,{}^{ij}-\nabla^2\B\nabla^2\B\Big)
+3\Big(\H^2-2\frac{a''}{a}\Big)
 \Big( 4\iphi^2-\B_{,i}\B_,{}^i -\iiphi \Big)\nn\\
& =4\pi Ga^2\Big(3\dPP+2(\rhob+\Pb)v_{1,}{}^iV_{1,i}\Big)\,.
\end{align}

The Klein-Gordon equation can be obtained at second order by using the same 
technique as at first order: comparing the energy-momentum tensor for the scalar
field to that of a perfect fluid and using energy conservation.
The equation for a canonical scalar field is
\begin{align}
&\dvp_2''+2\H\dvp_2'-\nabla^2\dvp_2+a^2U_{,\vp\vp}\dvp_2+a^2U_{,\vp\vp\vp}\dvp_1^2
+2a^2U_{,\vp}\iiphi-\vpb'(\nabla^2\BB+\iiphi')\nn\\
&+4\vpb'B_{1,k}\phi_{1,}{}^k+2(2\H\vpb'+a^2U_{,\vp})B_{1,k}B_{1,}{}^k
+4\iphi(a^2U_{,\vp\vp}\dvp_1-\nabla^2\dvp_1)+4\vpb'\iphi\iphi'\nn\\
&-2\dvp_1'(\nabla^2\B+\iphi')-4\dvp_{1,k}'B_{1,}{}^k=0\,,
\end{align}
where we are yet to use the field equations to remove the metric perturbations.
See Refs.\cite{Malik:2006ir, Huston:2009ac, Huston:2010by, Huston:2011vt} for 
the closed form of the Klein-Gordon equation at second order and for 
detailed work on the second order Klein-Gordon equation.

%%%%%%%%%%%%%%%%%
\subsection{Poisson Gauge}
%%%%%%%%%%%%%%%%%

In this section we present the second order equations in the Poisson gauge, in order
to connect with the literature which often uses this gauge (for example, Ref.~\cite{Acquaviva:2002ud}
presents the Einstein equations with scalar field matter in this gauge). The gauge is defined by $\widetilde{E}=0=\widetilde{B}$,
and then $\wt\phi=\Phi$ and $\wt\psi=\Psi$. In the absence of anisotropic stress, as is the case for 
this work, $\Psi_1=\Phi_1$ (though 
note that this does not hold true for the second order variables $\Phi_2$ and $\Psi_2$).
Note also that, in this gauge, $V=v$.

Energy conservation then becomes
\begin{align}
&\drhorho'+3\H(\drhorho+\dPP)+(\rhob+\Pb)\Big(\nabla^2 v_2-3\Psi_2'\Big)
+2(\drho+\dP)_{,i}v_{1}{}^i\nn\\
&+2(\drho+\dP)\Big(\nabla^2 v_1-3\Phi_1'\Big)
+2(\rhob+\Pb)\Big[2(v_{1,i}'+4\H v_{1,i})v_{1,}{}^i\nn\\
&+3\Phi_1\Phi_1'+\nabla^2 v_1\Phi_1
-v_{1,}{}^i\Phi_{1,i}\Big]=0\,,
\end{align}
while the momentum conservation equation is 
\begin{align}
&\Big[(\rhob+\Pb)v_{2,i}\Big]'+(\rhob+\Pb)\Big(\Phi_2+4\H v_2\Big)_{,i}+\delta P_{2,i}
+2\Big[v_{1,i}(\drho+\dP)\Big]'\nn\\
&+2(\drho+\dP)\Big(\Phi_1+4\H v_1\Big)_{,i}
-6(\rhob+\Pb)'\Phi_1 v_{1,i}\nn\\
&+2(\rhob+\Pb)\Big[v_{1,i}\Big(\nabla^2v_1-3\Phi_1'\Big)
+v_{1,}{}^jv_{1,ij}-\Phi_1\Big(v_1'+2\Phi_1+4\H v_1\Big)_{,i}\nn\\
&-2\Big(\Phi_1 v_{1,i}\Big)'
-8\H\Phi_1 v_{1,i}\Big]=0\,.
\end{align}

Then, the Einstein equations (where we here do not decompose the velocity into a scalar and divergenceless vector) are
\begin{align}
&3\H(\Psi_2'+\H\Phi_2)-\nabla^2\Psi_2-3\Phi_1'\Phi_1'-3\Phi_{1,}{}^i\Phi_{1,i}-8\nabla^2\Phi_1\Phi_1
-12\H^2\Phi_1^2\nn\\
&\qquad\qquad\qquad\qquad=-4\pi Ga^2\Big(2(\rhob+\Pb)v_{1}{}^kv_{1k}+\drhorho\Big)\,,
\end{align}
\begin{align}
&\Psi_{2,i}'+\H\Phi_{2,i}+4(\Phi_{1,i}\Phi)'-\Phi_{1,i}(8\H\Phi_1+2\Phi_1')
-4\Phi_{1,i}'\Phi_1\nn\\
&\qquad=-4\pi Ga^2\Big[(\rhob+\Pb)
(v_{2i}-6\Phi_1v_{1i})+2(\drho+\dP)v_{1i}\Big]\,,
\end{align}
and
\begin{align}
\label{eq:gijtrace2}
&\Psi_2''+\H(2\Psi_2+\Phi_2)'+\frac{1}{3}\nabla^2(\Phi_2-\Psi_2)+\Big(\frac{2a''}{a}-\H^2\Big)\Phi_2\nn\\
&+4\Phi_1^2\Big(\H^2-\frac{2a''}{a}\Big)
-2\Phi_{1,}{}^i\Phi_{1,i}
-8\H\Phi_1\Phi_1'-\frac{8}{3}\nabla^2\Phi_1\Phi_1-3(\Phi_1')^2\nn\\
&=4\pi Ga^2\Big(\dPP+\frac{2}{3}(\rhob+\Pb)v_{1}{}^iv_{1i}\Big)\,.
\end{align}
For completeness, we present the fourth field equation, obtained by applying the operator $\partial_i\partial^j$ to
the $i-j$ component of the Einstein equations, Eq.~(\ref{eq:gij_second}):
\begin{align}
\label{eq:gijtracereversed2}
& \Psi_2''+\H(2\Psi_2'+\Phi_2')+\Big(\frac{2a''}{a}-\H^2\Big)\Phi_2=4\pi Ga^2\delta P_2
+8\pi Ga^2(\rhob+\Pb)\nabla^{-2}\partial_i\partial^j(v_1{}^iv_{1j})\nn\\
&-\nabla^{-2}\Big\{2\Phi_{1,k}\Phi_{1,}{}^k{}'+4\Phi_{1,}{}^i{}_j\Phi_{1,i}{}^j-\nabla^2\Big[\Phi_1+\Phi_1''
+2\Phi_1^2\Big(\H^2-\frac{2a''}{a}\Big)\Big]\nn\\
&\qquad\qquad+\Phi_1'\Big(4\nabla^4\Phi_1'-3\nabla^2\Phi_1'+2\H\nabla^2\Phi_1\Big)\Big\}\,,
\end{align}
where $\nabla^{-2}$ is the inverse Laplacian operator. Finally, combining Eqs.~(\ref{eq:gijtrace2}) and 
(\ref{eq:gijtracereversed2}), we obtain
\begin{align}
&\nabla^2(\Psi_2-\Phi_2)=24\pi Ga^2(\rhob+\Pb)\Big[v_1^iv_{1i}-\nabla^{-2}\big(\partial_i\partial^j(v^iv_j)\big)\Big]
+12\Phi_1^2\Big(\H^2-\frac{2a''}{a}\Big)\nn\\
&-18\H\Phi_1\Phi_1'-3\nabla^{-2}\Big\{2\Phi_{1,k}{}'\Phi_{1,}{}^k{}'+4\Phi_{1,}{}^i{}_j\Phi_{1,i}{}^j
+\Phi_1'(4\nabla^4\Phi_1'-3\nabla^2\Phi_1'+2\H\nabla^2\Phi_1)\Big\}\nn\\
&-6\Phi_{1,i}\Phi_{1,}{}^i+\Phi_1\Phi_1''+(\Phi_1')^2+2\Phi_1^2\Big(\H^2-\frac{2a''}{a}\Big)\,.
\end{align}
which is the second order analogue of the equation which, at first order, tells us that the two Newtonian potentials
are identical in the absence of anisotropic stress.

%%%%%%%%%%%%%%
\section{Discussion}
%%%%%%%%%%%%%%%

In this chapter we have presented the dynamical equations for general scalar, vector and
tensor perturbations of a flat FRW spacetime at both linear and second order. The
case of linear perturbations is relatively simple since the different types of perturbation
decouple from one another. However, as one moves beyond linear order, this is no longer 
the case and so things necessarily become more involved. As we have seen, at second order the 
energy-momentum conservation equations, for example, not only depend upon the true second
order perturbations, but also involve terms quadratic in first order perturbations. So, while in this
chapter we have managed to solve the linear equations analytically in, for example, the uniform
curvature gauge, doing this at second order would be far more complicated. But with this 
complication comes great reward, since this coupling between lower order perturbations
provides a source which can result in qualitatively new results, 
and thus new observational phenomena. In the next chapter we will 
discuss one such example: vorticity generation at second order in perturbation theory.

%% file: vorticity.tex
% % % % % % % % % % % % % % % % % % % % % % % % % % % % 
% vorticity.tex 
% vorticity chapter
% % % % % % % % % % % % % % % % % % % % % % % % % % % % 
% Redefine CVSRevision for this section. 
% If you don't want to use CVS tags comment out this line
%\renewcommand{\CVSrevision}{\version$Id: chapter.tex,v 1.3 2009/12/17 18:16:48 ith Exp $}

% % % % % % % % % % % % % % % % % % % % % % % % % % % % % % % % 
% =========================================================== %
% % % % % % % % % % % % % % % % % % % % % % % % % % % % % % % % 
\chapter{Vorticity}
\label{ch:vorticity}
% % % % % % % % % % % % % % % % % % % % % % % % % % % % % % % % 
% =========================================================== %
% % % % % % % % % % % % % % % % % % % % % % % % % % % % % % % % 

Vorticity is a common phenomenon in situations involving fluids in the
`real world' (see e.g.~Refs. \cite{landau, Acheson:1990:EFD}). There has also been some interest
recently in studying vorticity in astrophysical scenarios, including
the inter galactic medium \cite{DelSordo:2010mt, Zhu2010}, but
relatively little attention has been paid to the role that vorticity
plays in cosmology and the early universe.

In this chapter we will consider vorticity in early universe cosmology. Starting
with a summary of vorticity in classical fluid dynamics as motivation, we show that vorticity is generated
by gradients in energy density and entropy. We then consider vorticity in cosmology, for which
we need to use general relativity and cosmological perturbation theory. At linear order, there is no source
term present in the evolution equation,
and any vorticity present in the early universe will decay with the universe expansion. At second order,
however, vorticity is induced by linear order perturbations. As mentioned in Chapter \ref{ch:intro}, while at first
order perturbations of different types decouple, this is no longer true at higher orders. Recent work
in the area of second order gravitational waves has exploited this fact \cite{Baumann:2007zm, Tomita, 
Ananda:2006af, Mollerach:2003nq, Matarrese:1997ay, Osano:2006ew, Assadullahi:2009nf, Sarkar:2008ii,
 Arroja:2009sh, Assadullahi:2009jc},
 as have recent studies
of induced vector perturbations \cite{Lu2008, Lu2009}. 
Though Ref.~\cite{Lu2009} assumed the restrictive condition
of adiabaticity which therefore could not source vorticity at any order, we show that, in analogy 
with the classical case, vorticity is sourced at second order in perturbation theory by a term
quadratic in energy density and entropy (or non-adiabatic pressure) perturbations. Finally, we
present a first estimate of the magnitude and scale dependence of this induced 
vorticity, using the expression for the energy density derived in Section \ref{sec:dynamicsflat}
as an input power spectrum along with a sensible ansatz for that of the non-adiabatic
pressure perturbation. We close the chapter with a discussion of the results, and highlight
some potential observational consequences. The results in the chapter have been published in 
Refs.~\cite{vorticity, vortest, Christopherson:2010dw}.

% % % % % % % % % % % % % % % % % % % % % % % % % % % % % % % % 
% =========================================================== %
% % % % % % % % % % % % % % % % % % % % % % % % % % % % % % % % 
\section{Introduction}
\label{sec:introvort}
% % % % % % % % % % % % % % % % % % % % % % % % % % % % % % % % 
% =========================================================== %
% % % % % % % % % % % % % % % % % % % % % % % % % % % % % % % % 

In classical fluid dynamics the evolution of an inviscid fluid in the
absence of body forces is governed by the Euler, or momentum, equation
\cite{landau}
\be
\label{eq:euler}
\frac{\p{\bm v}}{\p t}+({\bm v}\cdot{\bm \nabla}) {\bm v}=-\frac{1}{\rho}{\bm \nabla} P \,,
\ee
where ${\bm v}$ is the velocity vector, $\rho$ the energy density and $P$ the pressure of the fluid.
The vorticity, ${\bm \w}$, is a vector field and is defined as
\be
{\bm \w} \equiv {\bm \nabla}\times{\bm v}\,,
\ee
and can be thought of as the circulation per unit area at a point in the fluid flow.\footnote{The circulation, $\Gamma$, is defined  
as
\beb 
\Gamma = \oint_C {\bm v}\cdot {\bm dl}\,,
\eeb
where $C$ is the boundary of the surface $S$. Then, using Stokes theorem, this becomes
\beb
\Gamma = \int\!\!\int_S({\bm\nabla}\times{\bm v})\cdot{\bm dS}\,,
\eeb
which gives the result that the vorticity is the circulation per unit area at a point in the fluid flow. $\Box$}
An evolution 
equation for the vorticity can be obtained by taking the curl of \eq{eq:euler}, which gives
\be
\label{eq:vorclassical}
\frac{\p{\bm \omega}}{\p t}
={\bm \nabla}\times({\bm v}\times{\bm \omega})
+\frac{1}{\rho^2}{\bm\nabla}\rho\times{\bm \nabla}P\,.
\ee
The second term on the right hand side of \eq{eq:vorclassical}, often called the baroclinic term in the literature,
then acts as a source for the vorticity. Evidently, this term vanishes if lines of constant energy
and pressure are parallel, or if the energy density or pressure are constant. 
A special class of fluid for which the former is true is a barotropic fluid, defined such that
the equation of state is a function of the energy density only, i.e. $P\equiv P(\rho)$, and so
$(1/ \rho^2){\bm\nabla}\rho\times{\bm \nabla}P=0$.

For a barotropic fluid, the vorticity evolution equation, Eq.~(\ref{eq:vorclassical}), 
can then be written, by using vector calculus identities, as
\be
\label{eq:vorclass2}
\frac{D {\bm \omega}}{D t}\equiv\frac{\p {\bm \omega}}{\p t}+({\bm v}\cdot{\bm \nabla}){\bm \omega}
=({\bm \omega}\cdot{\bm \nabla}){\bm v}-{\bm \omega}({\bm \nabla}\cdot{\bm v})\,,
\ee
where $D/Dt$ denotes the convective, or material derivative, which is commonly used in fluid
dynamics. From \eq{eq:vorclass2} it is clear that the vorticity vector has no source, in this case, and so
${\bm \w={\bm 0}}$ is a solution.

Allowing for a more general perfect fluid
with an equation of state depending not only on the energy density, but of the form
 $P\equiv P(S,\rho)$ will mean that, in general, the baroclinic term is no longer vanishing,
and so acts as a source for the evolution of the vorticity. This is Crocco's theorem \cite{crocco} which states that
vorticity generation is sourced by gradients in entropy in classical fluid dynamics.

% % % % % % % % % % % % % % % % % % % % % % % % % % % % % % % % 
% =========================================================== %
% % % % % % % % % % % % % % % % % % % % % % % % % % % % % % % % 
\section{Vorticity in Cosmology}
\label{sec:vortcosmo}
% % % % % % % % % % % % % % % % % % % % % % % % % % % % % % % % 
% =========================================================== %
% % % % % % % % % % % % % % % % % % % % % % % % % % % % % % % % 

In General Relativity, the  vorticity tensor is defined as the projected anti symmetrised
covariant derivative of the fluid four velocity, that is \cite{ks}
\be
\label{eq:defomega}
\omega_{\mu\nu}= \P^{~\alpha}_{\mu}
\P^{~\beta}_{\nu} u_{[\alpha; \beta]}
\,,
\ee
where the projection tensor $\P_{\mu\nu}$ into the instantaneous fluid 
rest space is given by
\be
\label{eq:projection}
\P_{\mu\nu}=g_{\mu\nu}+u_{\mu}u_{\nu}.
\ee
Note that, in analogy with the classical case, it is possible to define
a vorticity vector as $\w_\mu=\frac{1}{2}\varepsilon_{\mu\nu\gamma}\w^{\nu\gamma}$,
where $ \varepsilon_{\mu\nu\gamma}\equiv u^\delta \varepsilon_{\delta\mu\nu\gamma}$ 
is the covariant permutation tensor in the fluid rest space
(see Refs.~\cite{Lu2009, hawkingellis}).
However, since this is less general we choose to work
with the vorticity tensor when deriving the equations, and only switch to using the vorticity vector when
solving the equation in Section~\ref{sec:solvevort}.

The vorticity tensor can then be decomposed in the usual way,
up to second
order in perturbation theory, as 
$\omega_{ij}\equiv\omega_{1ij} +\frac{1}{2}\omega_{2ij}$. 
Working in the uniform curvature gauge, and considering only
scalar and vector perturbations, we can obtain the components
of the vorticity tensor by substituting the expressions for the 
fluid four velocity, \eq{eq:fourvel1}, along with the metric tensor
 into \eq{eq:defomega}. At first order this gives us
\be
\label{eq:omegafirst}
\omega_{1ij}=aV_{1[i,j]}\,,
\ee
and at second order 
\be
\label{eq:omegasecond}
\w_{2ij}=aV_{2[i,j]}+2a\left[V_{1[i}^\prime V_{1j]}
+\phi_{1,[i}\left(V_{1}+B_{1}\right)_{j]}
-\phi_1B_{1[i,j]}\right] \,.
\ee
The first order vorticity is gauge invariant. In
order to see 
this we 
recall, from Eq.~(\ref{eq:transV}), that 
$V_{1i}$ transforms under a gauge transformation as
\be 
\widetilde{V_{1i}}=V_{1i}-\alpha_{1,i}\,,
\ee 
so the first order vorticity transforms as
\be 
\widetilde{\w_{1ij}}=a\widetilde{V_{1[i,j]}}=a(V_{1i,j}-\alpha_{1,ij}-V_{1j,i}+\alpha_{1,ji})
=\w_{1ij}\,,
\ee 
and is therefore gauge invariant.

\section{Vorticity Evolution}

In order to obtain an evolution equation for the vorticity at first order, we take the time derivative of
\eq{eq:omegafirst} to get
\be 
\label{eq:om1deriv}
\omega_{1ij}'=a'V_{1[i,j]}+aV_{1[i,j]}' \,.
\ee
Noting that, from \eq{eq:mmtmconflat}, 
\be 
V_{1[i,j]}'+\H(1-3\cs)V_{1[i,j]}+\left[\frac{\dP}{\rhob+\Pb}+\phi_1\right]_{,[ij]} =0\,,
\ee 
which gives
\be 
V_{1[i,j]}'=-\H(1-3\cs)V_{1[i,j]}=-\frac{1}{a}\H(1-3\cs)\w_{1ij}\,,
\ee
and so, from  \eq{eq:om1deriv}, 
\be 
\omega_{1ij}'-3\H\cs\w_{1ij}=0\,.
\ee
This reproduces the well known result that, during radiation domination, 
$|\w_{1ij}\w_1^{ij}|\propto a^{-2} $ in the absence of an anisotropic stress term \cite{ks}.\\

At second order things get somewhat more complicated. We now take the time derivative of 
\eq{eq:omegasecond}, to give
\begin{align}
\w_{2ij}'&=a' V_{2[i,j]}+aV_{2[i,j]}
+2a'\left[V_{1[i}^\prime V_{1j]}
+\phi_{1,[i}\left(V_{1}+B_{1}\right)_{j]}
-\phi_1B_{1[i,j]}\right] \nn\\
&+2a\Big[V_{1[i}^{\prime\prime} V_{1j]}+V_{1[i}^{\prime} V^\prime_{1j]}
+\phi^\prime_{1,[i}\left(V_{1}+B_{1}\right)_{j]}+\phi_{1,[i}\left(V_{1}+B_{1}\right)^\prime_{j]}\nn\\
&\qquad\qquad-\phi_1^\prime B_{1[i,j]}-\phi_1 B^\prime_{1[i,j]}\Big] \,.
\end{align}
Therefore we now must use the first order conservation and field equations to eliminate the first 
order metric perturbations as well as the
second order conservation equations in order to eliminate  the second order metric perturbation variables. 
This process involves simple algebra, but is very tedious and so we omit the intermediate steps
and instead quote the result. We arrive at the evolution equation
for the second order vorticity
\begin{align}
\label{eq:vorsecondevofull}
&\w_{2ij}^\prime-3\H\cs\w_{2ij}
+2\left[
\left(
\frac{\dP+\drho}{\rhob+\Pb}\right)^\prime+V^k_{1,k}-X^k_{1,k}
\right]\w_{1ij}  \\
&
+2\left(V^k_1-X^k_1\right)\w_{1ij,k}-2\left(X^k_{1,j}-V^k_{1,j}\right)\w_{1ik}
+2\left(X^k_{1,i}-V^k_{1,i}\right)\w_{1jk}\nn \\
&=\frac{a}{\rhob+\Pb}\Big\{3\H\left(V_{1i}\dPn_{1,j}-V_{1j}\dPn_{1,i}\right)\nn\\
&
\qquad\qquad\qquad+\frac{1}{\rhob+\Pb}\left(\drho_{,j}\dPn_{1,i}-\drho_{,i}\dPn_{1,j}\right)\Big\}\nn \,, 
\end{align}
where $X_{1i}$ is given entirely in terms of matter perturbations as 
\be
X_{1i}=\nabla^{-2}\left[\frac{4\pi G a^2}{\H}
\left(\drho_{,i}-\H(\rhob+\Pb)V_{1i}\right)\right]\,.
\ee
Eq.~(\ref{eq:vorsecondevofull}) then shows that the second order vorticity is sourced by terms
quadratic in linear order perturbations. 

In fact, even assuming zero first order vorticity, i.e. $\w_{1ij}=0$, the second order vorticity evolves as
\be 
\label{eq:vorsecondevolution}
\w_{2ij}^\prime -3\H\cs\w_{2ij}
=\frac{2a}{\rhob+\Pb}\left\{3\H V_{1[i}\dPn_{1,j]}
+\frac{\drho_{,[j}\dPn_{1,i]}}{\rhob+\Pb}\right\}\,,
\ee
 and so we see that there is a non zero source term for the vorticity at second order in perturbation theory which
 is, in analogy with classical fluid dynamics, made up of gradients in entropy and density perturbations. 
 Note that, in the
 absence of a non-adiabatic pressure perturbation, we recover the result of Ref.~\cite{Lu2009} that there is no
 vorticity generation.

% % % % % %  % % % % % % % % % % % % % % % %%  %% 
\section{Solving the Vorticity Evolution Equation}
\label{sec:solvevort}
% % % % %% % %  % % % % % % % % % % % % % % % %

Having derived an evolution equation for the second order vorticity in the previous section,
we now seek an analytic solution to this equation (or, more precisely, the power spectrum
of the vorticity governed by this evolution equation), using suitable, realistic approximations
for the input power spectra.

% % % % % % % % % % % % % % % % % % % % % %
\subsection{The Vorticity Power Spectrum}
 % % % % % % % % % % % % % % % % % % % % % % %

In order to keep our results conservative and our calculation as simple as
possible and hence analytically tractable, we assume that the source
term in \eq{eq:vorsecondevolution} is dominated by the second
term.\footnote{This is a reasonable assumptions since we are working on small scales
and the first term has a prefactor $1/k^2$.}
Then, choosing the radiation
era as our background in which $\cs=1/3$, the evolution equation
simplifies to
\begin{align}
\label{eq:vorevrad}
\w_{ij}^\prime -\H\w_{ij}
=\frac{9a}{8\rhob^2}{\delta\rho_{,[j}\dPn_{,i]}}
%\equiv S_{ij}
\,,
\end{align}
where we note that, for the remainder of this Chapter, we omit the subscripts denoting the order
of the perturbation in order to avoid notational ambiguities, and to keep the expressions as compact
and clear as possible: the vorticity is a second order quantity and the energy density and non-adiabatic pressure
perturbations are first order quantities. We define the right hand side of Eq.~(\ref{eq:vorevrad}) to be the
source term,
\be 
S_{ij}({\bm x},\eta)\equiv \frac{9a(\eta)}{8\rhob(\eta)^2}{\delta\rho_{,[j}\dPn_{,i]}}\,,
\ee
which on defining the function $f(\eta)$ as
\be 
f(\eta)=\frac{9 a}{16 \rhob^2}\,,
\ee
is written, for convenience, as
\be 
S_{ij}({\bm x},\eta)\equiv 2 f(\eta){\delta\rho_{,[j}\dPn_{,i]}}\,.
\ee

Since we want to solve the evolution equation we move to Fourier space, in which the 
source term becomes the convolution integral 
\be 
\label{eq:Sijdef}
S_{ij}({\bm k}, \eta)=-\frac{f(\eta)}{(2\pi)^{3/2}}
\int d^3\tilde{ k}(\tilde{k}_i k_j-\tilde{k}_j k_i)
\dPn(\tilde{\bm k}, \eta)\drho({\bm k}- \tilde{\bm k})\,,
\ee
where ${\bm k}$ is the wavevector, as usual. Instead of considering the vorticity tensor $\w_{ij}$,
it is easier, and more natural in this case, to work, in analogy with the classical case, with the 
vorticity vector. This is defined as\footnote{Though strictly this is a covector not a vector, since
we are working in a flat background the two are equivalent up to raising or lowering of indices by the
Kronecker delta. Therefore, we are slightly loose with terminology here, and do not 
differentiate between the two.}
\be 
\w_i({\bm x},\eta) = \epsilon_{ijk}\w^{jk}({\bm x},\eta)\,,
\ee
where $\epsilon_{ijk}$ is the totally antisymmetric tensor, and we can define a source vector in
an analogous way:
\be 
\label{eq:Sdef}
S_i({\bm x},\eta)=\epsilon_{ijk}S^{jk}({\bm x},\eta)\,.
\ee
Note that the vorticity is an axial vector (that is, it arises from the generalisation of the 
cross product), and so both $\w_i$ and $S_i$ are pseudovectors. Thus, under the transformation of the 
argument ${\bm x} \to -{\bm x}$ the vector vorticity transforms as $\w_i \to -\w_i$, and similarly for $S_i$.
The source vector is then Fourier transformed as
\begin{align}
\label{eq:sourcetransform}
S_i({\bm x},\eta)=\frac{1}{(2\pi)^{3/2}}\int d^3{\bm k}\,S_i({\bm k},\eta){\rm e}^{i{\bm k}\cdot{\bm x}}\,,
\end{align}
with the source vector in Fourier space then being split up as
\be 
\label{eq:sourcesplit}
S_i({\bm k},\eta)={\mathcal{S}}_A({\bm k},\eta)e^A_i\,,
\ee
where $\S_{A}$ are the amplitudes, with $A\in\{1,2,3\}$, and the basis vectors are
\be
\left\{e_i^{1}, e_i^{2}, e_i^{3}\right\}=
\left\{{ e}_i({\bm k}), {\bar{e}_i}({\bm k}), 
\frac{{k}_i}{|{\bm k}|}\equiv \hat{k}_i\right\}\,,
\ee
%
%so that 
%%
%\be 
%{\mathcal{S}}_A({\bm k},\eta)=S_i({\bm k},\eta)\, e_A^i\,.
%\ee
%
In order to keep a right handed orthonormal basis under the sign reversal of ${\bm k}$ (${\bm k}\to 
-{\bm k}$), the basis vectors must obey
\begin{align}
{e}_i(-{\bm k})&={e}_i({\bm k})\,,\\
\bar{e}_i(-{\bm k})&=-\bar{ e}_i({\bm k}) \, ;
\end{align}
the basis vectors are also cyclic:
\be 
\epsilon_{ijk}\, e_1^j e_2^k=e_{3i}\,.
\ee

Given these definitions, the evolution equation, Eq.~(\ref{eq:vorevrad}), can then be written as
\be
\label{eq:evpol}
\w_{A}'({\bm k}, \eta)-\H\w_{A}({\bm k},\eta)=\S_{A}({\bm k},\eta)\,,
\ee
for each basis state, $A$ (though we omit the subscript in the next few lines, since the evolution equations is 
the same for each polarisation).
The left hand side of Eq.~(\ref{eq:evpol}) can be expressed as an exact
derivative, giving
\be 
a \Bigg(\frac{\w({\bm k},\eta)}{a}\Bigg)'={\mathcal{S}}({\bm k},\eta)\,,
\ee
which, in radiation domination when $a=\eta$, becomes
\be 
\Bigg(\frac{\w({\bm k},\eta)}{\eta}\Bigg)'=\eta^{-1}{\mathcal{S}}({\bm k},\eta)\,.
\ee
This can then be integrated to give
\be 
\label{eq:timesolved}
\w({\bm k},\eta)=\eta \int_{\eta_0}^\eta\tilde{\eta}^{-1} \S({\bm k},\tilde{\eta})d\tilde{\eta}\,,
\ee
for some initial time $\eta_0$. Having solved the temporal evolution of the vorticity, we now move
on to considering the power spectrum. In analogy with the standard case for scalar perturbations,
we define the power spectrum of the vorticity ${\mathcal{P}}_\w$ as
\be
\label{eq:PSomega}
\left<{\w}^*({\bm k}_1, \eta){\w}({\bm k}_2, \eta)\right>
=\frac{2\pi}{k^3}\delta({\bm k}_1-{\bm k}_2){\mathcal{P}}_\w(k, \eta)\,,
\ee
where here the star denotes the complex conjugate, and $k=|{\bm k}|$ is the wavenumber, as usual.
On substituting Eq.~(\ref{eq:timesolved}) into Eq.~(\ref{eq:PSomega}), we can write the correlator 
for the vorticity as
\be
\label{eq:PSomegawithS}
\left<{\w}^*({\bm k}_1, \eta){\w}({\bm k}_2, \eta)\right>
=\eta^2\int_{\eta_0}^\eta \tilde{\eta}_1^{-1}
\int_{\eta_0}^\eta \tilde{\eta}_2^{-1} 
\left<{\S}^*({\bm k}_1, \tilde{\eta}_1){\S}({\bm k}_2, \tilde{\eta}_2)\right> 
d\tilde{\eta}_1d\tilde{\eta}_2\,,
\ee
and so, in order to obtain the vorticity power spectrum, we must calculate the 
correlator of the source term. Thus, we need to consider how the Fourier amplitudes, 
$\S_A({\bm k},\eta)$, behave under complex conjugation. From Eqs.~(\ref{eq:sourcetransform})
and (\ref{eq:sourcesplit}) we can write
\begin{align}
\label{eq:Sdef}
S_i({\bm x},\eta)=\frac{1}{(2\pi)^{3/2}}\int d^3{k}
\Big[\S_{1}({\bm k},\eta)\,e_i({\bm k})+\S_2({\bm k},\eta)\bar{e}_i({\bm k})
+\S_3({\bm k},\eta)\hat{k}_i\Big]{\rm e}^{i{\bm k}\cdot{\bm x}}\,,
\end{align}
whose complex conjugate is then
\begin{align}
\label{eq:Sdefconj}
S_i({\bm x},\eta)=\frac{1}{(2\pi)^{3/2}}\int d^3{k}
\Big[\S^*_{1}({\bm k},\eta)\,e_i({\bm k})+\S^*_2({\bm k},\eta)\bar{e}_i({\bm k})
+\S^*_3({\bm k},\eta)\hat{k}_i\Big]{\rm e}^{-i{\bm k}\cdot{\bm x}}\,.
\end{align}
Alternatively, under the change ${\bm k} \to -{\bm k}$, on which $S_i \to -S_i$, as 
mentioned above, Eq.~(\ref{eq:Sdef}) becomes
 \begin{align}
\label{eq:Sdefnegation}
S_i({\bm x},\eta)&=\frac{1}{(2\pi)^{3/2}}\int d^3{k}
\Big[-\S_{1}(-{\bm k},\eta)\,e_i(-{\bm k})-\S_2(-{\bm k},\eta)\bar{e}_i(-{\bm k})\nn\\
&\qquad\qquad\qquad\qquad\qquad
-\S_3(-{\bm k},\eta)\cdot(-\hat{k}_i)\Big]{\rm e}^{-i{\bm k}\cdot{\bm x}}\,.
\end{align}
Comparing Eqs.~(\ref{eq:Sdefconj}) and (\ref{eq:Sdefnegation}), we can then read off
the relationship between the Fourier amplitudes and their conjugates:
 \begin{align}
 \S_1^*({\bm k},\eta) &= -\S_1(-{\bm k},\eta)\,, \\
 \S_2^*({\bm k},\eta) &= \S_2(-{\bm k},\eta)\,,\\
 \S_3^*({\bm k},\eta) &= \S_3(-{\bm k},\eta)\,.
 \end{align}
 Then, from the definition of the vector source term, we obtain the Fourier amplitudes
 \begin{align}
\S_1({\bm k},\eta)&=-\frac{f(\eta)}{(2\pi)^{3/2}}\int d^3\tilde{{\bm k}}\cdot 2k\bar{e}_i\tilde{k}^i
\dPn({\bm k},\eta)\drho({\bm k}-\tilde{\bm k},\eta)\,,\\
\S_2({\bm k},\eta)&=\frac{f(\eta)}{(2\pi)^{3/2}}\int d^3\tilde{{\bm k}}\cdot 2k{e}_i\tilde{k}^i
\dPn({\bm k},\eta)\drho({\bm k}-\tilde{\bm k},\eta)\,,\\
\S_3({\bm k},\eta)&=0\,.
\end{align}
This last equation tells us that the projection of Eq.~(\ref{eq:Sdef}) onto the basis vector $k_i$ gives zero (since contracting
Eq.~(\ref{eq:Sdef}) with $k_i$ contracts two copies  of $k$ with the permutation symbol, automatically giving zero).
The complex conjugates are
\begin{align}
\S^*_1({\bm k},\eta)&=-\frac{f(\eta)}{(2\pi)^{3/2}}
\int d^3\tilde{\bm k} 2k\bar{e}_i\tilde{k}^i\dPn({\bm k},\eta)\delta\rho(-({\bm k} + \tilde{\bm k}),\eta)\,,\\
\S^*_2({\bm k},\eta)&=
\frac{f(\eta)}{(2\pi)^{3/2}}
\int d^3\tilde{\bm k} 2k{e}_i\tilde{k}^i\dPn({\bm k},\eta)\delta\rho(-({\bm k} + \tilde{\bm k}),\eta)\,.
\end{align}

\newpage
Having now obtained the required amplitudes and their conjugates, we can compute the 
correlator of the source terms for the $A=1$ mode in Eq.~(\ref{eq:PSomegawithS}).\footnote{Note
that we need only consider one orthogonal component of the source vector, since
we can make an appropriate choice such that its component in one direction is zero.}
Assuming that the fluctuations $\delta\rho$ and $\delta P_{\rm{nad}}$ are Gaussian, 
we can put the directional dependence into Gaussian random variables $\hat{E}({\bm k})$, 
which obey the relationships
\be
\left<\hat{E}^*({\bm k}_1)\hat{E}({\bm k}_2)\right>=\delta^3({\bm k}_1-{\bm k}_2)\,,
\hspace{1cm}
\left<\hat{E}({\bm k}_1)\hat{E}({\bm k}_2)\right>=\delta^3({\bm k}_1+{\bm k}_2)\,,
\ee
and write, for example,\footnote{In making this choice, we are assuming that $\delta\rho$ and $\delta P_{\rm nad}$
are completely correlated variables. This is perhaps not the most physically motivated assumption, since one might
expect some level of decorrelation between the two variables. However, this assumption will likely give the largest 
signal (the partially decorrelated case will, in its simplest form, require a new parameter less than one, which 
characterises how decorrelated the two variables are -- this parameter can be determined from the specific
model for the production of the non-adiabatic pressure perturbation), so is suitable as a first approximation. 
We leave the case where the two variables are decorrelated for future work.}
\be
\delta\rho({\bm k},\eta)=\delta\rho(k,\eta)\hat{E}({\bm k})\,.
\ee
The correlator then becomes
\begin{align}
\label{eq:PSbeforewicks}
\left< \S^*({\bm k}_1, \tilde{\eta}_1)\S({\bm k}_2, \tilde{\eta}_2)\right>&=
\frac{f_{\tilde{1}}f_{\tilde{2}}}{(2\pi)^3}
\int d^3\tilde{k}_1
2k_1\tilde{k}_{1i}\bar{e}_1^i\dPn(\tilde{ k}_1, \tilde{\eta}_1)\delta\rho(|{\bm k}_1+\tilde{\bm k}_1|, \tilde{\eta}_1)\nn\\
&\qquad
\times\int d^3\tilde{k}_2
2k_2\tilde{k}_{2i}\bar{e}_2^i\dPn(\tilde{ k}_2, \tilde{\eta}_2)
\delta\rho(|{\bm k}_2-\tilde{\bm k}_2|, \tilde{\eta}_2)\nn\\
&\qquad\times\left<\hat{E}(-\tilde{\bm k}_1)\hat{E}(-{\bm k}_1
-\tilde{\bm k}_1)\hat{E}(\tilde{\bm k}_2)\hat{E}({\bm k}_2-\tilde{\bm k}_2)\right>\,,
\end{align}
where we have introduced the notation $f_{\tilde{1}}\equiv f(\tilde{\eta}_1)$.
Wick's theorem (see, e.g., Ref.~\cite{Durrer:1127831}) then allows us to express the correlator 
in the above in terms of delta functions as
\begin{align}
\left<\hat{E}(\tilde{\bm k}_1)\hat{E}(-{\bm k}_1-\tilde{\bm k}_1)
\hat{E}(\tilde{\bm k}_2)\hat{E}({\bm k}_2-\tilde{\bm k}_2)\right>
&=\left<\hat{E}(\tilde{\bm k}_1)\hat{E}(-{\bm k}_1-\tilde{\bm k}_1)\right>
\left<\hat{E}(\tilde{\bm k}_2)\hat{E}({\bm k}_2-\tilde{\bm k}_2)\right>\nn\\
&+
\left<\hat{E}(\tilde{\bm k}_1)\hat{E}(\tilde{\bm k}_2)\right>
\left<\hat{E}(-{\bm k}_1-\tilde{\bm k}_1)\hat{E}({\bm k}_2-\tilde{\bm k}_2)\right>
\nn\\
&+
\left<\hat{E}(\tilde{\bm k}_1)\hat{E}({\bm k}_2-\tilde{\bm k}_2)\right>
\left<\hat{E}(-{\bm k}_1-\tilde{\bm k}_1)\hat{E}(\tilde{\bm k}_2)\right>\nn
\\
=\delta^3(\tilde{\bm k}_1+\tilde{\bm k}_2)\delta^3(-{\bm k}_1-\tilde{\bm k}_1+{\bm k}_2-\tilde{\bm k}_2)
\label{eq:wicksdelta}
&+\delta^3(\tilde{\bm k}_1+{\bm k}_2-\tilde{\bm k}_2)\delta^3(-{\bm k}_1-\tilde{\bm k}_1+\tilde{\bm k}_2)\,,
\end{align}
which gives 
\begin{align}
&\left< \S^*({\bm k}_1, \tilde{\eta}_1)\S({\bm k}_2, \tilde{\eta}_2)\right>
=
\frac{f_{\tilde{1}}f_{\tilde{2}}}{2\pi^3}
\int d^3\tilde{k}_1\int d^3\tilde{k}_2
\Big\{k_1\bar{e}_i\tilde{k}^i\dPn(\tilde{k}_1,\tilde{\eta}_1)
\delta\rho(|{\bm k}_1+\tilde{\bm k}_1|, \tilde{\eta}_1)\nn\\
&\qquad\qquad\qquad\qquad\qquad\qquad\qquad\times
k_2\bar{e}_i\tilde{k}_2^i\dPn(\tilde{k}_2, \tilde{\eta}_2)
\delta\rho(|{\bm k}_2-\tilde{\bm k}_2|, \tilde{\eta}_2)\Big\}\\
&\qquad\times\left\{\delta^3(\tilde{\bm k}_1+\tilde{\bm k}_2)\delta^3({\bm k}_2
-{\bm k}_1-\tilde{\bm k}_1-\tilde{\bm k}_2)
+\delta^3(\tilde{\bm k}_1-\tilde{\bm k}_2+{\bm k}_2)\delta^3(\tilde{\bm k}_2-{\bm k}_1-\tilde{\bm k}_1)
\right\}\,.
\end{align}

By integrating over the delta functions, and collecting terms, we arrive at
\begin{align}
\left< \S^*({\bm k}_1, \tilde{\eta}_1)\S({\bm k}_2, \tilde{\eta}_2)\right>
&=
\frac{f_{\tilde{1}}f_{\tilde{2}}}{2\pi^3}\delta^3({\bm k}_2-{\bm k}_1)k^2\int d^3\tilde{k}(\bar{e}_i \tilde{k}^i)^2
\dPn(\tilde{k},\tilde{\eta}_1)\drho(|{\bm k}+\tilde{\bm k}|,\tilde{\eta}_1)\nn\\
&\times\Big[\dPn(|{\bm k}+\tilde{\bm k}|, \tilde{\eta}_2)\drho(\tilde{k},\tilde{\eta}_2)
-\dPn(\tilde{k},\tilde{\eta}_2)\drho(|{\bm k}+\tilde{\bm k}|, \tilde{\eta}_2)\Big]\,,
\end{align}
from which we can read off the power spectrum:
\begin{align}
\mathcal{P}_\w(k,\eta)&=\frac{k^5\eta^2}{4\pi^4}\int\int f_{\tilde{1}}
f_{\tilde{2}}\tilde{\eta}_1^{-1}\tilde{\eta}_2^{-1}
d\tilde{\eta}_1 d\tilde{\eta}_2
\int d^3\tilde{k}(\bar{e}_i \tilde{k}^i)^2
\dPn(\tilde{k},\tilde{\eta}_1)\drho(|{\bm k}+\tilde{\bm k}|,\tilde{\eta}_1)\nn\\
&\qquad\times\Big[\dPn(|{\bm k}+\tilde{\bm k}|, \tilde{\eta}_2)\drho(\tilde{k},\tilde{\eta}_2)
-\dPn(\tilde{k},\tilde{\eta}_2)\drho(|{\bm k}+\tilde{\bm k}|, \tilde{\eta}_2)\Big]\,.
\end{align}

Now, as a first approximation for the source term, we can expand Eq.~(\ref{eq:drho1nonad}) to lowest
order in $k\eta$ to give
\begin{align}
\delta\rho(k,\eta)=\bar{A}k^\beta\,\eta^{-4}\,,\\
\dPn(k,\eta)=\bar{D}k^\alpha\,\eta^{-5}\,,
\end{align}
where $\bar{A}$ and $\bar{D}$ are yet unspecified amplitudes, and $\alpha$ and $\beta$ undetermined
powers. Using these approximations gives
\begin{align}
\label{eq:PSetaint}
\mathcal{P}_\w(k,\eta)&=\frac{81}{256}\frac{k^5\eta^2}{4\pi^4
}(AD)^2
\left[\ln\left({\frac{\eta}{\eta_0}}\right)\right]^2\nn\\
&\qquad\times\int d^3\tilde{k}(\bar{e}_i \tilde{k}^i)^2\tilde{k}^\alpha|{\bm k}+\tilde{\bm k}|^\beta
\Big(|{\bm k}+\tilde{\bm k}|^\alpha\tilde{k}^\beta-\tilde{k}^\alpha|{\bm k}+\tilde{\bm k}|^\beta\Big)
\end{align}
where we have performed the temporal integral by noting that
as mentioned above,
$a\propto\eta$ during radiation domination, and thus $\rhob\propto\eta^{-4}$.
To perform the $k$-space integral, we first move to spherical coordinates
oriented with the axis in the direction of ${\bm k}$.  Then, denoting
the angle between ${\bm k}$ and ${\tilde{\bm k}}$ as $\theta$, the
integral can be transformed as
\be
\int d^3\tilde{k} \to 2\pi\int_0^{k_{\rm c}}\tilde{k}^2d\tilde{k}\int_0^\pi\sin\theta d\theta\,,
\ee
where the prefactor comes from the fact that the integrand has no
dependence on the azimuthal angle, and $k_{\rm c}$ denotes a cut-off on small scales. 
This cut-off is chosen to be smaller than the typical separation of galaxies, and therefore
much smaller than the continuum limit, solely for the purpose of studying this toy model. 
Noting that, in this coordinate
system $\tilde{k}_i\bar{e}^i=\tilde{k}\sin\theta$
the integral in Eq.~(\ref{eq:PSetaint}) becomes
\begin{align}
\label{eq:int}
I(k)&=2\pi\int_0^{k_{\rm c}}\int_0^\pi\tilde{k}^{4+\alpha}\sin\theta \, d\theta\,
 d\tilde{k}\,\sin^2\theta
k^\beta\Big[1+(\tilde{k}/k)^2+2(\tilde{k}/k)\cos\theta\Big]^{\beta/2}\nn\\
&\times\Big(k^\alpha(1+(\tilde{k}/k)^2+2(\tilde{k}/k)\cos\theta)^{\alpha/2}\tilde{k}^\beta-
\tilde{k}^\alpha k^\beta (1+(\tilde{k}/k)^2+2(\tilde{k}/k)\cos\theta)^{\beta/2}\Big)
\end{align}
Finally, we change variables again to
dimensionless $u$ and $v$ defined as \cite{Ananda:2006af} (or similarly \cite{Brown:2010ms})
\begin{align}
v=\frac{\tilde{k}}{k}\,, \, \,
u=\sqrt{1+(\tilde{k}/{k})^2+2(\tilde{k}/{k})\cos\theta}=\sqrt{1+v^2+2v\cos\theta}\,,
\end{align}
for which the integral (\ref{eq:int}) becomes
\begin{align}
\label{integralIk}
I(k)
&=k^{2(\alpha+\beta)+5}\int_0^{{k_{\rm c}}/{k}}\int^{v+1}_{|v-1|}u\, du\, v^3\, dv\, u^\beta v^\alpha
\Big(1-\frac{1}{4v^2}(u^2-v^2-1)^2\Big)
\Big[u^\alpha v^\beta -v^\alpha u^\beta \Big]\,.
\end{align}
%

%%%%%%%%%%%%%%%%%%%%%%%%%%%%
\subsection{Evaluating the Vorticity Power Spectrum}
%%%%%%%%%%%%%%%%%%%%%%%%%%%

In order to perform the integral \eq{integralIk} derived above, we need
to specify the exponents for the power spectra of the energy density
and the non-adiabatic pressure $\alpha$ and $\beta$. The energy
density perturbation on slices of uniform curvature can be related to the curvature perturbation on
uniform density hypersurfaces, $\zeta$, during radiation domination
through \cite{MW2008}
\be
\delta\rho=-\frac{\rho_0'}{\H}\zeta=4\rhob\zeta\,,
\ee
and hence the initial power spectra can be related as
$\left<\delta\rho_{\rm ini}\delta\rho_{\rm ini}\right>
=16\rho_{0\gamma\rm ini}^2 \left<\zeta_{\rm ini}\zeta_{\rm
ini}\right>$, and we get the power spectrum of the initial
density perturbation 
\be
\delta\rho_{\rm ini}\propto\zeta_{\rm ini}
\propto\left(\frac{k}{k_0}\right)^{\frac{1}{2}(n_{\rm s}-1)}\,,
\ee
where $k_0$ is the {\sc{Wmap}} pivot scale and $n_{\rm s}$ the
spectral index of the primordial curvature perturbation \cite{WMAP7}.
This allows us to relate our ansatz for the density perturbation,
to the {\sc{Wmap}}-data
which gives
\be 
\delta\rho=\delta\rho_{\rm ini}\,\left(\frac{k}{k_0}\right)\left(\frac{\eta}{\eta_0}\right)^{-4}
=A_{\rm ini}\rho_{0{\rm ini}}
%\H_{{\rm ini}}^2\Omega_{\gamma{\rm ini}}
\left(\frac{k}{k_0}\right)^{\frac{1}{2}(n_{\rm s}+1)}
\left(\frac{\eta}{\eta_0}\right)^{-4}\,.
\ee
From this, we can read off that $\beta=\frac{1}{2}(n_{\rm s}+1)\simeq
1$ and the amplitude $A=A_{\rm ini}\rho_{0{\rm ini}}$.
% (since in the radiation
 % era $\Omega_{\gamma{\rm ini}}\sim1$). 
 We have some freedom in
choosing $\alpha$, however would expect the non-adiabatic pressure to
have a blue spectrum, though the calculation demands
$\alpha\neq\beta$.
Using the notation of Ref.~\cite{WMAP7} we get
\be 
A^2=\rho_{0  \rm ini}^2
{{P}}_{\mathcal{R}}(k_0)^2=
\rho_{0  \rm ini}^2k_0^{-6}
\Delta_{\mathcal{R}}^4\,, \hskip 1cm 
D^2=\rho_{0 \rm ini}^2
{{P}}_{\mathcal{S}}(k_0)^2
= \rho_{0 \rm ini}^2 k_0^{-6}
\Delta_{\mathcal{S}}^4
%{\mathcal{P}}_{\mathcal{S}}(k_0)\,,
\ee 
where we also have the ratio\footnote{The parameter $\alpha(k_0)$ is introduced in Refs.~\cite{WMAP7, Bean:2006qz}
in order to quantify the ratio of $\Delta_{\mathcal{S}}^2$ to $\Delta_{\mathcal{R}}^2$.}
\be 
\frac{\Delta_{\mathcal{S}}^2}{\Delta_{\mathcal{R}}^2}
%\frac{{\mathcal{P}}_{\mathcal{R}}(k_0) }{{\mathcal{P}}_{\mathcal{S}}(k_0)}
=\frac{\alpha(k_0)}{1-\alpha(k_0)}\,,
\ee
and therefore, 
\be 
(AD)^2
=\frac{\alpha(k_0)}{1-\alpha(k_0)}\Delta_{\mathcal{R}}^8\rho_{0 \rm ini}^4k_0^{-12}\,.
\ee
We can then substitute in numerical values for $\Delta_{\mathcal{R}}^2$ and
$\alpha(k_0)$ from Ref.~\cite{WMAP7} later on.

\begin{figure}
\center
%\subfigure[Small range of $k<k_{\rm c}$.]{
\includegraphics[width=0.9\textwidth]{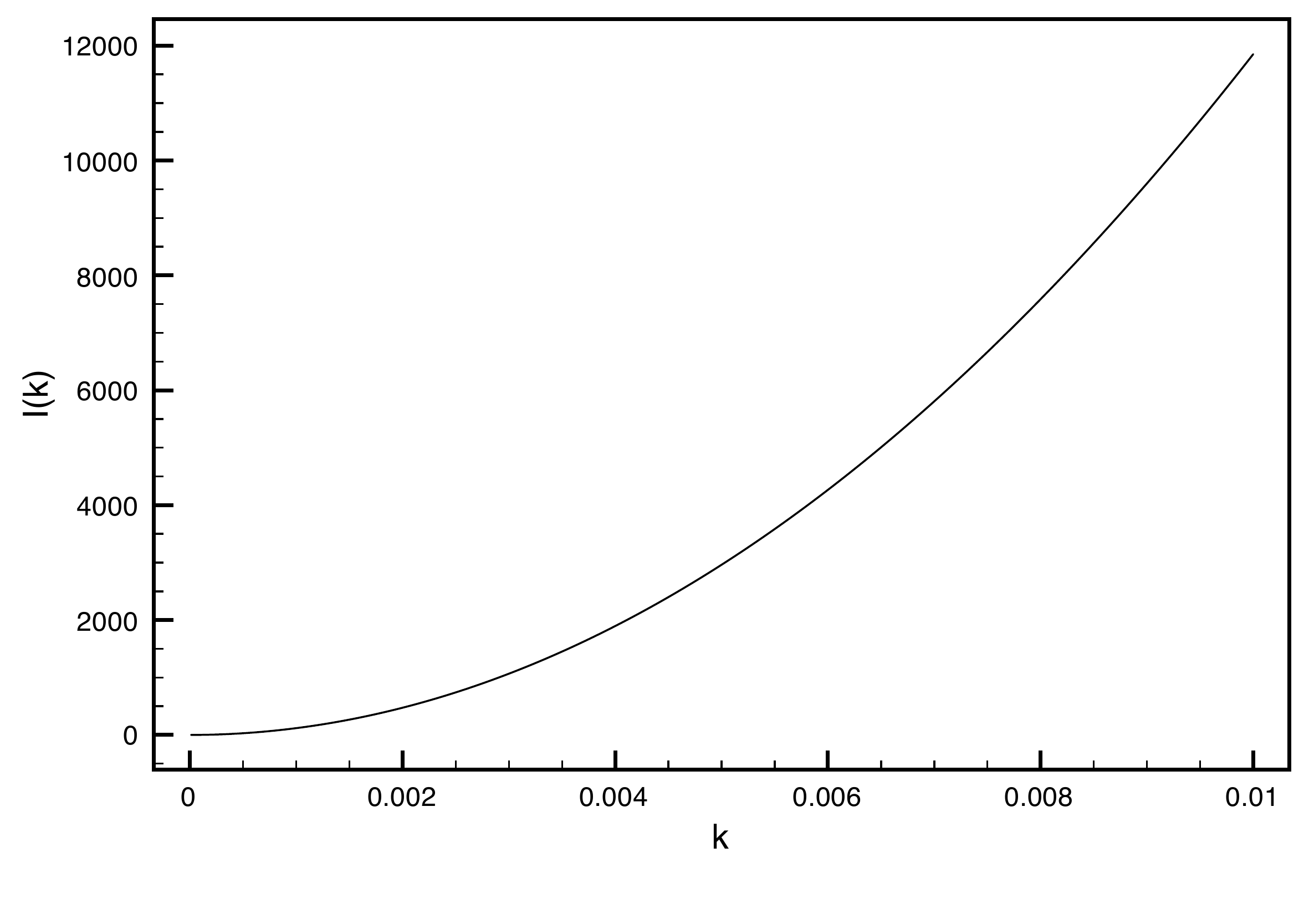}
\caption{Plot of $I(k)$, Eq.~(\ref{eq:Ik}), for the illustrative 
choice of $k_{\rm c}=10 {\rm Mpc}^{-1}$; small range of $k<k_{\rm c}$.}
\label{fig:integrand}

%\end{figure}
%\begin{figure}[ht]
%\center
%\subfigure[ Wide range of $k$ including $k>k_{\rm c}$.]{
\includegraphics[width=0.9\textwidth]{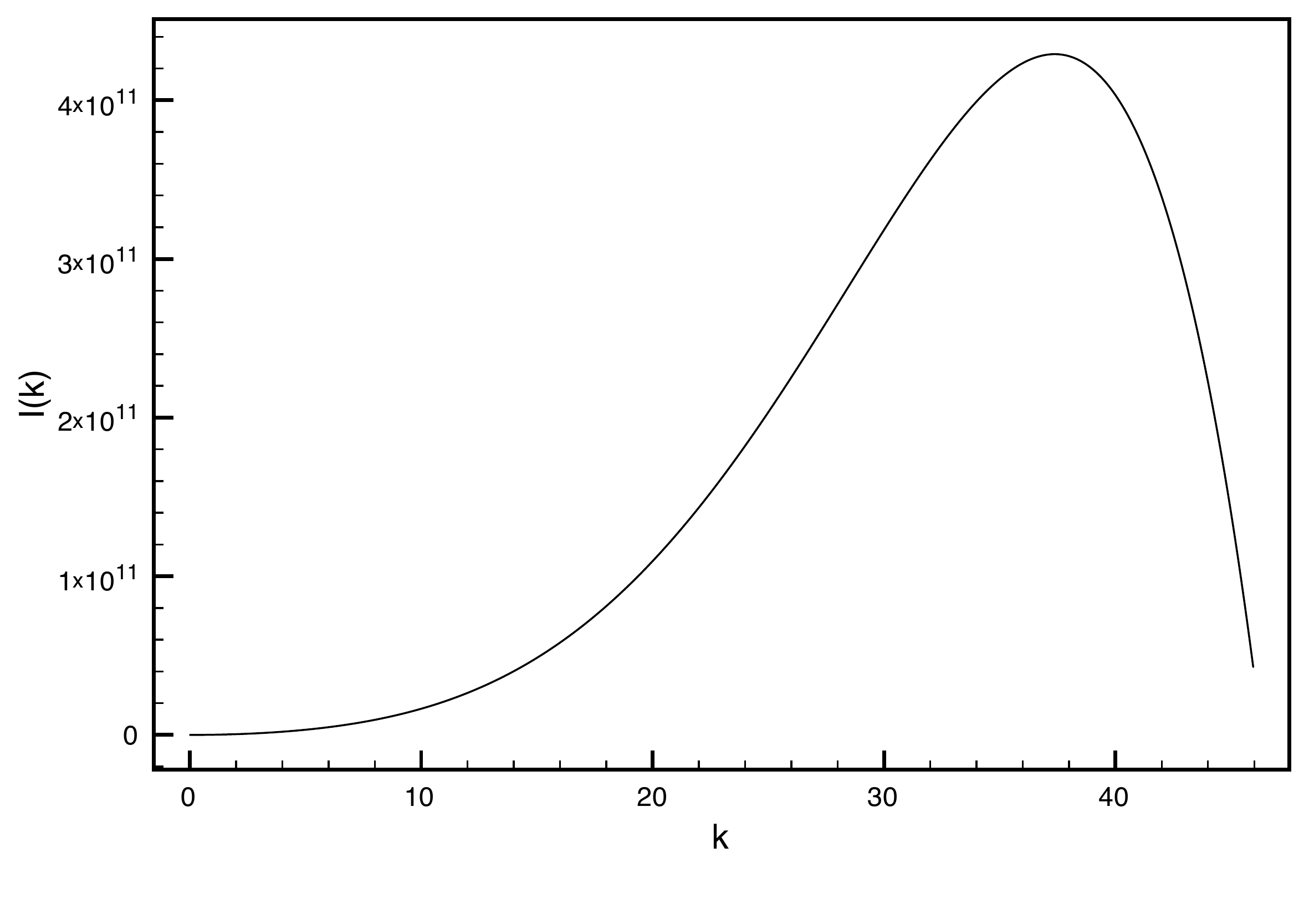}
\caption{Plot of $I(k)$ for the illustrative choice of $k_{\rm c}=10 {\rm Mpc}^{-1}$; 
wide range of $k$ including $k>k_{\rm c}$.}
\label{fig:integrand2}
%\caption{Plots of $I(k)$ for the illustrative choice of $k_{\rm c}=10$.}
\end{figure}

Then, making the choice $\alpha=2$, the input
power spectra are 
\be
\label{input_power_norm}
\delta\rho({k}, \eta)
=A \left(\frac{k}{k_0}\right)\left(\frac{\eta}{\eta_0}\right)^{-4}\,, \hspace{1cm}
\dPn({k}, \eta)= D\left(\frac{k}{k_0}\right)^2\left(\frac{\eta}{\eta_0}\right)^{-5}\,,
\ee
for which the integral \eq{integralIk} becomes
\begin{align}
\label{eq:Ik}
I(k)
&={k^{11}}
\int_0^{{k_{\rm c}}/{k}}\int^{v+1}_{|v-1|}u^2\, du\, v^5\, dv\, 
\Big(1-\frac{1}{4v^2}(u^2-v^2-1)^2\Big)
\Big[u^2 v -v^2 u\Big]\,.
\end{align}
We can then integrate this analytically  to give
\be 
\label{integrated}
I(k)=\frac{16}{135}k_{\rm c}^9k^2+\frac{12}{245}k_{\rm c}^7k^4-\frac{4}{1575}k_{\rm c}^5k^6\,,
\ee 
which clearly depends upon the small scale cut-off, as expected.\footnote{It should
be noted that if one were to reduce the assumption of 100\% correlation between
the energy density and entropy perturbation, this could soften this dependence. However, this is 
left for future investigation.}
 For illustrative 
purposes, we choose $k_{\rm c}=10 {\rm Mpc}^{-1}$ and plot the solution $I(k)$.
Fig.~\ref{fig:integrand} shows that the amplitude of the integral grows as the wavenumber
increases.  Fig.~\ref{fig:integrand2} shows a turn
around and a decrease in power at some wavenumber (in fact, for a non-specific cutoff,
this point is at $3.7375k_{\rm c}$). However, we note that this 
value is greater than our cutoff, and therefore not physical. As long as we 
consider values of $k$ less than the cutoff, our approximation will still be valid.

\begin{figure}
\center
\includegraphics[width=0.9\textwidth]{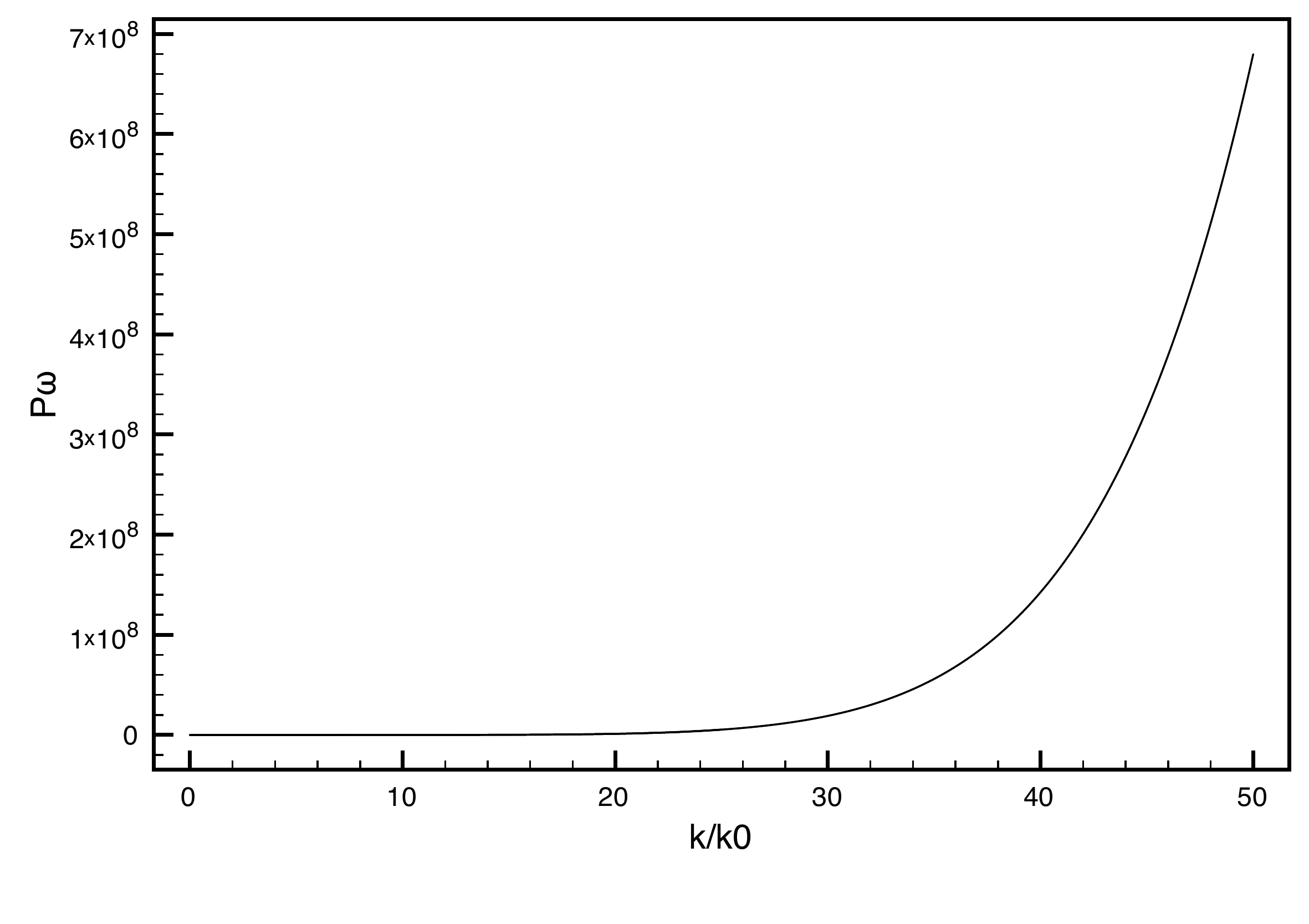}
\caption{Plot of ${\cal{P}}_\w$(k), i.e. the scale dependence of the vorticity power spectrum.}
\label{fig:ps}
\end{figure}

\begin{figure}
\center
\includegraphics[width=0.9\textwidth]{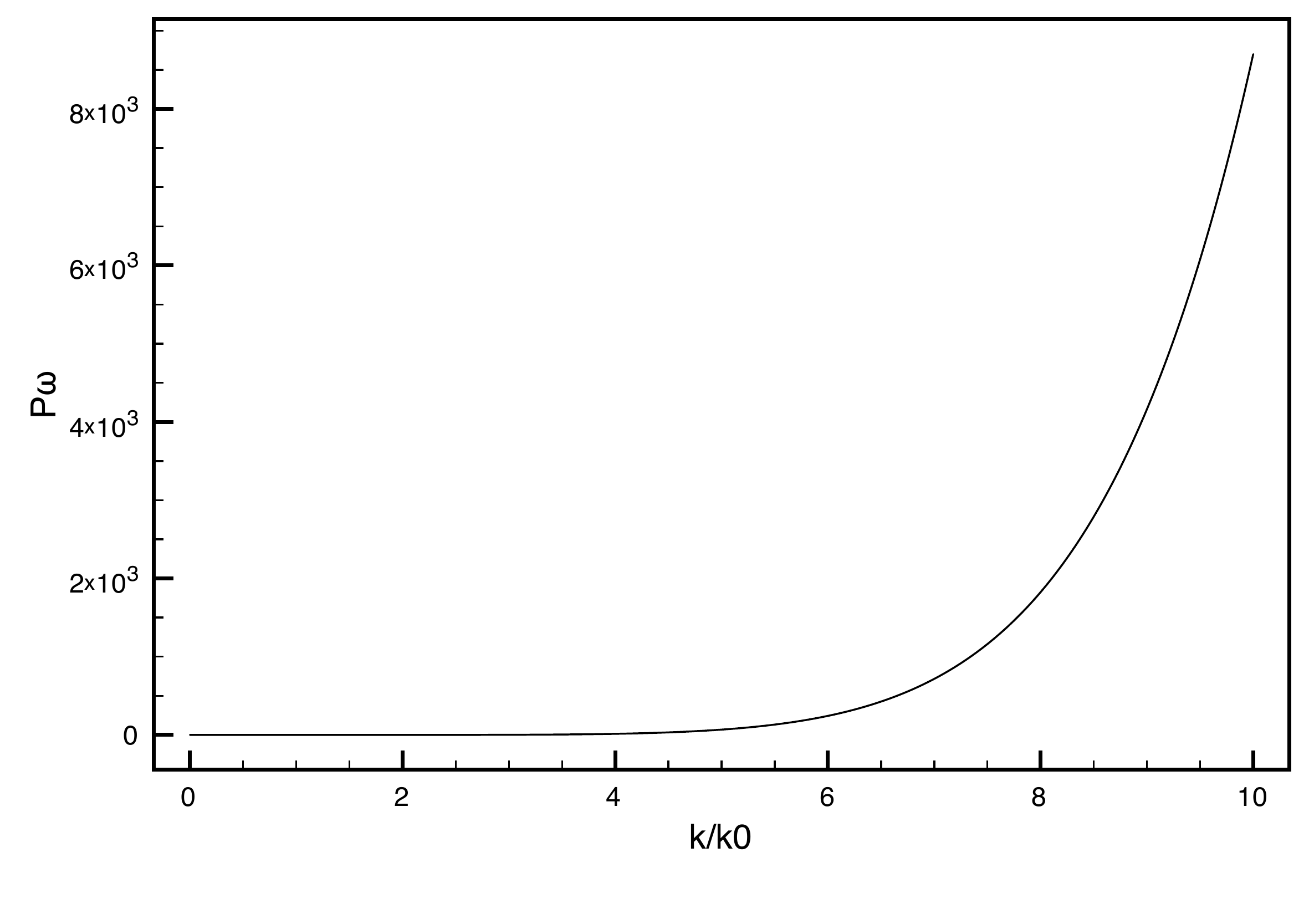}
\caption{Plot of ${\cal{P}}_\w$(k), for a narrower range of $k$ values than Figure~\ref{fig:ps}.}
\label{fig:ps2}
\end{figure}

Then, using the above, and noting that the input to the temporal integrals are
\be 
a\propto\left(\frac{\eta}{\eta_0}\right)\,,
\hskip 1cm
\rho_0=\rho_{0 \rm ini}\left(\frac{\eta}{\eta_0}\right)^{-4}\,,
\ee
we obtain the power spectrum for the vorticity, for general $k_{\rm c}$ as 
\begin{align}
{\mathcal{P}}_\w(k, \eta)&=
\frac{81}{256}k_0^5\frac{\eta^2}{4 \pi^4}\frac{1}{\rho_{0 \rm ini}^4}\ln^2\left(\frac{\eta}{\eta_0}\right)
\Bigg(\frac{\alpha(k_0)}{1-\alpha(k_0)}\Bigg)^2k_0^{-12}\Delta_{\mathcal{R}}^8\rho_{0 \rm ini}^4k_{\rm c}^5
\nn\\
&\qquad\qquad\times\Bigg[\frac{16}{135}\frac{k_{\rm c}^4}{k_0^4}\left(\frac{k}{k_0}\right)^7
+\frac{12}{245}\frac{k_{\rm c}^2}{k_0^2}\left(\frac{k}{k_0}\right)^9
-\frac{4}{1575}\left(\frac{k}{k_0}\right)^{11}\Bigg]\,.
\end{align}
Substituting in values of parameters from Ref.~\cite{WMAP7}, as presented in Table~\ref{tab:wmap}
taking a conservative estimate for $\alpha(k_0)$, being
10\% of the upper bound as reported by {\sc Wmap}7, we obtain the vorticity power spectrum
\begin{align}
{\mathcal{P}}_\w(k, \eta)
&=\eta^2\ln^2\left(\frac{\eta}{\eta_0}\right)\Bigg[0.87 \times 10^{-12} 
k_{\rm c}^9\left(\frac{k}{k_0}\right)^7
{\rm Mpc^{11}}
+3.73\times 10^{-18}k_{\rm c}^7\left(\frac{k}{k_0}\right)^9{\rm Mpc^{9}}\nn\\
&\qquad\qquad\qquad\qquad\qquad
-7.71\times 10^{-25}k_{\rm c}^5 \left(\frac{k}{k_0}\right)^{11}{\rm Mpc^{7}}\Bigg]
\,,
\end{align}
and for our above choice of $k_{\rm c}=10 {\rm Mpc}^{-1}$,
\begin{align}
{\mathcal{P}}_\w(k, \eta)&=\eta^2\ln^2\left(\frac{\eta}{\eta_0}\right)\Bigg[0.87\times 10^{-3}
\left(\frac{k}{k_0}\right)^7
+3.73\times 10^{-11}\left(\frac{k}{k_0}\right)^9\nn\\
&\qquad\qquad\qquad\qquad\qquad-7.71\times 10^{-20}\left(\frac{k}{k_0}\right)^{11}\Bigg]{\rm Mpc^{2}}\,.
\end{align}
\begin{table}[t]
\centering
\begin{tabular}{cc}
\hline\hline
Parameter & WMAP7 value \\
\hline
\\
$k_0$ & $0.002 \, {\rm Mpc}^{-1}$\\
$\Delta^2_{\cal{R}}(k_0)$ & $2.38 \times 10^{-9}$  \\
$\alpha(k_0)$ & $0.13$ (95\% CL)\\
\\
\hline
\end{tabular}
\caption{Parameter values from the WMAP seven year data \cite{WMAP7}.}
\label{tab:wmap}
\end{table}
This shows that, under our approximations, the vorticity spectrum has a non-negligible amplitude, with
 a huge amplification of power on small scales. We plot this, for illustrative purposes, in Figs.~\ref{fig:ps} and \ref{fig:ps2},
  where
 we have ignored the time dependence and focused only on the scale dependence of the spectrum.
 We should emphasise that we are studying the generation of vorticity in the wavenumber region $k_0<k<k_c$
 and do not expect the dynamics to be dominated by an ``inverse cascade'' (i.e. a feedback of power from smaller scales to larger scales).
 Therefore, the physics around the cut-off wavenumber cannot influence the vorticity generation. To study this
 phenomenon in detail, a much more detailed calculation including backreaction effects
 would have to be performed, beyond the scope of this thesis.

% % % % % % % % % % % % % % % % % % % 
\section{Discussion}
% % % % %  % % %% % %%  %% % % % % % % %  

In this Chapter we have studied the generation of vorticity in the early universe, showing that second 
order in cosmological perturbation theory vorticity is sourced by first order scalar and vector 
perturbations for a perfect fluid. This is an extension of Crocco's theorem to an expanding, dynamical
background, namely, a FRW universe. 
Whereas previous works assumed barotropic flows, allowing for entropy gives a qualitatively novel result.
This implies that the description of the cosmic fluid as a potential flow, which works exceptionally well
at first order in the perturbations, will break down at second order for non-barotropic flows. 
Similarly, in barotropic flow Kelvin's theorem guarantees conservation of vorticity. This is no
longer true if the flow is non-barotropic.

Having derived the qualitative result, we then obtained the first realistic calculation of 
the amount of vorticity generated at second order. As an input spectrum for the linear energy density, 
we used the solution obtained in Section~\ref{sec:dynamicsflat} approximated for small $k\eta$ and normalised to {\sc Wmap7}.
Then, making the ansatz that the non-adiabatic pressure perturbation has a bluer spectrum than that of the 
energy density in order to keep the non-adiabatic pressure sub-dominant  on all scales, we 
obtained an analytical result for the vorticity. Our results show that the vorticity power spectrum
has a non-negligible magnitude which depends on the cutoff, $k_{\rm c}$, and the chosen 
parameters. As this is a second order effect, the magnitude is somewhat surprising. 
We have also shown that the result has a dependence of the wavenumber 
to the power of at least seven for the choice $\alpha=2$, where $\alpha$ is the exponent of the wavenumber for the 
 non-adiabatic pressure input spectrum. Therefore the amplification due
to the large power of $k$ is huge, rendering the vorticity not only possibly observable, but also
important for the general understanding of the physical processes taking place in the early universe.

The consequences of this significant power are not immediately clear as
the model under consideration is only a toy model. Although vorticity is
not generated in standard cosmology (at linear order), the vorticity
generated at second order will not invalidate the standard predictions,
as it is, on large scales, very small and can only be of significant size
on small scales. However, any possible observational consequences will
depend on the wavenumber at which the power spectrum peaks (to be
determined by an actual model).

One prospect for observing early universe vorticity is in the B-mode polarisation of the CMB. 
Both vector and tensor perturbations produce B-mode polarisation, but at linear order such
vector modes decay with the expansion of the universe. However, vector modes produced by gradients in
energy density and entropy perturbations, such as those discussed in this chapter, will source B-mode 
polarisation at second order. Furthermore, it has recently been noted that vector perturbations
in fact generate a stronger B-mode polarisation than tensor modes with the same amplitude 
\cite{bellido}. Therefore, it is feasible for vorticity to be observed by future surveys such as the space-based 
{\sc CMBPol} \cite{CMBPol}, which is currently in the planning stage, or the ground based experiment
{\sc Polarbear} \cite{Errard:2010bn}, which is due to start making observations in 2011.

Finally, a non-zero vorticity at second order in perturbation theory has important consequences 
for the generation of magnetic fields, as it has been long known that vorticity and magnetic fields are 
closely related (see Refs.~\cite{biermann, harrison}).
Previous works either used momentum exchange between
multiple fluids to generate vorticity, as in Refs.~\cite{Matarrese:2004kq, Gopal:2004ut, Takahashi:2005nd, 
Ichiki:2006cd, Siegel:2006px, Kobayashi:2007wd, Maeda:2008dv}, 
or used intermediate steps to first generate vorticity for example by using shock fronts 
as in Ref.~\cite{Ryu:2008hi}. However, we do not require such additional steps. 
Therefore, an important extension to the work presented in this chapter is to
consider the magnetic fields generated by our mechanism which could be an important
step in answering the  question regarding the origin of the primordial
magnetic field. We will discuss more future prospects in the concluding chapter of this thesis.

%% file: third.tex
% % % % % % % % % % % % % % % % % % % % % % % % % % % % 
% third.tex AJC
%third order stuff
% % % % % % % % % % % % % % % % % % % % % % % % % % % % 
% Redefine CVSRevision for this section. 
% If you don't want to use CVS tags comment out this line
%\renewcommand{\CVSrevision}{\version$Id: chapter.tex,v 1.3 2009/12/17 18:16:48 ith Exp $}

% % % % % % % % % % % % % % % % % % % % % % % % % % % % % % % % 
% =========================================================== %
% % % % % % % % % % % % % % % % % % % % % % % % % % % % % % % % 
\chapter{Third Order Perturbations}
\label{ch:third}
% % % % % % % % % % % % % % % % % % % % % % % % % % % % % % % % 
% =========================================================== %
% % % % % % % % % % % % % % % % % % % % % % % % % % % % % % % % 

In the preceding chapters we have developed cosmological perturbation theory at both linear
and second order. However, one does not need to stop there: it is worthwhile and feasible to
consider perturbation theory even beyond second order. In this chapter we explore aspects
of cosmological perturbation theory at third order.\\

We have already seen that extending perturbation theory to second order reveals new
phenomena which arise due to quadratic source terms in general, and 
the coupling between different types of perturbation which
is not present at linear order.  At third
order a new coupling occurs in the energy conservation equation,
namely the coupling of scalar perturbations to tensor perturbations.
This will allow for the calculation of yet
another different observational signature, highlighting another aspect
of the underlying full theory.

There has already been some work on third order theory.
For example, Refs.~\cite{Hwang:2005he,Hwang:2007wq} considered third
order perturbations of pressureless irrotational fluids as ``pure''
general relativistic correction terms to second order quantities. The
calculations focused on the temporal comoving gauge, allowing the authors to
consider only second order geometric and energy-momentum
components, and neglected vector perturbations.  
Ref.~\cite{Lehners:2009ja} includes a study of third order
perturbations with application to the trispectrum in the two-field
ekpyrotic scenario in the large scale limit.
There has also been reference in the literature of the need to extend
perturbation theory beyond second order. For example, in
Ref.~\cite{Clarkson:2009hr} UV divergences in the Raychaudhuri
equation are found when considering backreaction from averaging
perturbations to second order. The authors state that these
divergences may be removed by extending perturbation theory to third,
or higher, orders.

In this chapter, we develop the essential tools for third order
perturbation theory, such as the gauge transformation rules for
different types of perturbation, and construct gauge invariant
quantities at third order. We 
consider perfect fluids
with non-zero pressure, including all types of perturbation, namely,
scalar, vector and tensor perturbations.  In particular allowing for
vector perturbations is crucial for realistic higher order studies, since
vorticity is generated 
at second order in all models employing non-barotropic fluids as shown in  Chapter \ref{ch:vorticity}.
 Hence studying irrotational fluids at higher order
will only give partial insight into the underlying physics.
We present the energy and
momentum conservation equations for such a fluid, and also give the
components of the perturbed Einstein tensor, up to third order. All
equations are given without fixing a gauge.  We also give the Klein-Gordon
 equation for a scalar field minimally coupled to gravity at
third order in cosmological perturbation theory. This work is published in Ref.~\cite{third}.\\

%%%%%%%%%%%%%%%%%%%%%
\section{Definitions}
\label{sec:def}
%%%%%%%%%%%%%%%%%%%%%%

As in Chapter \ref{ch:perturbations} we take the perturbed metric tensor with covariant components
\begin{align}
g_{00} = -a^2(1+2\phi) \,, \hskip1cm
g_{0i} = a^2B_{i} \,, \hskip1cm
g_{ij} = a^2(\dij+2C_{ij}) \,,
\end{align}
In order to obtain the 
contravariant metric components we impose the constraint Eq.~(\ref{eq:metricconstraint}),
 $g_{\mu\nu}g^{\nu\lambda}=\delta_{\mu}{}^{\lambda} $, up to third order. This gives
\begin{align}
\label{g00up}
g^{00}&=-\frac{1}{a^2}\Big(1-2\phi+4\phi^2-8\phi^3-B_kB^k+4\phi B_kB^k
+2B^{i}B^{j}C_{ij}\Big) \,, \\
g^{0i} &= \frac{1}{a^2}\Big(B^i-2\phi B^i-2B_k C^{ki}+4\phi^2B^i+4B_kC^{ki}\phi
+4C^{kj}C_j{}^iB_k-B^kB_kB^i\Big)\,, \\
g^{ij} &= \frac{1}{a^2}\Big(\delta^{ij}-2C^{ij}+4C^{ik}C_k{}^j-B^iB^j+2\phi B^iB^j
-8C^{ik}C^{jl}C_{kl}\nn\\
&\qquad\qquad\qquad\qquad\qquad
\qquad\qquad\qquad\qquad+2B^iC^{kj}B_k
+2B_kB^jC^{ik}\Big)\,.
\end{align}
Note that in this chapter we do not explicitly split terms up into first, second and third order parts
unless where necessary, since doing so would dramatically increase the size of the equations presented. For example, 
expanding the $0-0$ component fully
order by order gives
\begin{align}
g^{00}&=-\frac{1}{a^2}\Big(1-2\phi_1-\phi_2-\frac{1}{3}\phi_3+\phi_1^2+8\phi_1\phi_2-8\phi_1^3
-B_{1k}B_1{}^k-B_{2k}B_1{}^k\nn\\
&\qquad\qquad\qquad\qquad\qquad+4\phi_1B_1{}^kB_{1k}+2B_1{}^iB_1{}^jC_{1ij}\Big)\,,
\end{align}
which when compared to \eq{g00up} illustrates the increase in number of terms, and thus 
why we refrain from splitting perturbations up. \\

Furthermore, to third order in  perturbation theory, the fluid four velocity, defined in Eq.~(\ref{eq:fourveldef})
and satisfying the constraint 
\be 
u^\mu u_\mu =-1\,,\nn
\ee
 has components
\begin{align}
u^i &= \frac{1}{a}v^i \,, \\
u^0 &= \frac{1}{a}\left(1-\phi+\frac{3}{2}\phi^2-\frac{5}{2}\phi^3
+\frac{1}{2}v_kv^k+v_kB^k+C_{kj}v^kv^j-2\phi v^kB_k
-\phi v^kv_k \right)\,, \\
u_i &= a\left(v_i+B_i-\phi B_i+2C_{ik}v^k+\frac{3}{2}B_i \phi^2
+\frac{1}{2}B_iv^kv_k+B_iv^kB_k\right) \,, \\
u_0 &= -a\left(1+\phi-\frac{1}{2}\phi^2+\frac{1}{2}\phi^3
+\frac{1}{2}v^kv_k+\phi v_kv^k+C_{kj}v^kv^j\right)\,.
\end{align}

%%%%%%%%%%%%%%%%%%%%%%%%%
\section{Gauge Transformations}
\label{sec:gauge}
%%%%%%%%%%%%%%%%%%%%%%%%%

We firstly need to extend the gauge transformations derived earlier to third order
in cosmological perturbation theory. Expanding the exponential map, (\ref{eq:exmap})
\be
\widetilde{\bf T}=e^{\pounds_{\xi}}{\bf T}\,,
\ee
to third order gives
\be
\label{exp_expand}
\exp(\pounds_{\xi})=1+\epsilon\pounds_{\xi_1}
+\frac{1}{2}\epsilon^2\pounds_{\xi_1}^2
+\frac{1}{2}\epsilon^2\pounds_{\xi_2}
+\frac{1}{6}\epsilon^3\pounds_{\xi_3}
+\frac{1}{6}\epsilon^3\pounds_{\xi_1}^3
+\frac{1}{4}\epsilon^3\pounds_{\xi_1}\pounds_{\xi_2}
+\frac{1}{4}\epsilon^3\pounds_{\xi_2}\pounds_{\xi_1}
+\ldots
\ee
Splitting the tensor ${\bf T}$ up to third order and collecting terms of like order in $\ep$ we
find that tensorial quantities transform at
third order as
\begin{align}
\label{Ttrans3}
\widetilde {\bf \delta T}_3
&= {\bf \delta T}_3 
+\Big(\pounds_{\xi_3}+\pounds^3_{\xi_1} +\frac{3}{2}\pounds_{\xi_1}\pounds_{\xi_2}
+\frac{3}{2}\pounds_{\xi_2}\pounds_{\xi_1}
\Big){\bf T}_0 \nn\\
&\qquad\qquad\qquad\qquad\qquad+3\left(\pounds^2_{\xi_1}+\pounds_{\xi_2}\right)\dT
+3\pounds_{\xi_1} \dTT\,.
\end{align}

Then, by expanding the coordinate transformation, Eq.~(\ref{defcoordtrans}) to third order we
obtain the relationship
\begin{align}
\label{coordtrans3}
{x^\mu}(q) &= x^\mu(p)+\epsilon\xi_1^{\mu}(p)
+\frac{1}{2}\epsilon^2\left(\xi^{\mu}_{1,\nu}(p)\xi_1^{~\nu}(p)
+ \xi_2^{\mu}(p)\right)\nonumber\\
&+\frac{1}{6}\epsilon^3\left[\xi_3^\mu(p)+
\left(\xi^{\mu}_{1,\lambda\beta}\xi_1^\beta
+\xi^{\mu}_{1,\beta}\xi_{1,\lambda}^\beta
\right) \xi_1^{~\lambda}(p)\right]
+\frac{1}{4}\epsilon^3\left(
\xi^{\mu}_{2,\lambda}(p)\xi_1^\lambda(p)
+\xi^{\mu}_{1,\lambda}(p)\xi_2^\lambda(p)
\right)  \,,
\end{align}
relating the coordinates of the points $p$ and $q$.

In the rest of this section, we will derive the gauge transformations at third
order in an analogous way to those derived for first and second order in section
\ref{sec:gaugetrans}.

%%%%%%%%%%%%%%%%%%%%%%%%%%%%%%%%%%%%%%%%%%%%%%%%%%%
\subsection{Four Scalars}
\label{gauge_scalar_sec}
%%%%%%%%%%%%%%%%%%%%%%%%%%%%%%%%%%%%%%%%%%%%%%%%%%%

At third order we also split the generating vector
$\xi_3^\mu$ into a scalar temporal and scalar and vector spatial part,
as
\be
\label{def_xi3}
\xi_3^\mu=\left(\alpha_2,\beta_{2,}^{~~i}+\gami3\right)\,,
\ee
where the vector part is again divergence-free ($\p_k\gamk3=0$).
We then find from Eqs.~(\ref{Ttrans3})  that the
energy density transforms as
\begin{align}
\widetilde{\delta\rho_3}
&=\delta \rho_3
+\Big(\pounds_{\xi_3}+\pounds^3_{\xi_1} +\frac{3}{2}\pounds_{\xi_1}\pounds_{\xi_2}
+\frac{3}{2}\pounds_{\xi_2}\pounds_{\xi_1}
\Big){\rho}_0 \nn\\
&\qquad\qquad\qquad
\qquad\qquad\qquad+3\left(\pounds^2_{\xi_1}+\pounds_{\xi_2}\right)\delta\rho_1
+3\pounds_{\xi_1} \delta\rho_2\,,
\end{align}
which gives
\begin{align}
\label{rhotransform3}
\widetilde{\delta\rho_3}
&=\delta \rho_3+\rho_0'\alpha_3 
+\rho_0'''\alpha_1^3+3\rho_0''\alpha_1\alpha_{1,\lambda}\xi_1^\lambda
+\rho_0'\left(
\alpha_{1,\lambda\beta}\xi_1^\lambda
+\alpha_{1,\lambda}\xi_{1,~\beta}^\lambda\right)\xi_1^\beta\nonumber\\
&+3\rho_0''\alpha_1\alpha_2
+\rho_0'\frac{3}{2}\left(
\alpha_{2,\lambda}\xi_1^\lambda+\alpha_{1,\lambda}\xi_2^\lambda
\right)
+3\left(\delta\rho_{1,\lambda\beta}\xi_1^\lambda
+\delta\rho_{1,\lambda}\xi_{1,~\beta}^\lambda\right)\xi_1^\beta\nn\\
&+3\delta\rho_{1,\lambda}\xi_2^\lambda
+3\delta\rho_{2,\lambda}\xi_1^\lambda\,.
\end{align}
Similar to the second order case, we need to specify the time slicings
(at all orders), and also the spatial gauge or threading at first and
second order, in order to render the third order density perturbation
gauge-invariant.

%%%%%%%%%%%%%%%%%%%%%%%%%%%%%%%%%%%%%%%%%%%%%%%%%%%%%%%
\subsection{The Metric Tensor}
%%%%%%%%%%%%%%%%%%%%%%%%%%%%%%%%%%%%%%%%%%%%%%%%%%%%%%%

We now give the transformation behaviour of the metric tensor at third order. 
The starting point is again the Lie derivative, which for a covariant
tensor is given by Eq.~(\ref{eq:lietensor}).

As above in the case of the transformation behaviour of a four scalar
at third order, the change under a gauge transformation of a
two-tensor can be found applying the same methods as at second
order. We therefore find that the metric tensor transforms at third
order, from \eqs{Ttrans3} and (\ref{eq:lietensor}), as
\begin{align}
\label{general_gmunu3}
\wt{\delta g^{(3)}_{\mu\nu}}&=\delta g^{(3)}_{\mu\nu}
+g^{(0)}_{\mu\nu,\lambda}\xi^\lambda_3
+g^{(0)}_{\mu\lambda}\xi^\lambda_{3~,\nu}
+g^{(0)}_{\lambda\nu}\xi^\lambda_{3~,\mu}\nn\\
&+3\Bigg[
\delta g^{(1)}_{\mu\nu,\lambda}\xi^\lambda_2
+\delta g^{(1)}_{\mu\lambda}\xi^\lambda_{2~,\nu}
+\delta g^{(1)}_{\lambda\nu}\xi^\lambda_{2~,\mu}
+\delta g^{(2)}_{\mu\nu,\lambda}\xi^\lambda_1 
 +\delta g^{(2)}_{\mu\lambda}\xi^\lambda_{1~,\nu}
+\delta g^{(2)}_{\lambda\nu}\xi^\lambda_{1~,\mu}\nn\\
&+\delta g^{(1)}_{\mu\nu,\lambda\alpha}\xi^\lambda_1\xi^\alpha_1
+\delta g^{(1)}_{\mu\nu,\lambda}\xi^\lambda_{1~,\alpha}\xi^\alpha_1
 +2\Big[
\delta g^{(1)}_{\mu\lambda,\alpha} \xi^\alpha_1\xi^\lambda_{1~,\nu}
+\delta g^{(1)}_{\lambda\nu,\alpha} \xi^\alpha_1\xi^\lambda_{1~,\mu}
+\delta g^{(1)}_{\lambda\alpha}  \xi^\lambda_{1~,\mu} \xi^\alpha_{1~,\nu}
\Big]
\nonumber \\
&+\delta g^{(1)}_{\mu\lambda}\left(
\xi^\lambda_{1~,\nu\alpha}\xi^\alpha_1
+\xi^\lambda_{1~,\alpha}\xi^\alpha_{1,~\nu}
\right)
+\delta g^{(1)}_{\lambda\nu}\left(
\xi^\lambda_{1~,\mu\alpha}\xi^\alpha_1
+\xi^\lambda_{1~,\alpha}\xi^\alpha_{1,~\mu}
\right)\Bigg]\nonumber \\
&+\frac{3}{2}\Bigg[
2g^{(0)}_{\mu\nu,\lambda\beta}\xi_2^\lambda\xi_1^\beta
+g^{(0)}_{\mu\nu,\lambda}\left(
\xi_{2,\beta}^\lambda\xi_1^\beta+\xi_{1,\beta}^\lambda\xi_2^\beta
\right)
+2g^{(0)}_{\lambda\beta}\left(
\xi_{2,\mu}^\lambda\xi_{1,\nu}^\beta+\xi_{2,\nu}^\lambda\xi_{1,\mu}^\beta
\right)
\nonumber \\
&+\left(
g^{(0)}_{\mu\lambda,\beta}\xi_{2,\nu}^\lambda
+g^{(0)}_{\lambda\nu,\beta}\xi_{2,\mu}^\lambda
+g^{(0)}_{\mu\lambda}\xi_{2,\beta\nu}^\lambda
+g^{(0)}_{\lambda\nu}\xi_{2,\beta\mu}^\lambda
\right)\xi_1^\beta
+\left(g^{(0)}_{\beta\nu,\lambda}\xi_2^\lambda
+g^{(0)}_{\lambda\nu}\xi_{2,\beta}^\lambda
\right)\xi_{1,\nu}^\beta \nn\\
&+\left(g^{(0)}_{\beta\nu,\lambda}\xi_2^\lambda
+g^{(0)}_{\lambda\nu}\xi_{2,\beta}^\lambda
\right)\xi_{1,\mu}^\beta
%\nonumber \\ 
%
%&&
+\left(
g^{(0)}_{\mu\lambda,\beta}\xi_{1,\nu}^\lambda
+g^{(0)}_{\lambda\nu,\beta}\xi_{1,\mu}^\lambda
+g^{(0)}_{\mu\lambda}\xi_{1,\beta\nu}^\lambda
+g^{(0)}_{\lambda\nu}\xi_{1,\beta\mu}^\lambda
\right)\xi_2^\beta\nn\\
&+\left(g^{(0)}_{\beta\nu,\lambda}\xi_1^\lambda
+g^{(0)}_{\lambda\nu}\xi_{1,\beta}^\lambda
\right)\xi_{2,\nu}^\beta
+\left(g^{(0)}_{\beta\nu,\lambda}\xi_1^\lambda
+g^{(0)}_{\lambda\nu}\xi_{1,\beta}^\lambda
\right)\xi_{2,\mu}^\beta
\Bigg]\nonumber \\
&+\Bigg\{
g^{(0)}_{\mu\nu,\alpha\beta\lambda}\xi_{1}^\alpha\xi_{1}^\beta\xi_{1}^\lambda
+3g^{(0)}_{\mu\nu,\lambda\beta}\xi_{1}^\lambda\xi_1^\alpha\xi_{1,\alpha}^\beta
+g^{(0)}_{\mu\nu,\lambda}\xi_{1,\alpha\beta}^\lambda\xi_{1}^\alpha\xi_{1}^\beta
+g^{(0)}_{\mu\nu,\lambda}\xi_{1,\beta}^\lambda\xi_{1,\alpha}^\beta\xi_{1}^\alpha
\nonumber \\
&
+3g^{(0)}_{\mu\beta,\lambda}
\left(\xi_{1}^\lambda\xi_{1,\alpha}^\beta\xi_{1,\nu}^\alpha
+\xi_{1}^\alpha\xi_{1}^\lambda\xi_{1,\alpha\nu}^\beta\right)
+3g^{(0)}_{\mu\alpha,\lambda}
\xi_{1}^\beta\xi_{1,\beta}^\lambda\xi_{1,\nu}^\alpha
+3g^{(0)}_{\mu\alpha,\beta\lambda}
\xi_{1}^\beta\xi_{1}^\lambda\xi_{1,\nu}^\alpha
\nonumber \\
&+g^{(0)}_{\mu\lambda}\Bigg[
\xi_{1}^\alpha\xi_{1}^\beta\xi_{1,\alpha\beta\nu}^\lambda
+\xi_{1,\alpha}^\beta\xi_{1,\beta}^\lambda\xi_{1,\nu}^\alpha
+\xi_{1}^\alpha\xi_{1,\alpha}^\beta\xi_{1,\beta\nu}^\lambda
+2\xi_{1}^\beta\xi_{1,\alpha\beta}^\lambda\xi_{1,\nu}^\alpha
+\xi_{1}^\beta\xi_{1,\alpha}^\lambda\xi_{1,\beta\nu}^\alpha\Bigg]
\nonumber \\
%%%%%%%%%%%%%%%%%%%%%%%%%%%%%%%%%%%%%%%%%%%
&+3g^{(0)}_{\nu\beta,\lambda}
\left(\xi_{1}^\lambda\xi_{1,\alpha}^\beta\xi_{1,\mu}^\alpha
+\xi_{1}^\alpha\xi_{1}^\lambda\xi_{1,\alpha\mu}^\beta\right)
+3g^{(0)}_{\nu\alpha,\lambda}
\xi_{1}^\beta\xi_{1,\beta}^\lambda\xi_{1,\mu}^\alpha
+3g^{(0)}_{\nu\alpha,\beta\lambda}
\xi_{1}^\beta\xi_{1}^\lambda\xi_{1,\mu}^\alpha
\nonumber \\
&+g^{(0)}_{\nu\lambda}\Bigg[
\xi_{1}^\alpha\xi_{1}^\beta\xi_{1,\alpha\beta\mu}^\lambda
+\xi_{1,\alpha}^\beta\xi_{1,\beta}^\lambda\xi_{1,\mu}^\alpha
+\xi_{1}^\alpha\xi_{1,\alpha}^\beta\xi_{1,\beta\mu}^\lambda
+2\xi_{1}^\beta\xi_{1,\alpha\beta}^\lambda\xi_{1,\mu}^\alpha
+\xi_{1}^\beta\xi_{1,\alpha}^\lambda\xi_{1,\beta\mu}^\alpha\Bigg]
\nonumber \\
&+6g^{(0)}_{\alpha\beta,\lambda}\xi_{1}^\lambda
\xi_{1,\nu}^\beta
+3g^{(0)}_{\alpha\lambda}\Bigg[
\xi_{1,\beta}^\lambda\left(
\xi_{1,\mu}^\alpha\xi_{1,\nu}^\beta+\xi_{1,\nu}^\alpha\xi_{1,\mu}^\beta
\right)
+\xi_{1}^\beta\left(
\xi_{1,\mu}^\alpha\xi_{1,\beta\nu}^\lambda
+\xi_{1,\nu}^\alpha\xi_{1,\beta\mu}^\lambda
\right)\Bigg]
\Bigg\}
\,.
\end{align}
However, in this case it becomes even more obvious than in Section~\ref{sec:def}
 that the expressions at third order are of not inconsiderable
size. This will also be clear from the Einstein tensor components and
the evolution equations given below in Section~\ref{sec:equ}.

Now, following along the same lines as at second order,
\eq{general_gmunu3} gives the transformation for the spatial part of
the metric at third order,
\bea
\label{Cij3trans}
2\widetilde C_{3ij}&=&2C_{3ij}+2\H\alpha_3 \delta_{ij}
+2\xi_{3(i,j)}
+\X_{3ij}\,,
\eea
where $\X_{3ij}$ contains terms cubic in the first order
perturbations. Extracting the curvature perturbation gives
\be
\label{transpsi3}
\wt\psi_3=\psi_3-\H\alpha_3-\frac{1}{4}\X^k_{3~k}
+\frac{1}{4}\nabla^{-2} \X^{ij}_{3~,ij}\,.
\ee
This expression is general, including scalar,
vector, and tensor perturbations and is valid on all scales.
However,  we shall detail here only the expression valid for
scalar perturbations and large scales and find that $\X_{3ij}$ takes then the 
simple form
\begin{align}
\label{X3ijdef}
\X_{3ij}&\equiv
2a^2\delta_{ij}\Bigg\{
-3\Big[\alpha_2\psi_1'+\frac{1}{2}\alpha_1\psi_2'
+\alpha_1\alpha_1'\left(\psi_1'+2\H\psi_1\right)
+\alpha_1^2\left(\psi_1''+4\H\psi_1'\right)\nn\\
&+2\H\alpha_2\psi_1+\H\alpha_1\psi_2
+2\left(\frac{a''}{a}+\H^2\right)\alpha_1^2\psi_1
\Big]+\left(\frac{a'''}{a}+3\H\frac{a''}{a}\right)\alpha_1^3\nn\\
&+3\left(\frac{a''}{a}+\H^2\right)\alpha_1^2\alpha_1'
+\H\alpha_1\left(\alpha_1''\alpha_1+{\alpha_1'}^2\right)\nn\\
&+\frac{3}{2}\left[\H\left(\alpha_1\alpha_2'+\alpha_1'\alpha_2\right)
+2\left(\frac{a''}{a}+\H^2\right)\alpha_1\alpha_2
\right]\Bigg\}\,. 
\end{align}
Hence we finally get for the transformation of $\psi_3$ 
\begin{align}
\label{transpsi3ls}
-\wt{\psi_3}&=-\psi_3+\H\alpha_3
+\left(\frac{a'''}{a}+3\H\frac{a''}{a}\right)\alpha_1^3
+3\left(\frac{a''}{a}+\H^2\right)\alpha_1^2\alpha_1'\nonumber \\
&+\H\alpha_1\left(\alpha_1''\alpha_1+{\alpha_1'}^2\right)
+\frac{3}{2}\left[\H\left(\alpha_1\alpha_2'+\alpha_1'\alpha_2\right)
+2\left(\frac{a''}{a}+\H^2\right)\alpha_1\alpha_2
\right] \nn\\
&-3\Big[\alpha_2\psi_1'+\frac{1}{2}\alpha_1\psi_2'
+\alpha_1\alpha_1'\left(\psi_1'+2\H\psi_1\right)
+\alpha_1^2\left(\psi_1''+4\H\psi_1'\right)
+2\H\alpha_2\psi_1\nn\\
&+\H\alpha_1\psi_2
+2\left(\frac{a''}{a}+\H^2\right)\alpha_1^2\psi_1
\Big]
\,.
\end{align}
%

%%%%%%%%%%%%%%%%%%%%%%%%%%%%%%%%%%%%%%
\section{Gauge Invariant Variables}
\label{sec:invariant}
%%%%%%%%%%%%%%%%%%%%%%%%%%%%%%%%%%%%%%

In the previous section we have described how perturbations transform
under a gauge shift. We can now use these results to construct
gauge-invariant quantities, in particular the curvature perturbation
on uniform density hypersurfaces, $\zeta$. In this section, as before,
we consider only scalar perturbations, and restrict ourselves to the
large scale limit.\\

We first define hypersurfaces in different gauges as in Section~\ref{sec:gaugechoice}.
From
Eq.~(\ref{rhotransform3})
we find the time slicing defining uniform density hypersurfaces
at  third order in the large scale limit as
\begin{align}
\alpha_{3\udg}&=-\frac{\delta\rho_3}{\rho_0'}
+\frac{1}{2{\rho_0'}^2}\Big[
3\left(\delta\rho_1\delta\rho_2'+\delta\rho_1'\delta\rho_2\right)
-\frac{\delta\rho_1''{\delta\rho_1}^2}{\rho_0'}
-4{\delta\rho_1'}^2\frac{\delta\rho_1}{\rho_0'}
+\rho_0''\delta\rho_1'\left(\frac{\delta\rho_1}{\rho_0'}\right)^2
\Big]\,.
\end{align}
Similarly, the temporal gauge transformation on uniform curvature hypersurfaces is
 defined by evaluating
(\ref{transpsi3ls}) and gives, at third
order,
\begin{align}
\alpha_{3\fg}&=\frac{\psi_3}{\H}+\frac{1}{2\H^2}\Big[3\psi_1'\psi_2
+\frac{\psi_1^2\psi_1''}{\H}+6\H\psi_1\psi_2
+\frac{4\psi_1\psi_1^2}{\H}\Big]\nn\\
&\qquad\qquad\qquad\qquad\qquad
-\frac{\psi_1^2\psi_1'}{\H^4}\Big(\frac{a''}{a}-\frac{37}{2}\H^2\Big)+\frac{8\psi_1^3}{\H}\,.
\end{align}

We can now combine the results found so far to get gauge invariant
quantities, and as before choose the curvature perturbation on uniform
density slices as well as the density perturbation on uniform curvature hypersurfaces
as examples.

One gauge invariant matter quantity of interest is the perturbation to
the energy density on uniform curvature hypersurfaces. This is
obtained by substituting the temporal gauge transformation components
in the uniform density gauge  into
the appropriate transformation equation,
Eq.~(\ref{rhotransform3}). This gives, at third order,
\begin{align}
\widetilde{\delta\rho_{3\fg}} &=
\drhorhorho+\rhob'\frac{\psi_3}{\H}+\frac{3\rhob'}{2\H^2}(2\psi_2\psi_1'+\psi_2'\psi_1)
+3\frac{\psi_1^2}{\H^3}\Big[2(\rhob'\psi_1+\psi_1'\rhob'')+\psi_1''\Big]\nn\\
&+3\frac{\psi_2\psi_1}{\H^2}\Big[\rhob''+2\rhob'\H-\rhob'\frac{a''}{a}\Big]
-9\rhob'\frac{\psi_1^2\psi_1'}{\H^3}\Big(\frac{a''}{a}-\H\Big)\nn\\
&+\frac{\psi_1^3}{\H^3}\Big[\rhob'''-3\rhob''\Big(\frac{a''}{a}+3\H\Big)+3\rhob'\H\Big(3\frac{a''}{a}-\H\Big)
+\rhob'\Big(\left(\frac{a''}{a}\right)^2-\frac{a'''}{a}\Big)\Big]
 \,.
\end{align}

The curvature perturbation on uniform density hypersurfaces, $\zeta$, as introduced in Eq.~(\ref{eq:zeta1}), is 
defined as
\be
-\zeta\equiv\widetilde{\psi_{\udg}}\,.
\ee
This is obtained by substituting the temporal gauge transformation
components in the uniform curvature gauge into the appropriate transformation equation,
Eq.~(\ref{transpsi3ls}).  Evaluating this on spatially flat hypersurfaces
then gives, at third order
\begin{align}
\label{eq:z3cm}
\zeta_3 &= -\H\frac{\drhorhorho}{\rhob'}+\frac{3\H}{\rhob'^2}(\drhorho'\drho+\drho'\drhorho)
-\frac{3}{\rhob'^2}\drhorho\drho\Big[\frac{\H\rhob''}{\rhob'}-\Big(\frac{a''}{a}+\H^2\Big)\Big]\nn\\
&-\frac{3\H}{\rhob'^3}\drho^2\drho''-\frac{6\H}{\rhob'^3}\drho'^2\drho
-\frac{3}{\rhob'^3}\drho^2\drho'\Big[2\Big(\frac{a''}{a}+\H^2\Big)-3\H\frac{\rhob''}{\rhob'}\Big]\nn\\
&-\frac{\drho^3}{\rhob'^3}\Big[3\H\left(\frac{\rhob''}{\rhob'}\right)^2-\H\frac{\rhob'''}{\rhob'}
+\frac{a'''}{a}+3\H \frac{a''}{a}-3\frac{\rhob''}{\rhob'}\Big(\frac{a''}{a}+\H^2\Big)\Big]\,.
\end{align}

There are different definitions of the curvature perturbation present
in the literature, depending on different decompositions of the
spatial part of the metric tensor. A different definition to the
one above, as discussed in e.g.~Ref.~\cite{MW2008}, was used by
Maldacena in Ref.~\cite{Maldacena} to calculate the non-gaussianity
from single field inflation, and was introduced by Salopek {\it et al.} in
Refs.~\cite{Salopek:1990jq, Salopek:1988qh}. 
They define the local scale factor
$\tilde{a}\equiv e^{\alpha}$, then
\be
e^{2\alpha}=a^2(\eta)e^{2\zeta}=a^2(\eta)(1+2\zeta_{\rm{SB}}+2\zeta_{\rm{SB}}^2+\frac{4}{3}\zeta_{\rm{SB}}^3)\,.
\ee
Comparing to the expansion from perturbation theory
\be
e^{2\alpha}=a^2(\eta)(1+2\zeta)\,,
\ee
one can obtain the relationship
\be
\zeta=\zeta_{\rm{SB}}+\zeta_{\rm{SB}}^2+\frac{2}{3}\zeta_{\rm{SB}}^3\,.
\ee
Splitting this up order by order gives, at second order
\be
\label{zeta2SB}
\zeta_{\rm{2SB}}=\zeta_2-2(\zeta_1)^2\,,
\ee
and, at third order,
\be
\label{zeta3sb}
\zeta_{3\rm{SB}}=\zeta_{3}-6\zeta_{2}\zeta_1+8(\zeta_1)^3\,.
\ee
Note that it is this definition, (\ref{zeta3sb}), of the curvature
perturbation which occurs in Ref.~\cite{Lehners:2009ja}, though with
different pre-factors since their perturbative expansion is defined
differently. It is perhaps worth mentioning that $\zeta_{\rm{SB}}$,
  the variable first introduced by Salopek and Bond and then employed
  for non-gaussianity calculations by Maldacena, is extremely Gaussian
  after slow-roll inflation, as opposed to other $\zeta$ variables
  which exhibit non-gaussianity, as can be seen e.g.~from
  \eq{zeta2SB}.

%%%%%%%%%%%%%%%%%%%%%%%%
\section{Governing Equations}
\label{sec:equ}
%%%%%%%%%%%%%%%%%%%%%%%%

Having constructed gauge invariant quantities up to third order in the
previous section, we now turn to the evolution and the field
equations. The equations presented in this section in full generality are new; Refs.~\cite{Hwang:2007wq,Hwang:2005he}
previously considered some governing equations at third order in perturbation theory, however
they focussed on pressureless, irrotational fluids.

%%%%%%%%%%%%%%
\subsection{Fluid Conservation Equation}
%%%%%%%%%%%%%%%

In this section, we give the energy momentum conservation equations
for a fluid with non-zero pressure and in the presence of scalar,
vector and tensor perturbations. The latter generalisation is
important since at orders above linear order, all types of
perturbation are coupled. 

As at linear and second order, presented in the previous chapters, energy momentum
conservation
\be
\label{eq:EMconservation}
\nabla_{\mu}T^{\mu}{}_{\nu}=0,
\ee
gives us evolution equations.
Substituting
the definition for the energy momentum tensor, Eq.~(\ref{eq:emdeffluid}) expanded to third order,
into Eq.~(\ref{eq:EMconservation}) gives energy conservation (the 0-component)
\begin{align}
\label{eq:energycons}
&\delta\rho' +3\H(\delta\rho+\delta P)+(\rhob+\Pb)(C^i{}_{i}'+v_{i,}{}^i)+(\delta\rho+\delta P)(C^i{}_{i}'+v_{i,}{}^i)
+(\delta\rho+\delta P)_{,i}v^i \nn \\
&+(\rhob+\Pb)\Big[(B^i+2v^i)(v_i'+B_i')+v^i{}_{,i}\phi-2C_{ij}'C^{ij}
+v^i(C^j{}_{j,i}+2\phi_{,i})+4\H v^i(B_i+2v_i)\Big] \nn \\
&+ (\delta\rho+\delta P)\Big[(B^i+2v^i)(v_i'+B_i') - 2C_{ij}'C^{ij}+v^i(C^j{}_{j,i}
+2\phi_{,i}+4\H(v_i+B_i))+v^i{}_{,i}\phi\Big]  \nn\\
& +(\delta\rho'+\delta P')v^i(B_i+v_i)+(\delta\rho+\delta P)_{,i}\phi v^i +(\rhob+\Pb)(2C^{ij}v_iv_j-B_iv^i\phi+v^iv_i\phi) \nn\\
&
-(\rhob+\Pb)\Big[2C^{ij}(C_{ij,k}v^k-2v_i'v_j +B_iB_j'-2C_i{}^kC_{jk}') 
+\frac{1}{2}v^i{}_{,i}(\phi^2-v^jv_j)\nn \\
& +(v^i+B^i)(2B_i'\phi-C^j{}_j'v_i) 
 -v^j\Big\{B^iB_{i,j}+v^iv_{i,j}+\phi(v_j+2v_j'+3\H v_j-2\phi_{,j}+C^i{}_{i,j}) \nn \\
 &-2\phi'B_j+3C_{ij}'v^i\Big\}
+B^j(B^iC_{ij}'+B_j\phi'+v_j'\phi)\Big] =0\,,
\end{align}
and momentum conservation (the $i-$component)
\begin{align}
&\Big[(\rhob+\Pb)(v_i+B_i)\Big]'+(\rhob+\Pb)(\phi_{,i}+4\H(v_i+B_i))
+\delta P_{,i}+\Big[(\delta\rho+\delta P)(v_i+B_i)\Big]'\nn \\
&+(\delta\rho+\delta P)(\phi_{,i}+4\H(v_i+B_i)) -(\rhob'+\Pb')\Big[(2B_i+v_i)\phi -2C_{ij}v^j\Big]\nn\\
&
+(\rhob+\Pb)\Big[(v_i+B_i)(C^j{}_{j}'+v^j{}_{,j})-B_i(\phi'+8\H\phi)
+v^j(B_{i,j}-B_{j,i}+v_{i,j}+8\H C_{ij}) \nn \\
& +(2C_{ij}v^j)'-\phi(v_i'+2B_i'+2\phi_{,i}+4\H v_i)\Big]
+(\rhob'+\Pb')\Big[v^j(B_{i}v_j+B_jv_i+2B_iB_j+\frac{1}{2}v_iv_j\nn\\
&-2C_{ij}\phi)
+(\frac{3}{2}v_i+4B_i)\phi^2\Big]
+(\rhob+\Pb)\Big[\phi^2\left(4B_i'+\frac{3}{2}v_i'+2\H(8B_i+3v_i)+4\phi_{,i}\right) \nn \\
& 
+(v_i+2B_i)(v_j'B^j-C^j{}_j'\phi)+(v_i+B_i)(B_j'B^j-C_{jk}'C^{jk})
+v^j\Big\{2B_i(B_j'+v_j') \nn \\
& +B_j(2B_i'+2\phi_{,i}+v_i')+2\H(v_i+2B_i)(2B_j+v_j)
+2C_{ij}(C^k{}_{k}'+v^k{}_{,k}-4\H\phi)\nn \\
&+C^k{}_{k,j}(B_i+v_i) +v_i(B_j'+\phi_{,j})+v_j(B_i'+\phi_{,i})+v^k(2C_{ik,j}-C_{jk,i})\nn\\
&
+(B_{j,i}-B_{i,j}-2C_{ij}')\phi+2C_{ik}v^k{}_{,j}\Big\}
+\frac{1}{2}(v_iv_jv^j)'-2C_{ij}v^j{}'\phi+B_i(4\phi'\phi-v_{j,}{}^j\phi)\Big] =0\,.
\end{align}

As emphasised earlier, 
Eq.~(\ref{eq:energycons}) highlights the coupling between tensor and
scalar perturbations which occurs only at third order (and higher) in
perturbation theory. At both linear and second order, no such coupling
exists, since the only terms coupling the spatial metric perturbation,
$C_{ij}$, to scalar perturbations contain either the trace or the
divergence of $C_{ij}$ and the tensor perturbation, $h_{ij}$ is, by
definition, transverse and trace-free. However, at third order, terms
like $\delta\rho C_{ij}'C^{ij}$ occur in the energy conservation equation
which, on splitting up order by order and decomposing $C_{ij}$ becomes
$\drho h_{1ij}'h_1{}^{ij}$. It is clear that this term only shows up
at third order and beyond. Thus, as mentioned earlier, third order is
the lowest order at which all the different types of perturbations
couple to one another in the evolution equations, which will produce
another physical signature of the full theory.

%%%%%%%%%%%%%%%
\subsubsection{Scalars only}
%%%%%%%%%%%%%%%%%

It will be useful to have energy and momentum conservation equations
for only scalar perturbations. These equations are obtained by making
the appropriate substitutions $C_{ij}=-\dij\psi+E_{,ij}$,
$v_i=v_{,i},$ and $B_i=B_{,i}$ into the above expressions.  On doing
so, we obtain the energy conservation equation
\begin{align}
&\delta\rho' +3\H(\delta\rho+\delta P)+(\rhob+\Pb)(\nabla^2v+\nabla^2E-3\psi')+(\delta\rho+\delta P)(\nabla^2v+\nabla^2E-3\psi')
 \nn \\
&+(\delta\rho+\delta P)_{,i}v_,{}^i+(\rhob+\Pb)\Big[(B^i+2v_,{}^i)(v_{,i}'+B_{,i}')+\nabla^2v\phi-2\Big(\psi'(3\psi-\nabla^2E)\nn\\
&
-\psi\nabla^2E'+E_{,ij}'E_,{}^{ij}\Big)+v_,{}^i(2\phi_{,i}+\nabla^2E_{,i}-3\psi_{,i})+4\H v_,{}^i(B_{,i}+2v_{,i})\Big] \nn \\
&+ (\delta\rho+\delta P)\Big[(B_,{}^i+2v_,{}^i)(v_{,i}'+B_{,i}') -2\Big(\psi'(3\psi-\nabla^2E)-\psi\nabla^2E'+E_{,ij}'E_,{}^{ij}\Big)\nn\\
&+v_,{}^i(2\phi_{,i}-3\psi_{,i}+\nabla^2E
+4\H(v_{,i}+B_{,i}))+\nabla^2v\phi\Big] +(\delta\rho'+\delta P')v_,{}^i(B_{,i}+v_{,i})\nn\\
& +(\delta\rho+\delta P)_{,i}\phi v_,{}^i  +(\rhob+\Pb)(v_,{}^iv_{,i}\phi-B_{,i}v_,{}^i\phi-6\psi v_,{}^jv_{,j}+E_,{}^{ij}v_{,i}v_{,j}) \nn\\
&
-(\rhob+\Pb)\Big[2E_,{}^{ij}(v_,{}^kE_{,ijk}-2v_{,i}'v_{,j}+B_{,i}B_{,j}'-2E_{,i}{}^kE_{,kj})-2\nabla^2E\psi_{,k}v_,{}^k\nn\\
&
-2\psi\Big(v_,{}^k(\nabla^2E_{,k}-3\psi_{,k})-2v_{,j}'v_,{}^j+B_{,j}'B_,{}^j+6\psi\psi'\Big)+\frac{1}{2}\nabla^2v(\phi^2-v_,{}^jv_{,j})\nn\\
&
 +(v_,{}^i+\nabla^2E'+B_,{}^i)(2B_{,i}'\phi+3\psi'v_{,i})  -v_,{}^j\Big\{B_,{}^iB_{,ij}+v_,{}^iv_{,ij}
 -2\phi'B_{,j}-3\psi'v_{,j}\nn \\
& 
+\phi(v_{,j}+2v_{,j}'+3\H v_{,j}-2\phi_{,j}-3\psi_{,j}+\nabla^2E_{,j}) +E_{,ij}'v_,{}^i\Big\}\nn\\
&
+B_,{}^j(B_{,j}\psi'+E_{,ij}'v_,{}^i+B_{,j}\phi'+v_{,j}'\phi)\Big] =0\,,
\end{align}

\ni and the momentum conservation equation

\begin{align}
& \Big[(\rhob+\Pb)(v_{,i}+B_{,i})\Big]'+(\rhob+\Pb)(\phi_{,i}+4\H(v_{,i}+B_{,i}))
+\delta P_{,i}\nn\\
&+\Big[(\delta\rho+\delta P)(v_{,i}+B_{,i})\Big]'
+(\delta\rho+\delta P)(\phi_{,i}+4\H(v_{,i}+B_{,i}))\nn\\
&-(\rhob'+\Pb')\Big[(2B_{,i}+v_{,i})\phi+2\psi v_{,i}-2E_{,ij}v_,{}^j\Big]\nn\\
&
+(\rhob+\Pb)\Big[(v_{,i}+B_{,i})(\nabla^2v-3\psi'+\nabla^2E')-B_{,i}(\phi'+8\H\phi)\nn\\
&
+v_,{}^j\Big(v_{,ij}-8\H (\psi\dij-E_{,ij}) \Big)
-2(\psi v_{,i}+E_{,ij}v_,{}^j)'-\phi(v_{,i}'+2B_{,i}'+2\phi_{,i}+4\H v_{,i})\Big]\nn\\
& +(\rhob'+\Pb')\Big[v_,{}^j(B_{,i}v_{,j}+B_{,j}v_{,i}+2B_{,i}B_{,j}
 +\frac{1}{2}v_{,i}v_{,j}+2\phi(\psi\dij-E_{,ij}))\nn\\
&+(\frac{3}{2}v_{,i}+4B_{,i})\phi^2\Big] 
 +(\rhob+\Pb)\Big[(v_{,i}+2B_{,i})\Big(v_{,j}'B_,{}^j-(3\psi'-\nabla^2E')\phi\Big)\nn\\
 &
+\phi^2\left(4B_{,i}'+\frac{3}{2}v_{,i}'+2\H(8B_{,i}+3v_{,i})+4\phi_{,i}\right)+(v_{,i}+B_{,i})\Big(B_{,j}'B_,{}^j-3\psi'\psi
\nn \\
& +\psi'\nabla^2E+\psi\nabla^2E'-E_,{}^{jk}E_{,jk}'\Big) +v_,{}^j\Big\{2B_{,i}(B_{,j}'+v_{,j}')+B_{,j}(2B_{,i}'+2\phi_{,i}+v_{,i}')\nn\\
&+2\H(v_{,i}+2B_{,i})(2B_{,j}+v_{,j})
 -(3\psi_{,j}-\nabla^2E_{,j})(B_{,i}+v_{,i})\nn\\
 &-2(\psi\dij-E_{,ij})(\nabla^2v+\nabla^2E'-3\psi'-4\H\phi)+v_{,i}(B_{,j}'+\phi_{,j})\nn\\
&+v_{,j}(B_{,i}'+\phi_{,i})-2\psi_{,j}v_{,i}+E_{,ikj}v_,{}^k+\psi_{,i}v_{,j}
+2(\psi'\dij-E_{,ij}))\phi-2\psi v_{,ij}\nn\\
&+2E_{,ik}v_,{}^k{}_j\Big\}
+\frac{1}{2}(v_{,i}v_{,j}v_,{}^j)'
+2\psi v_{,i}'\phi-2E_{,ij}v_,{}^j{}'\phi+B_i(4\phi'\phi-\nabla^2v\phi)\Big] =0\,.
\end{align}

Considering the large scale limit, in which spatial gradients vanish, the energy conservation equation 
becomes
\begin{align}
&\delta\rho' +3\H(\delta\rho+\delta P)-3\psi'(\rhob+\Pb)-3\psi'(\delta\rho+\delta P)\nn\\
&-6\psi\psi' (\rhob+\Pb) -6\psi\psi'\ (\delta\rho+\delta P) +12\psi^2\psi'(\rhob+\Pb)=0\,.
\end{align}
Splitting up perturbations order by order, this becomes
\begin{align}
&\drhorhorho'+3\H(\drhorhorho+\dPPP)-3\psi_3'(\rhob+\Pb)-9\psi_2'(\drho+\dP)
-9\psi_1'(\drhorho+\dPP)\nn\\
&\quad-18(\rhob+\Pb)(\psi_2\psi_1'+\psi_1\psi_2')+72\psi_1^2\psi_1'(\rhob+\Pb)=0\,.
\end{align}
In the uniform curvature gauge, where $\psi=0$, this is
\begin{align}
&\delta\rho_{3\fg}'+3\H(\delta\rho_{3\fg}+\delta P_{3\fg})=0\,,
\end{align}
and in the uniform density gauge, where $\delta\rho=0$,
\begin{align}
\label{eq:udgzetaev}
&3\H\delta P_{3\udg}+3\zeta_3'(\rhob+\Pb)+9\zeta_2'\delta P_{1\udg}
+9\zeta_1'\delta P_{2\udg}\nn\\
&-18(\rhob+\Pb)(\zeta_2\zeta_1'+\zeta_1\zeta_2')
-72\zeta_1^2\zeta_1'(\rhob+\Pb)=0\,,
\end{align}
with $\zeta$ as defined above. This can be recast in the more familiar form by introducing the (gauge invariant) non-adiabatic pressure perturbation. At linear order the pressure perturbation can be expanded as, from Eq.~(\ref{eq:dPexpand}),
\begin{align}
\delta P_1 &=\frac{\p P}{\p S}\delta S_1+\frac{\p P}{\p \rho}\delta\rho_1\equiv\delta P_{\rm{nad}1}+\cs\drho\,.
\end{align}
This can be extended to second order \cite{nonad} and higher by simply not truncating the Taylor series:
\begin{align}
\dPn_2 &= \dPP-\cs\drhorho-\frac{\p \cs}{\p \rho}\drho^2\,,\\
\dPn_3&= \dPPP-\cs\drhorhorho-3\frac{\p \cs}{\p \rho}\drhorho\drho-\frac{\p^2\cs}{\p\rho^2}\drho^3\,.
\end{align}
Thus, in the uniform density gauge, the pressure perturbation is equal
to the non-adiabatic pressure perturbation at all orders. Then,
Eq.~(\ref{eq:udgzetaev}) becomes
\be
\label{zeta3_cons}
\zeta_3'+\frac{\H}{\rhob+\Pb}\dPn_3=
6(\zeta_2\zeta_1'+\zeta_1\zeta_2')+24\zeta_1^2\zeta_1'-\frac{3}{\rhob+\Pb}(\zeta_2'\dPn_1+\zeta_1'\dPn_2)\,.
\ee
In the case of a vanishing non-adiabatic pressure perturbation,
$\zeta_1'$ and $\zeta_2'$ are zero and hence we see that $\zeta_3$ is
also conserved, on large scales. This was also found in
Ref.~\cite{Lehners:2009ja}, and previously in Ref.~\cite{Enqvist:2006fs}.
%

%%%%%%%%%%%%%%%%%%%%%%%
\subsection{Klein-Gordon Equation}
%%%%%%%%%%%%%%%%%%%%%%%%

The energy momentum tensor for a canonical scalar field minimally
coupled to gravity is easily obtained by treating the scalar field as a
perfect fluid with energy-momentum tensor (c.f. Chapter \ref{ch:perturbations})
\be
T^{\mu}{}_{\nu} = g^{\mu\lambda}\vp_{,\lambda}\vp_{,\nu}
-\delta^{\mu}{}_{\nu}\left(\frac{1}{2}g^{\alpha\beta}\vp_{,\alpha}\vp_{,\beta}+U(\vp)\right) \,,
\ee
where the scalar field $\vp$ is split to third order as 
\be
\vp(\eta,x^i)=\vpb(\eta)+\dvph[1](\eta,x^i)+\frac{1}{2}\dvph[2](\eta,x^i)
+\frac{1}{3!}\dvph[3](\eta,x^i) \,,
\ee
and the potential $U$ similarly as 
\be
U(\vp)=U_{0}+\dU[1]+\frac{1}{2}\dU[2]+\frac{1}{3!}\dU[3]\,,
\ee
where we define
\begin{align}
\dU[1] = U_{,\vp}\dvph[1] \,, \qquad
\dU[2] =  U_{,\vp\vp}\dvph[1]^2 + U_{,\vp}\dvph[2] \,, \nn\\
\dU[3] = U_{,\vp\vp\vp}\dvph[1]^3+2U_{,\vp\vp}\dvph[1]\dvph[2]
+U_{,\vp}\dvph[3]\,,
\end{align}
and making use of  the shorthand notation $U_{,\vp}\equiv\frac{\p U}{\p\vp}$. Then, Eq. (\ref{eq:EMconservation})
gives the Klein-Gordon equation
\begin{align}
\label{eq:KG}
&\scriptstyle\dvph[3]''-\nabla^2\dvph[3]+4\H\dvph[3]'+\frac{\vpb''}{\vpb'}\dvph[3]'
-\frac{3\dvph[2]''}{\vpb'}\left(2\vpb'\phi-\dvph[1]'\right)
-\frac{3}{\vpb'}\left(\nabla^2\dvph[2]\dvph[1]'+\dvph[2]'\nabla^2\dvph[1]\right)  \nn\\
&\scriptstyle-\frac{6\dvph[2]'}{\vpb'}\left(\phi\vpb''-2\H\dvph[1]'+\vpb\phi'+\vpb'B^{i}{}_{,i}-C^i{}_{i}'\vpb'+4\H\vpb'\phi\right)
-\frac{6(\dvph[1]')^2}{\vpb'}\left(\phi'+B^{i}{}_{,i}+4\H\phi-C^i{}_{i}'\right)\nn \\
&\scriptstyle
-\frac{6\dvph[1]''}{\vpb'}\left(2\phi\dvph[1]'-4\vpb'\phi^2+2\vpb'\phi-\frac{1}{2}\dvph[2]'+B^iB_i\vpb'\right)
-\frac{6\dvph[1]'}{\vpb'}\Big[\vpb''(2\phi-4\phi^2+B^iB_i) \nn\\
&\scriptstyle+8\phi\vpb'(\H-2\H\phi-2\phi') +B^i(\vpb'B_i'+2\delta\vp_{1,i}'+4\H B_i)
+2\vpb'\Big\{B^i{}_{,i}+\phi'-C^i{}_{i}'(1-2\phi)\nn\\
&\scriptstyle-2C^{ij}(B_{j,i}-C_{ij}')+B^iC^j{}_{i,j}-2C^{ij}{}_{,j}B_i
-B^i\phi_{,i}-2B^i{}_{,i}\phi\Big\}-2C^{ij}\delta\vp_{1,ij}\nn\\
&\scriptstyle+\delta\vp_{1,i}(B^i{}'+2\H B^i+2C^{ij}{}_{,j}+\phi_,{}^i)\Big]-3\delta\vp_{2,i}\left(B^i{}'+C^j{}_{j,}{}^i+\phi_,{}^i-2C^{ij}{}_{,j}+2\H B^i\right)\nn \\
&\scriptstyle-6\delta\vp_{1,i}\Big[B^i{}'+C^j{}_{j,}{}^i-2C^{ij}{}_{,j}+\phi_{,}{}^i+\frac{C^j{}_{j,}{}^i}{\vpb'}
+2\H B^i-2B_jC^{ij}{}'+B^iC^j{}_{j}'+B^jB_{j,}{}^i-B^jB^i{}_{,j} \nn\\
&\scriptstyle-B^iB^j{}_{,j}-2C^{ij}B_{j}'-2C^{ij}C^k{}_{k,j}+4C^{ij}C_{jk,}{}^k-2C^{ij}\phi_{,j}-2B^i{}'\phi-B^i\phi'-2\phi_,{}^i\phi
-4\H B_jC^{ij} \nn\\
&\scriptstyle-4\H B^i\phi+4C^{kj}C_j{}^i{}_{,k}-2C^{kj}C_k{}^i{}_{,j}\Big]
-12B^i\left(\delta\vp_{1,i}'+\frac{1}{2}\delta\vp_{2,i}'\right)
+24\delta\vp_{1,i}'(B^i\phi+C^{ij}B_j)\nn\\
&\scriptstyle
+6\delta\vp_{2,ij}C^{ij}
-6\delta\vp_{1,ij}(4C^{kj}C_{k}{}^{i}-B^iB^j)
-24\H\vpb'\phi(1-2\phi+4\phi^2-\phi'+3\phi'\phi) \nn\\
&\scriptstyle-6\vpb''\Big(2\phi(1-2\phi+4\phi^2)+B^iB_i-4B^iB_i\phi-2B^iB^jC_{ij}\Big)
-6C^{ij}{}_{,j}\vpb'(B_i\phi-2B_i)
 \nn\\
&\scriptstyle
+6C^i{}_{i}'\vpb'(1-2\phi+4\phi^2-B^jB_j)
-6C^j{}_{j,i}\vpb'(B^i-2C^{ik}B_k-2B^i\phi)
-6C^{ij}{}'\vpb'(2C_{ij}-B_iB_j\nn \\
&\scriptstyle-4C_{ij}\phi-4C_{kj}C^k{}_i)
-12\vpb'C^{ij}\Big[2B_iC_{jk,}{}^k+2B^kC_{ki,j}-B_{j,i}-B_iB_j'+B_i\phi_{,j}
-B_kC_{ji,}{}^k\nn \\
&\scriptstyle+2C_{ik}B^k{}_{,j}+2B_{j,i}\phi-4\H B_iB_j\Big]
-6\vpb'\Big[\phi+B^i{}_{,i}+B^iB_i'-B^iB_iB^j{}_{,j}-B^iB_jB^j{}_{,i}-4B^iB_i'\phi \nn\\
&\scriptstyle+2\H B^iB_i-2B^iB_i\phi'-8\H B^iB_i\phi+4B^i\phi_{,i}-2B^i{}_{,i}\phi+4B^i{}_{,i}\phi^2-B^i\phi_{,i}\Big]
 +6U_{,\vp}a^2=0\,.
\end{align}
One can again see the coupling between first order tensor and scalar perturbations. For example, 
the $\delta\varphi_{1,i}C^{ij}\phi_{,j}$ contains a term that looks like $\delta\varphi_{1,i}h_1{}^{ij}\phi_{1,j}$, which 
occurs only at third order and beyond.

Again, we refrain from splitting up the perturbations order by order for ease of presentation. Once split up,
one can then replace the metric perturbations by using the appropriate order field equations.
We present the Einstein tensor at third order in the next section. Note also that Eq.~(\ref{eq:KG}) implicitly contains the
Klein-Gordon equations at first and second order. We refer the reader to, for example, Ref.~\cite{Malik:2006ir}, for
a detailed exposition of the second order Klein-Gordon equation.

%%%%%%%%%%%%%%%%%%%%%%
\subsection{Einstein Tensor}
%%%%%%%%%%%%%%%%%%%%%%%

The Einstein tensor, which describes the geometry of the universe, is defined 
(as shown in Section~\ref{sec:standardcosmology}) as
\be
G^\mu{}_\nu = R^\mu{}_\nu-\frac{1}{2}\delta^\mu{}_\nu R\,,
\ee
where $R^\mu{}_\nu$ is the Ricci curvature tensor and $R$ is the Ricci scalar. 
Here, we give the components of the Einstein tensor up to third order:

\begin{align}
\scriptstyle
a^2G^{0}{}_{0} &
\scriptstyle = -3\H^2+\nabla^2C^{j}{}_{j}-C_{ij,}{}^{ij}
+2\H( -C^{i'}{}_{i}+B^{i}{}_{,i}+3\H\phi)
+C^{j}{}_{j,i}(\frac{1}{2}C^{k}{}_{k,}{}^{i}-2C^{ik}{}_{,k}) +C_{ij}^{'}(\frac{1}{2}C^{ij'}-B^{j}{}_{,}{}^{i})\nn \\
&\scriptstyle +B^{i}\Big[ C^{j'}{}_{j,i}-C_{ij,}^{'}{}^{j}
+\frac{1}{2}\left(\nabla^2B_{i}-B_{j,i}{}^{j}\right)+2\H\left(C^{j}{}_{j,i}-2C_{ij,}{}^j-\phi_{,i}\right)\Big]
 +2C^{ij}\Big[2C_{jk,i}{}^{k}-C^{k}{}_{k,ij}-\nabla^2C_{ij} \nn \\
&\scriptstyle 
+2\H(C_{ij}^{'}-B_{i,j})\Big]
+C_{jk,i}(C^{ik}{}_{,}{}^{j}-\frac{3}{2}C^{jk}{}_{,}{}^{i}) +C^{i'}{}_{i}(B_{j,}{}^{j}-\frac{1}{2}C^{j'}{}_{j}+4\H\phi)
+2C^{ij}{}_{,i}C_{jk,}{}^{k}\nn \\
&\scriptstyle
+\frac{1}{4}B_{j,i}(B^{i}{}_{,}{}^{j}+B^{j}{}_{,}{}^{i})  -3\H^2(4\phi^2-B_{i}B^{i})-\frac{1}{2}B^{i}{}_{,i}B_{j,}{}^{j}
-4\H B^{i}{}_{,i}\phi + \mathbb{G}^0{}_0 \,,
\end{align}

\begin{align}
\scriptstyle
a^2G^{0}{}_{i} &\scriptstyle=  C^{k}{}_{k,i}^{'}
-C_{ik,}^{'}{}^{k} -\frac{1}{2}\left(B_{k,i}{}^k-\nabla^2 B_i\right)-2\H \phi_{,i}
+8\H\phi_{,i}\phi +C_{ij}^{'}\left(2C^{kj}{}_{,k}-C^{k}{}_{k,}{}^{j}+\phi_{,}{}^{j}\right)
-C^{j'}{}_{j}\phi_{,i} \nn \\
& \scriptstyle+2C^{kj}\left[C_{ik,j}^{'}-C_{jk,i}^{'}+\frac{1}{2}\left(B_{k,ij}-B_{i,kj}\right)\right]
+B^{j}\left(C_{kj,i}{}^{k}-C^{k}{}_{k,ij}+C_{ik,k}{}^{j}-\nabla^2C_{ij}-2\H B_{j,i}\right)\nn \\
&\scriptstyle-\frac{1}{2}\Big(B_{i,j}+B_{j,i}\Big)\phi_{,}{}^{j}  +\left(B_{i,j}-B_{j,i}\right)\left(\frac{1}{2}C^{k}{}_{k,}{}^{j}-C^{jk}{}_{,k}\right)
-C_{ik,j}\left(B^{k}{}_{,}{}^{j}-B^{j}{}_{,}{}^{k}\right)
+B^{j}{}_{,j}\phi_{,i} \nn\\
&\scriptstyle +\phi\left[B_{j,i}{}^{j}-\nabla^2B_{i}+2\left(C_{ij,}^{'}{}^{j}-C^{j'}{}_{j,i}\right)\right] -C^{kj'}C_{kj,i}
+\mathbb{G}^0{}_i \,,
\end{align}

\begin{align}
\scriptstyle a^2G^{i}{}_{j} &\scriptstyle= C^{i''}{}_{j}+2\H C^{i}{}_{j}^{'}-\frac{1}{2}(B^{i'}{}_{,j}+B_{j,}{}^{i'})
-C^{l}{}_{l,j}{}^{i}+C^{i}{}_{l,j}{}^{l}-\phi_{,}{}^{i}{}_{j}-\nabla^2C^{i}{}_{j} 
+C_{jl,}{}^{il}-\H\left(B^{i}{}_{,j}+B_{j,}{}^{i}\right)\nn \\
& \scriptstyle+\delta^{i}{}_{j}\left\{\left(\H^2-\frac{2a''}{a}\right)\left(1-2\phi\right)
+2\H\left(B^{k}{}_{,k}-C^{k'}{}_{k}+\phi^{'}\right) +B^{k'}{}_{k}-C^{kl}{}_{,kl}-C^{k''}{}_{k}
+\nabla^2\left(\phi+C^{l}{}_{l}\right)\right\} \nn\\
&\scriptstyle +B^{k}\Big[C_{jk,}{}^{i'}+C^{i'}{}_{k,j}-2C^{i'}{}_{j,k}+2\H(C_{jk,}{}^{i}+C^{i}{}_{k,j}-C^{i}{}_{j,k})
+\frac{1}{2}\left(B_{j,}{}^{i}{}_{k}+B^{i}{}_{,jk}-2B_{k,}{}^{i}{}_{j}\right)\Big]\nn\\
&\scriptstyle
+(C^{k}{}_{k}^{'}-\phi^{'}-B^{k}{}_{,k})(C^{i}{}_{j}^{'}
-\frac{1}{2}\left(B^{i}{}_{,j}+B_{j,}{}^{i}\right))+C^{ik'}\left(B_{j,k}-2C_{kj}^{'}\right)+C_{kj}^{'}B^{i}{}_{,}{}^{k} +\phi_{,}{}^{i}\phi_{,j}   \nn \\
& \scriptstyle+(B^{k'}-2C^{kl}{}_{,l}+C^{l}{}_{l,}{}^{k}+\phi_{,}{}^{k})(C_{jk,}{}^{i}+C^{i}{}_{k,j}-C^{i}{}_{j,k})
+ \frac{1}{2}B^{i}(B_{k,j}{}^{k}-\nabla^2B_{j}+4\H\phi_{,j}-2C^{k'}{}_{k,j}+2C^{'}_{kj,}{}^{k})
\nn \\
& \scriptstyle+2C^{ik}\Big[\frac{1}{2}\left(B_{j,k}^{'}+B_{k,j}^{'}\right)-C_{kj}^{''}+\phi_{,jk} -C_{kl,j}{}^{l}-C_{jl,k}{}^{l}
+\nabla^2C_{kj}+C^{l}{}_{l,jk}+\H\left(B_{j,k}+B_{k,j}-2C_{kj}^{'}\right)\Big] \nn \\
& \scriptstyle
-\frac{1}{2}\left(B_{k,}{}^{i}B^{k}{}_{,j}+B_{j,}{}^{k}B^{i}{}_{,k}\right) +\phi\Big[(B_{j,}{}^{i'}+B^{i}{}_{,j}^{'}+2\phi_{,}{}^{i}{}_{j}+2\H(B_{j,}{}^{i}+B^{i}{}_{,j})-2C^{i}{}_{j}^{''}
 -4\H C^{i}{}_{j}^{'}\Big]\nn \\
& \scriptstyle+2\left(C^{i}{}_{k,l}C^{k}{}_{j,}{}^{l}-C^{l}{}_{j,}{}^{k}C^{i}{}_{k,l}+C^{kl}{}_{,j}C_{kl,}{}^{i}\right)
+2C^{kl}\Big[C_{kl,j}{}^{i}-C_{jl,}{}^{i}{}_{k}-C^{i}{}_{l,jk}+C^{i}{}_{j,kl}\Big] +\mathbb{G}_o^i{}_j\nn \\
%%%%%
& \scriptstyle+\delta^{i}{}_{j}\left\{\left(\H^2-\frac{2a''}{a}\right)(4\phi^2-B_{k}B^{k})
 +2\phi\Big[C^{k''}{}_{k}-B^{k'}{}_{k}-\nabla^2\phi+2\H(C^{k'}{}_{k}-2\phi^{'}-B^{k}{}_{,k})\Big] \right. \nn \\
&\scriptstyle
\quad +B^{k}\Big[2C^{l'}{}_{l,k}-2C_{kl,}^{'}{}^{l}+\nabla^2B_{k}-B_{l,k}{}^{l}+2\H(B_{k}^{'}-\phi_{,k}-2C^{l}{}_{k,l}
+C^{l}{}_{l,k})\Big]+C^{kl'}\left(\frac{3}{2}C_{kl}^{'}-B_{l,k}\right) \nn \\
& \scriptstyle
\quad +2C^{kl}\Big[C_{kl}^{''}-\nabla^2C_{kl}+2\H C_{kl}^{'}+2C_{lm,k}{}^{m}-C^{m}{}_{m,kl}
-2\H B_{l,k}-B_{l,k}^{'}-\phi_{,kl}\Big] +2B^{k'}(C^{l}{}_{l,k}-C_{kl,}{}^{l})\nn \\
& \scriptstyle\quad +C^{k'}{}_{k}\left(B^{l}{}_{,l}-\frac{1}{2}C^{l'}{}_{l}\right)
+2C^{kl}{}_{,k}C_{lm,}{}^{m} +C_{lm,k}\left(C^{km}{}_{,}{}^{l}-\frac{3}{2}C^{lm}{}_{,}{}^{k}\right)
-C^{l}{}_{l,k}\left(2C^{k}{}_{m,}{}^{m}-\frac{1}{2}C^{m}{}_{m,}{}^{k}\right)\nn \\
&\scriptstyle \quad 
+\phi^{'}\left(C^{k'}{}_{k}-B^{k}{}_{,k}\right) \left. -\frac{1}{4}\left(2B^{k}{}_{,k}B_{l,}{}^{l}-B_{l,k}B^{k}{}_{,}{}^{l}
-3B^{l}{}_{,k}B_{l,}{}^{k}\right) + \phi_{,k}\left(C^l{}_{l,}{}^k-2C^{lk}{}_{,l}-\phi_{,}{}^k\right) +\mathbb{G}_d^i{}_j \right\} \,,
\end{align}
where $\mathbb{G}^0{}_0, \mathbb{G}^0{}_i, \mathbb{G}^i{}_j$ are the
third order corrections (the latter split into a diagonal part
$\mathbb{G}^i_o{}_j$, and an off diagonal part $\mathbb{G}^i_d{}_j$)
which we give in the appendix as Eqs. (\ref{eq:300}), (\ref{eq:30i}),
(\ref{eq:3ijo}) and (\ref{eq:3ijd}), respectively. Note that, in
calculating the third order components given above, we have implicitly
obtained the full second order Einstein tensor components for fully
general perturbations (i.e.~including all scalar, vector and tensor
perturbations).

%%%%%%%%%%%%%%%%%%%%
\section{Discussion}
\label{sec:dis}
%%%%%%%%%%%%%%%%%%%%

In this chapter we have developed the essential tools for cosmological
perturbation theory at third order. Starting with the definition of
the active gauge transformation we have extended the work presented in
 Section~\ref{sec:gaugetrans} to third order, and derived gauge invariant
variables, namely the curvature perturbation on uniform density
hypersurfaces, $\zeta_3$, and the density perturbation on uniform
curvature hypersurfaces. We also relate the curvature perturbation
$\zeta_3$, obtained using the spatial metric split of
Ref.~\cite{MW2008} to that introduced by Salopek and Bond
\cite{Salopek:1990jq}, which is also popular at higher order.

We have then calculated the energy and momentum conservation equations
for a general perfect fluid at third order, including all scalar,
vector and tensor perturbations. The Klein-Gordon equation for a
canonical scalar field minimally coupled to gravity is also
presented. We highlight the coupling in these conservation equations
between scalar and tensor perturbations which only occurs at third
order and above. 
Finally, we have presented the Einstein tensor components to third
order. No large scale approximation is employed for the tensor
components or the conservation equations. All equations are given
without specifying a particular gauge, and can therefore immediately
be rewritten in whatever choice of gauge is desired. However, as
examples to illustrate possible gauge choices, we give the energy
conservation equation on large scales (and only allowing for scalar
perturbations) in the flat and the uniform density gauge. This gives
an evolution equation for the curvature perturbation $\zeta_3$,
\eq{zeta3_cons}. As might be expected from fully non-linear
calculations \cite{Lyth:2004gb} and second order perturbative calculations
\cite{Malik2004}, the curvature perturbation is also conserved at third
order on large scales in the adiabatic case. It is worth noting that 
higher order perturbation theory, as discussed in this chapter,
has the advantage of being valid on all scales whereas fully non-linear
methods, such as separate universe approaches are gradient expansions (in powers of $k/aH$),
and so are only valid on superhorizon scales.

Another application of our third order variables and equations, in
particular the Klein-Gordon equation (\ref{eq:KG}), is the calculation of
the trispectrum by means of the field equations. Whereas calculations
of the trispectrum so far derive the trispectrum from the fourth order
action, it should also be possible to use the third order field
equations instead. The equivalence of the two approaches for
calculating the bispectrum, using the third order action or the second
order field equations, has been shown in Ref.~\cite{Seery:2008qj}. 
Having included tensor as well as scalar perturbations it will be in
particular interesting to see and be an important consistency check
for the theory whether we arrive at the same result as
Ref.~\cite{Seery:2008ax}.

A final advantage of extending perturbation theory to third order is
that, in doing so, one obtains a deeper insight into the second order
theory. Also second order perturbation theory, despite remaining
challenging, becomes less daunting having explored some of the third
order theory.

%% file: conclusions.tex
% % % % % % % % % % % % % % % % % % % % % % % % % % % % 
% conclusions.tex 
% Sample chapter layout
% % % % % % % % % % % % % % % % % % % % % % % % % % % % 
% Redefine CVSRevision for this section. 
% If you don't want to use CVS tags comment out this line
%\renewcommand{\CVSrevision}{\version$Id: chapter.tex,v 1.3 2009/12/17 18:16:48 ith Exp $}

% % % % % % % % % % % % % % % % % % % % % % % % % % % % % % % % 
% =========================================================== %
% % % % % % % % % % % % % % % % % % % % % % % % % % % % % % % % 
\chapter{Discussion and Conclusions}
\label{ch:conc}
% % % % % % % % % % % % % % % % % % % % % % % % % % % % % % % % 
% =========================================================== %
% % % % % % % % % % % % % % % % % % % % % % % % % % % % % % % % 

%blahdy lahdy blah
\section{Summary}

The main focus of this thesis has been the study of cosmological perturbations beyond linear order.
 In Section~\ref{sec:standardcosmology} we introduced the standard cosmological model, giving the background
 evolution and constraint equations and briefly discussing inflationary cosmology.
In Chapter~\ref{ch:perturbations} we introduced the theory of  cosmological perturbations
up to second order, presenting the perturbed metric tensor and energy momentum tensor for both a 
perfect fluid, including all types of scalar, vector and tensor perturbations,
 and a scalar field. We then considered the behaviour of the perturbations under a gauge
transformation in the active approach,
 using this behaviour to define gauges and construct gauge invariant variables. Next, we discussed
the thermodynamics of a perfect fluid and defined the non-adiabatic pressure perturbation,  
closing the chapter by considering how non-adiabatic pressure perturbations can arise naturally
in multiple-component systems.

In Chapter~\ref{ch:dynamics} we continued the discussion of the foundations of cosmological perturbation theory
by presenting the dynamic and constraint equations, from energy momentum conservation and the Einstein
field equations, up to second order in the perturbations.
Starting with the linear theory, we gave the governing equations for scalar,
vector and tensor perturbations of a perfect fluid in a gauge dependent form, i.e. without fixing a 
gauge. We then presented the gauge invariant form of the equations for three different gauges:
the uniform density, the uniform curvature, and the longitudinal gauges, solving the equations for the latter
two in the case of scalar perturbations. We then presented the Klein-Gordon equation for a scalar field, to linear
order for both a canonical and non-canonical action,
and highlighted the important difference between the adiabatic sound speed and the speed with
which perturbations travel in a scalar field system (the phase speed). Finally, we investigated the perturbations of a 
system containing both dark energy and dark matter. 

Having laid the foundations of the linear order theory
we then discussed the second order theory, presenting the governing equations for a perfect fluid derived,
as at first order, from  energy momentum conservation and the Einstein field equations. We then presented all
equations in the uniform curvature gauge, which we then used in Chapter~\ref{ch:vorticity}, and for scalars
only in the Poisson gauge, in order to connect with the literature. \\

In Chapter~\ref{ch:vorticity} we used the tools developed in the previous chapters to investigate non-linear 
vector perturbations in the early universe. Using
the qualitative difference between the linear theory and the
higher order theory we showed that, at second order in perturbation theory, vorticity is sourced by a  
coupling quadratic in linear energy density and entropy perturbations, extending Crocco's theorem from a classical
framework to an expanding,
cosmological background. In order to show this, we first defined the vorticity tensor in General Relativity
and calculated the vorticity tensor at first and second order in cosmological perturbation
theory using the metric tensor and fluid four velocity presented earlier. We then computed the evolution of
the vorticity tensor by taking the time derivative, and using the governing equations from 
Chapter~\ref{ch:dynamics} to simplify the expressions and replace the metric perturbation variables. 
We found that at linear order the vorticity is not sourced in the absence of anisotropic stress,
 in agreement with the previous results
known in the literature. However, at second order we obtained the novel result that there
exists a non-zero source term for a fluid with a general equation of state (depending on both the 
energy density and entropy) which is quadratic in linear 
energy density and non-adiabatic pressure perturbations.

 Having derived this qualitative  result we then gave the first quantitative 
solution, estimating the magnitude and scale dependence of the induced vorticity using simple
input power spectra: the energy density derived in Chapter~\ref{ch:dynamics} approximated for small $k\eta$, and an
ansatz for the non-adiabatic pressure perturbation. We found that the resulting spectrum has a surprisingly
large magnitude, given that it is a second order effect, and a dependence on the wavenumber to the power
of at least seven, given our assumptions. Thus, this spectrum is hugely amplified on small scales, rendering the vorticity not
only possibly observable, but also important for the general understanding of the physical
processes taking place in the early universe. \\

In Chapter~\ref{ch:third} we extended the formalism of cosmological perturbation theory
from the second order theory to third order,
starting with the gauge transformation rules and defining gauge invariant variables. Then, considering
perfect fluids and scalar, vector and tensor perturbations we presented the energy and momentum conservation
equations and the Klein-Gordon equation for a scalar field without fixing a gauge. Finally, we gave the components 
of the Einstein tensor at third order also in a gauge dependent form.

\section{Future Directions}

The work presented in this thesis can naturally be extended in several directions. One clear extension is to 
study the vorticity generation in specific models, moving beyond the simple ansatz for the non-adiabatic
pressure perturbation used in Chapter~\ref{ch:vorticity}.
As discussed in Section~\ref{sec:entropyinfl}, non-adiabatic pressure perturbations
naturally arise in any system consisting of more than one component, such as a multiple fluid, or multiple scalar fields 
model. Even in the case of
zero intrinsic non-adiabatic pressure perturbation, there exists a relative non-adiabatic pressure perturbation 
between the different
components of the system which is proportional to the relative entropy perturbation, i.e.
\be 
\delta P_{\rm nad}\propto{\mathcal S}_{IJ}\equiv
3\H\Bigg(\frac{\delta\rho_J}{\rho_{0J}'}-\frac{\delta\rho_I}{\rho_{0I}'}\Bigg)\,.\nn
\ee
This can then be written in terms of field variables using the definitions in Chapter~\ref{ch:perturbations} in order
to obtain an expression for the relative entropy perturbation in multi-field inflation models.

Another way in which this work can be extended is to exploit the potential for this mechanism to generate magnetic fields.  
As mentioned in Chapter~\ref{ch:vorticity}, magnetic fields and vorticity are intimately related. This relationship has
been studied in some detail in classical fluid mechanics and in astrophysical situations (see, e.g., Ref.~\cite{Widrow:2002ud} 
for a review), though there is
still much work to be done on incorporating magnetic fields into cosmological perturbation theory. Since the
energy density is related to the magnetic field, $b^i$, through the expression 
\be 
\label{eq:mag}
\rho_{b} \sim b^2\,,
\ee
when  considering linear perturbations of the energy density one often considers perturbations of the
 magnetic field to `half order'. That is, one assumes that the magnetic energy density as shown above to be 
 of order $b^2$, is formally the same order as the scalar density perturbation.
ensuring that the perturbed version of Eq.~(\ref{eq:mag}) holds at linear order (see, 
e.g., Ref.~\cite{Brown:2006wv}). However, one does not need to use this technique and, in fact, when considering
higher order perturbations it is not immediately clear how this will work. Instead, one can consider and develop
cosmological perturbation theory to consistently include magnetic fields at integer order. Having done that, it will be possible
to obtain estimates of the primordial magnetic field produced by the vorticity generated using cosmological perturbations
as shown in this thesis by using the simple ansatz considered in Chapter~\ref{ch:vorticity} \cite{inprep}.

A further extension to this work will be to consider the primordial magnetic field generated from the relative non-adiabatic
pressure perturbation in a system with multiple components, such as a hybrid inflation model.

\section{Outlook}

Cosmological perturbation theory has matured over the last few decades and has been incredibly successful
in making predictions that agree with observations. However, with the data sets available to us continually
growing in their size and quality it is now a realistic aim to use perturbation theory even beyond
linear order to make predictions which are observationally testable. 

The main observable with which we can constrain our cosmological models is the CMB, of which we have, to date,
collected much information, predominantly from the  successful {\sc Wmap} experiment.
 The {\sc Planck} satellite \cite{Planck} will greatly improve the temperature measurements of the CMB and
together with the proposed {\sc CMBPol} satellite \cite{CMBPol}, will measure the polarisation of the background
radiation. Since the CMB is not affected by the astrophysics of the late universe,  usually one prefers the use
of CMB data over other techniques for the study of higher order observables.

%\begin{figure}
%\begin{center}
%\includegraphics[width=\textwidth]{web.jpg}
%\end{center}
%\caption{Cosmic Web}
%\end{figure}

However, with the recent technological advances, Large Scale Structure (LSS) surveys, such as the Sloan Digital Sky Survey (SDSS)
\cite{SDSS} and the proposed 21cm anisotropy maps are attracting more attention as a way to probe the evolution
of the universe at different epochs of its history. The 21cm signal, generated by neutral Hydrogen left over after the Big 
Bang, can probe the era after decoupling but before galaxy formation, i.e.~between redshift 200 and 30, while 
LSS surveys probe out to around redshift 1. The 21cm anisotropy maps contain much more
data than the CMB \cite{Loeb:2003ya}, though it should be noted that
 it is still not clear whether the foregrounds can be removed with enough accuracy to enable reliable results. 
 
 This is but one area where calculations at higher order  can be tested against observations. 
 Thus, the future study of cosmological perturbation theory  will greatly increase our understanding,
 serving to broaden and deepen our knowledge of the universe in which we live.

%% file: secondij.tex
% % % % % % % % % % % % % % % % % % % % % % % % % % % % 
% chapter.tex - 
% Sample chapter layout
% % % % % % % % % % % % % % % % % % % % % % % % % % % % 

% % % % % % % % % % % % % % % % % % % % % % % % % % % % % % % % 
% =========================================================== %
% % % % % % % % % % % % % % % % % % % % % % % % % % % % % % % % 
\chapter{Second Order {$i-j$} Einstein Equation}
\label{ch:AppB}
% % % % % % % % % % % % % % % % % % % % % % % % % % % % % % % % 
% =========================================================== %
% % % % % % % % % % % % % % % % % % % % % % % % % % % % % % % % 

In this appendix we give the second order 
$(i-j)$ component of the Einstein Equation omitted from Section \ref{sec:eqnsecond}:
\begin{align}
\label{eq:gij_second}
&\scriptstyle C_2^{i''}{}_{j}+2\H C_2^{i}{}_{j}^{'}-\frac{1}{2}(B_2^{i'}{}_{,j}+B_{2j,}{}^{i'})
-C_2^{l}{}_{l,j}{}^{i}+C_2^{i}{}_{l,j}{}^{l}-\nabla^2C_2^{i}{}_{j} 
+C_{2jl,}{}^{il}-\phi_{2,}{}^{i}{}_{j}-\H\left(B_2^{i}{}_{,j}+B_{2j,}{}^{i}\right)\nn\\
&\scriptstyle 
 +\delta^{i}{}_{j}\Big\{2\left(\frac{2a''}{a}-\H^2\right)\phi_2
+2\H\left(B_2^{k}{}_{,k}-C_2^{k'}{}_{k}+\phi_2^{'}\right) 
+B_2^{k'}{}_{k}-C_2^{kl}{}_{,kl}-C_2^{k''}{}_{k}
+\nabla^2\left(\phi_2+C_2^{l}{}_{l}\right)\Big\} \nn\\
&\scriptstyle +B_1^{k}\Big[C_{1jk,}{}^{i'}+C_1^{i'}{}_{k,j}-2C_1^{i'}{}_{j,k}+2\H(C_{1jk,}{}^{i}+C_1^{i}{}_{k,j}-C_1^{i}{}_{j,k})
+\frac{1}{2}\left(B_{1j,}{}^{i}{}_{k}+B_1^{i}{}_{,jk}-2B_{1k,}{}^{i}{}_{j}\right)\Big]\nn\\
&\scriptstyle
+(C_1^{k}{}_{k}^{'}-\phi_1^{'}-B_1^{k}{}_{,k})(C_1^{i}{}_{j}^{'}
-\frac{1}{2}\left(B_1^{i}{}_{,j}+B_{1j,}{}^{i}\right))+C_1^{ik'}\left(B_{1j,k}-2C_{1kj}^{'}\right)+C_{1kj}^{'}B_1^{i}{}_{,}{}^{k} +\phi_{1,}{}^{i}\phi_{1,j}   \nn \\
& \scriptstyle +(B_1^{k'}-2C_1^{kl}{}_{,l}+C_1^{l}{}_{l,}{}^{k}+\phi_{1,}{}^{k})(C_{1jk,}{}^{i}+C_1^{i}{}_{k,j}-C_1^{i}{}_{j,k})
+ \frac{1}{2}B_1^{i}(B_{1k,j}{}^{k}-\nabla^2B_{1j}+4\H\phi_{1,j}-2C_1^{k'}{}_{k,j}+2C_1^{'}{}_{kj,}{}^{k})
\nn \\
&\scriptstyle +2C_1^{ik}\Big[\frac{1}{2}\left(B_{1j,k}^{'}+B_{1k,j}^{'}\right)-C_{1kj}^{''}+\phi_{1,jk} -C_{1kl,j}{}^{l}-C_{1jl,k}{}^{l}
+\nabla^2C_{1kj}+C_1^{l}{}_{l,jk}+\H\left(B_{1j,k}+B_{1k,j}-2C_{1kj}^{'}\right)\Big] \nn \\
&\scriptstyle 
-\frac{1}{2}\left(B_{1k,}{}^{i}B_1^{k}{}_{,j}+B_{1j,}{}^{k}B_1^{i}{}_{,k}\right) +\iphi\Big[(B_{1j,}{}^{i'}+B_1^{i}{}_{,j}^{'}
+2\phi_{1,}{}^{i}{}_{j}+2\H(B_{1j,}{}^{i}+B_1^{i}{}_{,j})-2C_1^{i}{}_{j}^{''}
 -4\H C_1^{i}{}_{j}^{'}\Big]\nn \\
&\scriptstyle +2\left(C_1^{i}{}_{k,l}C_1^{k}{}_{j,}{}^{l}-C_1^{l}{}_{j,}{}^{k}C_1^{i}{}_{k,l}+C_1^{kl}{}_{,j}C_{1kl,}{}^{i}\right)
+2C_1^{kl}\Big[C_{1kl,j}{}^{i}-C_{1jl,}{}^{i}{}_{k}-C_1^{i}{}_{l,jk}+C_1^{i}{}_{j,kl}\Big] \nn \\
%%%%%
&\scriptstyle +\delta^{i}{}_{j}\Big\{\left(\H^2-\frac{2a''}{a}\right)(4\phi_1^2-B_{1k}B_1^{k})
 +2\iphi\Big[C_1^{k''}{}_{k}-B_1^{k'}{}_{k}-\nabla^2\iphi+2\H(C_1^{k'}{}_{k}-2\iphi^{'}-B_1^{k}{}_{,k})\Big]  \nn \\
&\scriptstyle\quad +B_1^{k}\Big[2C_1^{l'}{}_{l,k}-2C_{1kl,}^{'}{}^{l}+\nabla^2B_{1k}-B_{1l,k}{}^{l}+2\H(B_{1k}^{'}-\iphi_{,k}
-2C_1^{l}{}_{k,l}
+C_1^{l}{}_{l,k})\Big]+C_1^{kl'}\left(\frac{3}{2}C_{1kl}^{'}-B_{1l,k}\right) \nn \\
&\scriptstyle \quad +2C_1^{kl}\Big[C_{1kl}^{''}-\nabla^2C_{1kl}+2\H C_{1kl}^{'}+2C_{1lm,k}{}^{m}-C_1^{m}{}_{m,kl}
-2\H B_{1l,k}-B_{1l,k}^{'}-\phi_{1,kl}\Big] +2B_1^{k'}(C_1^{l}{}_{l,k}-C_{1kl,}{}^{l})\nn \\
&\scriptstyle \quad +C_1{}^{k'}{}_{k}\left(B_1^{l}{}_{,l}-\frac{1}{2}C_1^{l'}{}_{l}\right)
+2C_1^{kl}{}_{,k}C_{1lm,}{}^{m} +C_{1lm,k}\left(C_1^{km}{}_{,}{}^{l}-\frac{3}{2}C_1^{lm}{}_{,}{}^{k}\right)
-C_1^{l}{}_{l,k}\left(2C_1^{k}{}_{m,}{}^{m}-\frac{1}{2}C_1^{m}{}_{m,}{}^{k}\right)\nn \\
&\scriptstyle \quad 
+\iphi^{'}\left(C_1{}^{k'}{}_{k}-B_1^{k}{}_{,k}\right) -\frac{1}{4}\left(2B_1^{k}{}_{,k}B_{1l,}{}^{l}-B_{1l,k}B_1^{k}{}_{,}{}^{l}
-3B_1^{l}{}_{,k}B_{1l,}{}^{k}\right) + \phi_{1,k}\left(C_1{}^l{}_{l,}{}^k-2C_1{}^{lk}{}_{,l}-\iphi_{,}{}^k\right)  \Big\} \nn\\
& \scriptstyle
=8\pi G a^2 \Big\{\delta P_2\delta^i{}_j+2(\rhob+\Pb)v_{1}{}^i(v_{1j}+B_{1j})\Big\}\,.
\end{align}

%% file: thirdeins.tex
% % % % % % % % % % % % % % % % % % % % % % % % % % % % 
% chapter.tex - Ian Huston
% Sample chapter layout
% % % % % % % % % % % % % % % % % % % % % % % % % % % % 
% Redefine CVSRevision for this section. 
% If you don't want to use CVS tags comment out this line
%\renewcommand{\CVSrevision}{\version$Id: chapter.tex,v 1.3 2009/12/17 18:16:48 ith Exp $}

% % % % % % % % % % % % % % % % % % % % % % % % % % % % % % % % 
% =========================================================== %
% % % % % % % % % % % % % % % % % % % % % % % % % % % % % % % % 
\chapter{Third Order Einstein Tensor}
\label{ch:appA}
% % % % % % % % % % % % % % % % % % % % % % % % % % % % % % % % 
% =========================================================== %
% % % % % % % % % % % % % % % % % % % % % % % % % % % % % % % % 

Here, we give the third order corrections to the components of the Einstein tensor. We do not split up perturbations
order by order.

\begin{align}
\label{eq:300}
\scriptstyle\mathbb{G}^{0}{}_{0}&\scriptstyle =
2C^{ij}\Big[2C_{ik}(C^l{}_{l,j}{}^k-2C^{kl}{}_{,jl}+\nabla^2C_j{}^k-2\H C'_j{}^k)
+(2C_{jk,i}-C_{ij,k})(C^l{}_{l,}{}^k-2C^k{}_{l,}{}^l)+C_{il,k}(3C_j{}^l{}_,{}^k-C_j{}^k{}_{,l}) \nn \\
& \scriptstyle+C^k{}_{k,i}(2C^j{}_{l,}{}^l-\frac{1}{2}C^l{}_{l,j})+B^i\Big\{C'_{jk,}{}^k
-C^k{}_{k,j}'+\frac{1}{2}(B_{k,j}{}^k-\nabla^2B_j)+\H(4C_{jk,}{}^k+2\phi_{,j}-3\H B_j  -2C^k{}_{k,j})\Big\}
\nn \\
&\scriptstyle
+C_{ik}'(B^k{}_{,j}+B_{j,}{}^k) 
-2C_{ik,}{}^kC_{jl,}{}^l-(B^k{}_{,k}+\phi)(C_{ij}'-B_{j,i})-\frac{1}{4}B_{k,i}(B^k{}_{,j}+2B_{j,}{}^k)
-\frac{1}{4}B_{i,k}B_{j,}{}^k-C^k{}_{k}'B_{j,i}\nn \\
& \scriptstyle
+8\H C_{ik}B^k{}_{,j}\Big] + 2C^{kj}B^i\Big[C_{ik,j}'-C_{jk,i}'+\frac{1}{2}(B_{k,ij}-B_{i,jk})+2\H(2C_{ik,j}-C_{kj,i})\Big]
+C^{ij}{}'\Big[2B^k(C_{ki,j}-C_{ji,k}) \nn \\
& \scriptstyle+B_i(2C_{jk,}{}^k-C^k{}_{k,j})+B_i\phi_{,j}+(2B_{j,i}-C_{ij}')\phi\Big]+8\H\phi^2(B^i{}_{,i}-C^i{}_{i}')
+(B_{i,j}-B_{j,i})(\frac{1}{2}B^iC^k{}_{k,}{}^j-B^iC^{jk}{}_{,k}) \nn \\
& \scriptstyle-\frac{1}{2}(B_{i,j}+B_{j,i})(\phi B^{j}{}_{,}{}^{i}+B^i\phi_{,}{}^j)-2\H B^iB^jB_{j,i}
+B^j{}_{,j}\Big[B^i(2C_{ik,}{}^k-C^k{}_{k,i}+\phi_{,i}-2\H B_i)+B^i{}_{,i}\phi-2C^i{}_{i}'\phi\Big] \nn \\
& \scriptstyle+B_{k,j}B^i(C^{kj}{}_{,i}-2C_i{}^k{}_,{}^j) +2\H B^i\Big[C^j{}_{j}'B_i+2\phi(2C_{ij,}{}^j-C^j{}_{j,i}+2\phi_{,i}
-3\H B_i)\Big] +C^i{}_{i}'C^j{}_{j}'\phi\nn \\
& \scriptstyle+B^iB^j(2C_{jk,i}{}^k-C^k{}_{k,ij}-\nabla^2C_{ij}) +B^iC^j{}_{j}'(C^k{}_{k,i}-2C_{ik,}{}^k-\phi_{,i})+B^i\phi(B^j{}_{,ij}+2C_{ij,}'{}^j-2C^j{}_{j,i}'-\nabla^2B_i) \,,
\end{align}
\begin{align}
\label{eq:30i}
\scriptstyle\mathbb{G}^{0}{}_{i} &
\scriptstyle= 
 C_{kj}\left[2C^{jl}\left(C^k{}_{l,i}{}'-C_{il,}{}^{k'}\right)
+C_i{}^j\left(C^l{}_{l,}{}^k-2C^{kl}{}_{,l}-2\phi_{,}{}^k\right)+C'_{il}\left(C^{jk}{}_,{}^l-2C^{kl}{}_,{}^j\right)
+2C^{jl}{}'C^k{}_{l,i}\right] \nn \\
& \scriptstyle+C^{kj}\Big[(B_{j,i}-B_{i,j})(C^{l}{}_{l,k}-C_{kl,}{}^l)+(B_{l,i}-B_{i,l})(C_{jk,}{}^l-2C_k{}^l{}_{,j})
+2(B_{j,l}-B_{l,j})(C_{ik,}^l-C_i{}^l{}_{,k})+(B_{j,i}+B_{i,j})\phi_{,k}  \nn \\
& \scriptstyle+ 2(C_{jk}'-B_{k,j})\phi_{,i}+2\phi(2C_{jk,i}'-2C_{ik,j}'+B_{i,kj}-B_{k,ij}) \Big]
+2B_j\Big[C^{kj}(C^l{}_{l,ik}-C_{kl,i}{}^l-C_{il,k}{}^l+\nabla^2C_{ik}) \nn \\
& \scriptstyle+C^{kl}(C_{kl,i}{}^j-C^j{}_{l,ik}-C_{il,k}{}^j+C_{i}{}^j{}_{,kl})
+(C^l{}_{l,}{}^k-C^{kl}{}_{,l})(C^j{}_{k,i}+C_{ik,}{}^j-C^j{}_{i,k})
+C_{il,k}(C^{jl}{}_{,}{}^k-C^{jk}{}_{,}{}^l) \nn \\
&\scriptstyle +\frac{1}{2}C_{kl,i}C^{kl}{}_{,}{}^{j}\Big]+2C_{ij}'\phi(C^k{}_{k,}{}^j-2C^{kj}{}_{,k})
+B^j\Big[B_{j}(C_{ik,}'{}^k-C^k{}_{k,i}'+\frac{1}{2}(B_{k,i}{}^k-\nabla^2B_i))
+C_{ik}'B_{j,}{}^k \nn \\
&\scriptstyle +B_{j,i}(B^k{}_{,k}-C^k{}_{k}') +\frac{1}{2}B_{j,k}(2C_i'{}^k-B^k{}_{,i}B_{i,}{}^k)\Big]
+2\phi\Big[(B_{i,j}-B_{j,i})(C^{kj}{}_{,k}-\frac{1}{2}C^k{}_{k,}{}^j)+\phi_,{}^j(B_{i,j}+B_{j,i}) \nn \\
&\scriptstyle +C_{kj}'C^{kj}{}_{,i}+2C^j{}_j'\phi_{,i}-2C_{ij}'\phi_,{}^j-B_{k,j}(C_i{}^j{}_,{}^k-C_i{}^k{}_,{}^j)
-2B_{j,}{}^j\phi_{,i}\Big] +2C_{kj}C^{jl}(B_{i,l}{}^k-B_{l,i}{}^k) \nn \\
&\scriptstyle +B^j\Big[\phi_{,i}(C^k{}_{k,j}-2C_{jk,}{}^k+4\H B_j)
+2\phi(C^k{}_{k,ij}-C_{jk,}{}^k{}_i-C_{ik,j}{}^k+\nabla^2C_{ij}+4\H B_{j,i}) +2\H(B^kC_{jk,i}+2C_{kj}B^k{}_{,i}) \nn \\
&\scriptstyle +\phi_,{}^k(C_{ij,k}-C_{ik,j}+C_{jk,i})\Big] +2\phi^2(2C^j{}_{j,i}'-2C_{ij,}'{}^j-B_{j,i}{}^j+\nabla^2B_i-12\H\phi_{,i})\,,
\end{align}

\begin{align}
\label{eq:3ijo}
\scriptstyle
\mathbb{G}^{i}_{\rm{o}}{}_{j} & \scriptstyle= 
2C^{ik}C_{kl}\Big[2C^l{}_j''-B^l{}_{,j}'-B_{j,}{}^l{}'-2C^m{}_{m,j}{}^l+2C^l{}_{m,j}{}^m
+2C_{jm,}{}^{lm}-2\nabla^2C^l{}_j-2\phi_{,j}{}^l\Big]\nn\\
& \scriptstyle+4C^{ik}C^{lm}(C_{km,jl}-C_{lm,kj}+C_{jm,kl}-C_{jk,lm})
+4C^{km}C_{kl}(C_{jm,}{}^{il}-C^l{}_{m,}{}^i{}_j+C^i{}_{m,j}{}^l-C^i{}_{j,}{}^l{}_m)\nn\\
& \scriptstyle+C^{ik}\Big[2(2C^{lm}{}_{,m}-C^m{}_{m,}{}^l-\phi_,{}^l-B^l{}')(C_{kl,j}+C_{jl,k}-C_{jk,l})
+4C_{jm,l}(C_k{}^l{}_,{}^m-C_k{}^m{}_,{}^l)-2C_{lm,j}C^{lm}{}_{,k}\nn\\
&\scriptstyle-2C_{jk}'(C^l{}_l'-B^l{}_{,l}-\phi')+2C_{jl}'(2C^l{}_k'-B_{k,}{}^l)+C^l{}_l'(B_{k,j}+B_{j,k})
-2C_{kl}'B_{j,}{}^l-(\phi'+B^l{}_{,l})(B_{k,j}+B_{j,k})\nn\\
&\scriptstyle+B_{l,j}B^l{}_{,k}+B_{j,l}B_{k,}{}^l-2\phi_{,k}\phi_{,j}-2\phi(B_{k,j}'+B_{j,k}'+\phi_{,jk}
+4\H C_{kl}(2C^l{}_j'-B^l{}_{,j}-B_{j,}{}^l)-4\H B_k\phi_{,j}\Big]\nn\\
&\scriptstyle+C^{kl}\Big[2(C^m{}_{m,l}-2C_{lm,}{}^m)(C^i{}_{j,k}-C^i{}_{k,j}-C_{jk,}{}^i)
+2(2C^m{}_{l,k}-C_{kl,}{}^m)(C_{jm,}{}^i+C^i{}_{m,j}-C^i{}_{j,m})-4C_{km,}{}^iC^m{}_{l,j}\nn\\
&\scriptstyle +4(C_{jl,m}-C_{jm,l})(C^{im}_{,k}-C^i{}_{k,}{}^m)-2C^i{}_j'C_{kl}'+4C^i{}_k'C_{jl}'
+(B^i{}_{,j}+B_{j,}{}^i)(C_{kl}'-B_{l,k})-2\phi_,{}^l(C_{jk,}{}^i+C^i{}_{k,j}-C^i{}_{j,k})\nn\\
&\scriptstyle +2C^i{}_j'B_{l,k}-2C^i{}_k'B_{j,l}-2C_{jk}'B^i{}_{,l}+B_{k,}{}^iB_{l,j}
-2B_k'(C_{jl,}{}^i+C^i{}_{l,j}-C^i{}_{j,l})\Big]
-B^kB_k\Big[C^i{}_j''-\frac{1}{2}(B^i{}_{,j}'+B_{j,}{}^i{}')\nn\\
&\scriptstyle -\phi_{,}{}^i{}_j-\H(B^i{}_{,j}+B_{j,}{}^i-2C^i{}_j')\Big]
+B^iB^k{}'\Big[C^l{}_{l,jk}-C^l{}_{k,jl}-C^l{}_{j,kl}+\nabla^2C_{jk}+2\H B_{k,j}\Big]
\nn\\
&\scriptstyle +B^kB^l\Big[C_{kl,}{}^i{}_j-C_{jl,}{}^i{}_k-C^i{}_{l,jk}+C^i{}_{j,kl}\Big]
+B^i\Big[C_{jk}'(C^l{}_{l,}{}^k-2C^{kl}{}_{,l}-\phi_,{}^k)+C_{kl}'C^{kl}{}_{,j}
+\frac{1}{2}\phi_{,}{}^k(B_{k,j}+B_{j,k})\nn\\
&\scriptstyle+\left(\frac{1}{2}C^l{}_{l,}{}^k-C^{kl}{}_{,l}\right)(B_{k,j}-B_{j,k})+B_{l,k}(C^l{}_{j,}{}^k-C^k{}_{j,}{}^l)
-\phi_{,j}(B^k{}_{,k}-C^k{}_{k}')\Big]+4\H\phi^2(2C^i{}_j'-B^i{}_{,j}-B_{j,}{}^i)\nn\\
&\scriptstyle+B^k\Big[C^{i}{}_k(2C^l{}_{l,j}'-2C_{jl,}{}^l{}'-B^l{}_{,jl}+\nabla^2B_j)
+C^{il}(4C_{jl,k}'-2C_{kl,j}'-C_{jk,l}'+2B_{k,jl}-B_{l,kj}-B_{j,kl})\nn\\
&\scriptstyle+C^l{}_k(4C^i{}_{j,l}'-2C^i{}_{l,j}'-C_{jl,}{}^i{}'+2B_{l,j}{}^i-B_{j,l}{}^i-B^i{}_{,jl})
+C^i{}_j'(2C_{kl,}{}^l-B_k'-C^l{}_{l,k}+\phi_{,k})+2C^{il}{}'(C_{jl,k}-C_{jk,l})\nn\\
&\scriptstyle+2C_{jl}'(C^{il}{}_{,k}-C^i{}_{k,}{}^l)+\frac{1}{2}(B_k'+C^l{}_{l,k}-2C_{kl,}{}^l)\left(B_{j,}{}^i+B^i{}_{,j}\right)
+(C^l{}_{j,}{}^i+C^{il}{}_{,j}-C^i{}_{j,}{}^l)(B_{k,l}-B_{l,k}-2C_{kl}'-4\H C_{kl})\nn\\
&\scriptstyle+(C_{jk,}{}^i+C^i{}_{k,j}-C^i{}_{j,k})(C^l{}_l'-\phi'-B^l{}_{,l})
+(B^i{}_,{}^l-4\H C^{il})(C_{jk,l}-C_{jl,k})+(B^l{}_,{}^i-4\H C^{il})C_{kl,j}+B^l{}_{,j}C_{kl,}{}^i\nn\\
&\scriptstyle-\frac{1}{2}\phi_{,k}(B^i{}_{,j}+B_{j,}{}^i)+B_{j,l}(C^i{}_{k,}{}^l-C^{il}{}_{,k})+B_{k,}{}^i\phi_{,j}+B_{k,j}\phi_,{}^i\Big]
+2\phi\Big[B^i\Big(C^k{}_{k,j}'-C_{jk,}{}^k{}'+\frac{1}{2}(\nabla^2B_j-B^k{}_{,kj})\Big)\nn\\
&\scriptstyle+B^k\Big(2C^i{}_{j,k}'-C^i{}_{k,j}'-C_{jk,}{}^i{}'+B_{k,j}{}^i-\frac{1}{2}(B_{j,k}{}^i+B^i{}_{,jk})\Big)
+(B^k{}'+\phi_{,}{}^k+4\H B^k)(C^i{}_{j,k}-C^i{}_{k,j}-C_{jk,}{}^i)\nn\\
&\scriptstyle+C^i{}_k'(2C^k{}_j'-B_{j,}{}^k)+2C^{ik}C_{jk}''-C_{jk}'B^i{}_,{}^k
+\frac{1}{2}(B^i{}_{,j}+B_{j,}{}^i)(C^k{}_k'-2\phi'-B^k{}_{,k})+C^i{}_j'(B^k{}_{,k}-C^k{}_k'+2\phi')\nn\\
&\scriptstyle+4\phi_{,j}(2\H B^i-\phi_{,}{}^i)-4\H C^{ik}(2C_{kj}'+B_{k,j}+B_{j,k})
+\frac{1}{2}\left(B_{k,}{}^iB^k{}_{,j}+B^i{}_{,k}B_{j,}{}^k\right)\Big]\,, 
\end{align}

\begin{align}
\label{eq:3ijd}
\scriptstyle\mathbb{G}^{i}_{\rm{d}}{}_{j} &\scriptstyle=
4C^{kl}C_{km}\Big[\nabla^2C_l{}^m-C^m{}_l''+C^n{}_{n,l}{}^m-2C^{mn}{}_{,nl}
+B^m{}_{,l}'+\phi_{,l}{}^m+2\H(B^m{}_{,l}-C^m{}_{l}')\Big]\nn\\
&\scriptstyle
+C^{kl}\Big[4C_{mn}(C^{mn}{}_{,kl}-2C_l{}^n{}_{,k}{}^m)+2C_{km}'(B^m{}_{,l}+B_{l,}{}^m-3C^m{}_l')
+2(C^n{}_{n,}{}^m-2C^{mn}{}_{,n})(2C_{lm,k}-C_{kl,m})\nn\\
& \scriptstyle+C^{kl}\Big[
+B^m{}'(4C_{ml,k}-C_{kl,m})+2B_k'(2C_{lm,}{}^m-C^m{}_{m,l})
+C^m{}_{m,k}(4C_{ln,}{}^n-C^n{}_{n,l})+C_{mn,k}(3C^{mn}{}_{,l}-4C^{n}{}_{l,}{}^m)\nn\\
&\scriptstyle+C_{kn,m}(6C_l{}^n{}_,{}^m-2C_l{}^m{}_,{}^n)+4C_{lm,k}\phi_,{}^m-2C^m{}_{m,k}\phi_{,l}
-2C_{kl,m}\phi_,{}^m+2\phi_{,k}\phi_{,l}+4C_{km,}{}^m(\phi_{,l}-C_{ln,}{}^n) \nn \\
&\scriptstyle +2(B_{l,k}-C_{kl}')(B^m{}_{,m}+\phi'+4\H\phi)+4\phi(\phi_{,kl}+B_{l,k}'-C_{kl}'')
+B_{m,k}(B_{l,}{}^m-\frac{3}{2}B^m{}_{,l})\nn\\
& \scriptstyle+2B_{k}\Big(2C_{lm,}{}^k-2C^m{}_{m,l}'+B^m{}_{,lm}-\nabla^2B_l-2\H(B_l'
+C^m{}_{m,l}-2C_{lm,}{}^m-\phi_{,l})\Big)-\frac{3}{2}B_{k,m}B_{l,}{}^m\Big]\nn\\
&\scriptstyle +C^k{}_k'\Big[2C^{ml}(C_{ml}'-B_{l,m})+\phi C^l{}_l'-8\H\phi^2-4\phi\phi'-2\phi B^l{}_{,l}
+B^l(B_l'+C^m{}_{m,l}-2C_{lm,}{}^m-\phi_{,l}+2\H B_l)\Big]\nn\\
&\scriptstyle +B^k\Big[B_k\Big(C^l{}_l''-B^l{}_{,l}'-\nabla^2\phi-2\H(2\phi'+B^l{}_{,l})\Big)
+B^l\Big(2C_{lm,k}{}^m-C^m{}_{m,kl}-\nabla^2C_{kl}-2\H(C_{kl}'+B_{l,k})\Big) \nn \\
& \scriptstyle-\left(\frac{2a''}{a}-\H^2\right)(2C_{kl}+4\phi)
+2C^{lm}\Big(2C_{km,l}'-2C_{lm,k}'+B_{m,lk}-B_{k,ml}-2\H(C_{lm,k}-2C_{km,l}\Big)\Big]\nn\\
&\scriptstyle -4\phi^2\Big[C^k{}_k''-B^k{}_{,k}'-\nabla^2\phi+2\H(\H\phi-3\phi'-B^k{}_{,k})\Big]
+C^{kl}{}'\phi\Big(2B_{l,k}-3C_{kl}'\Big)+\frac{1}{2}B_{l,k}\phi(B^k{}_,{}^l-3B^l{}_,{}^k)\nn\\
&\scriptstyle+B^k{}_{,k}\phi(4\phi'+B^l{}_{,l})+2\phi(C_{kl,}{}^k\phi_,{}^l-C^l{}_{l,k}\phi_,{}^k+2\phi_{,k}\phi_,{}^k)
+2B^k{}'(2C_{kl,}{}^l\phi-\phi C^l{}_{l,k})\nn\\
&\scriptstyle+B^k\Big[C^m{}_{m,}{}^l(B_{k,l}-B_{l,k}-2C_{kl}')+C^{lm}{}_{,m}(4C_{kl}'-2B_{k,l})
+B^l{}_{,l}(2C_{km,}{}^m-C^m{}_{m,k}-B_k'+\phi_{,k})\nn\\
&\scriptstyle +B_{m,l}(C^l{}_{k,}{}^m-3C^m{}_{k,}{}^l+C^{lm}{}_{,k})+C_{lm}'(C_k{}^m{}_,{}^l-C^{lm}{}_{,k})
-\phi'(C^l{}_{l,k}-2C_{kl,}{}^l)+B^l{}_{,k}C_{ml,}{}^m \nn \\
& \scriptstyle+2\phi\Big(2C_{kl,}{}^l{}'-2C^l{}_{l,k}'+B_{l,k}{}^l-\nabla^2B_k
-4\H(2B_k'+C^l{}_{l,k}-2C_{kl,}{}^l-2\phi_{,k})\Big)\Big].
\end{align}